\documentclass[article,nofootinbib,10pt,floats,aps]{revtex4}

\usepackage{amsmath}
\usepackage{amssymb}
\usepackage{graphicx}
\usepackage{epsfig}
\usepackage{appendix}
\usepackage{bm}
\usepackage{units}
\usepackage{subfigure}
\usepackage{hyperref}

\renewcommand{\Re}[1]{\hbox{Re} \left( #1 \right)}

\newcommand{\be}{\begin{equation}}
\newcommand{\ee}{\end{equation}}
\newcommand{\bea}{\begin{eqnarray}}
\newcommand{\eea}{\end{eqnarray}}
\newcommand{\ben}{\begin{enumerate}}
\newcommand{\een}{\end{enumerate}}
\newcommand{\bit}{\begin{itemize}}
\newcommand{\eit}{\end{itemize}}

\newcommand{\la}[1]{\label{#1}}

\newcommand{\Eq}[1]{Eq.~(\ref{#1})}
\newcommand{\Eqs}[2]{Eqs.~(\ref{#1}, \ref{#2})}

\newcommand{\Sec}[1]{Sec.~\ref{#1}}

\newcommand{\Fig}[1]{Fig.~\ref{#1}}

\def\nn{\nonumber \\ }
\def\T{\mathrm T}

\newcommand{\vv}[1]{\mathbf #1}						% 3-vector
\newcommand{\MM}[1]{\mathsf #1}						% Matrix

\newcommand{\bert}{\raise-0.45mm\hbox{\Large$\Box$}}	%D'Alembertian

\begin{document}

\preprint{1109.6640 [physics.class-ph]}

\title{Self-oscillation}
 
\author{Alejandro Jenkins}\email{jenkins@hep.fsu.edu}

\affiliation{High Energy Physics, 505 Keen Building, Florida State University, Tallahassee, FL 32306-4350, USA}

%%%%%%%%%%
%%% ABSTRACT
%%%%%%%%%%

\begin{abstract}

Physicists are very familiar with forced and parametric resonance, but usually not with self-oscillation, a property of certain dynamical systems that gives rise to a great variety of vibrations, both useful and destructive.  In a self-oscillator, the driving force is controlled by the oscillation itself so that it acts in phase with the velocity, causing a negative damping that feeds energy into the vibration: no external rate needs to be adjusted to the resonant frequency.  The famous collapse of the Tacoma Narrows bridge in 1940, often attributed by introductory physics texts to forced resonance, was actually a self-oscillation, as was the swaying of the London Millennium Footbridge in 2000.  Clocks are self-oscillators, as are bowed and wind musical instruments.  The heart is a ``relaxation oscillator,'' i.e., a non-sinusoidal self-oscillator whose period is determined by sudden, nonlinear switching at thresholds.  We review the general criterion that determines whether a linear system can self-oscillate.  We then describe the limiting cycles of the simplest nonlinear self-oscillators, as well as the ability of two or more coupled self-oscillators to become spontaneously synchronized (``entrained'').  We characterize the operation of motors as self-oscillation and prove a theorem about their limit efficiency, of which Carnot's theorem for heat engines appears as a special case.  We briefly discuss how self-oscillation applies to servomechanisms, Cepheid variable stars, lasers, and the macroeconomic business cycle, among other applications.  Our emphasis throughout is on the energetics of self-oscillation, often neglected by the literature on nonlinear dynamical systems. \\



\end{abstract}

\maketitle

\tableofcontents

\newpage

\section{Introduction}
\la{sec:intro}

{\scriptsize
\begin{tabbing}
\hspace{0.65 \textwidth}

\= My pulse, as yours, doth temperately keep time, \\
\> And makes as healthful music: it is not madness  \\
\> That I have utter'd \\
\` ---{\it Hamlet}, act 3, scene 4
\end{tabbing}}

\subsection{What and why}
\la{sec:what-why}

Self-oscillation is the generation and maintenance of a periodic motion by a source of power that lacks a corresponding periodicity:  the oscillation itself controls the phase with which the power source acts on it.  Self-oscillation is also known as ``maintained,'' ``sustained,'' ``self-excited,'' ``self-induced,'' ``spontaneous,'' ``autonomous,'' and (in certain contexts) ``hunting'' or ``parasitic'' vibration.\footnote{The variants ``self-maintained,'' ``self-sustained,'' ``self-sustaining,'' and ``self-exciting'' also occur.  This diversity of terminology probably reflects the lack of an authoritative textbook treatment of the subject from the point of view of elementary classical physics and wave mechanics (see Appendix \ref{sec:sources}), so that researchers tend to use whichever term is more prevalent in their own field.  The term ``self-oscillation'' (also translated as ``auto-oscillation'') was coined by Soviet physicist Aleksandr Aleksandrovich Andronov (1901--1952) \cite{Andronov,math-encyclopedia}.  Andronov, a student of Leonid Mandelstam, was a professor at Gorky State University and, later in life, a deputy of the Supreme Soviet of the USSR.  He and his associates made important contributions to the mathematical theory of the stability of nonlinear dynamical systems \cite{Andronov-history}.}  The main purpose of this article is to bring self-oscillation to the attention of theoretical physicists, to whom it is not usually taught in any systematic way.  We shall, therefore, emphasize not only its practical importance, including its applications in mechanical engineering, acoustics, electronics, and biomechanics (perhaps even in finance and macroeconomics) but also why it is conceptually fascinating.

Self-oscillators are distinct from resonant systems (including both forced and parametric resonators), in which the oscillation is driven by a source of power that is modulated {\it externally}.  Many textbooks in both introductory classical mechanics and acoustics treat resonant systems at length but fail to adequately characterize self-oscillators, in some cases even labeling as resonant phenomena that are actually self-oscillatory.  The most notorious instance of such a mischaracterization concerns the wind-powered galloping of a suspension bridge, as immortalized by the video footage of the torsional motion of the Tacoma Narrows Bridge that caused in to collapse in 1940. \cite{Tacoma-movie}

Many important and familiar natural phenomena, such as the heartbeat, the firing of neurons, ocean waves, and the pulsation of variable stars, are self-oscillatory.  Furthermore, self-oscillation has long been an essential aspect of human technology.  Turbines, clocks, many musical instruments (including the human voice), heat engines, and lasers are self-oscillators.  Indeed, as we shall see, only a self-oscillator can generate and maintain a regular mechanical periodicity without requiring a similar external periodicity to drive it.  Nonetheless, the theoretical question of how a steady source of power can give rise to periodic oscillation was not posed in the context of Newtonian mechanics until the 19th century and the literature on the subject has remained disjointed and confusing.

In general, the possibility of self-oscillation can be diagnosed as an {\it instability} of the linearized equation of motion for perturbations about an equilibrium.  But the linear equations then yield an oscillation whose amplitude grows exponentially with time.  It is therefore necessary to take into account nonlinearities in order to determine the form of the {\it limiting cycle} attained by the self-oscillator.  The study of self-oscillation is therefore, to a large extent, an application of the theory of nonlinear vibration, a subject which has been much more developed in mathematics and in engineering than in theoretical physics.

The linear instability of a self-oscillator is usually caused by a {\it positive feedback} between the oscillator's motion and the action of the power source attached to that oscillator.  For example, in the case of the Tacoma Narrows disaster, the bridge's large torsional motion resulted from a feedback between that motion and the formation of turbulent vortices in the wind flowing past the bridge.  The study of self-oscillators therefore connects naturally with the treatment of feedback and stability in control theory, a rich field in applied mathematics of great relevance to modern technology.

The main difference between our own approach in this review and the treatments of self-oscillation that exist in the literature is that we will focus on the {\it energetics} of self-oscillators, rather than simply characterizing the solutions to the corresponding nonlinear equations of motion.  By studying the flow of energy that powers self-oscillators we hope to develop a more concrete physical understanding of their operation, to make the subject more interesting and accessible to physicists, and to obtain a few original results.

\subsection{Plan for this review}
\la{sec:plan}

Section \ref{sec:perpetual} introduces the subject by revisiting the hoary question of perpetual motion.  The earliest mathematical treatment of self-oscillation (at least as far as we have been able to determine) appears in a brief paper published in 1830 by G.~B.~Airy, in which he characterized the vibration of the human vocal chords as a form of ``perpetual'' motion compatible with the laws of Newtonian physics.  A discussion of Airy's work gives us the opportunity to introduce the key concept of {\it negative damping}

Section \ref{sec:resonance} stresses, both theoretically and through concrete examples, the distinction between self-oscillation and the phenomena of forced and parametric resonance.  Meanwhile, \Sec{sec:feedback} underlines the role of feedback in self-oscillators and illustrates important qualitative features of their performance.  Sections \ref{sec:perpetual} -- \ref{sec:feedback} form a self-contained presentation in which the main features of self-oscillation are reviewed at a level that should be accessible and interesting to an advanced undergraduate physics student.

Section \ref{sec:control} characterizes self-oscillation with greater mathematical precision, covering both the linear instability at equilibrium and the nonlinear limiting cycles.  Although it is also self-contained, the purpose of this section is not to provide a thorough recapitulation of the mathematical theory of self-oscillating systems, but rather to illuminate the close connection of the study of self-oscillation with control theory.  We also explore how a more physical approach to self-oscillators and related dynamical systems can complement the purely mathematical treatment that has dominated the literature.

Section \ref{sec:motors} treats turbines and motors as self-oscillators, focusing on their ability to convert energy inputted at one frequency (usually zero) into work outputted at another, well-defined frequency.  We state and prove a general result about the maximum efficiency attainable by motors, of which Carnot's theorem for heat engines appears as a special case.  The arguments made in \Sec{sec:Carnot} are, as far as we know, largely original and might provide a somewhat novel perspective on certain aspects of thermodynamics.

Section \ref{sec:applications} covers other specific instances of self-oscillation.  This is an eclectic selection, intended to underline the broad applicability of the concepts covered in this article.  The description of lasers as self-oscillators in \Sec{sec:lasers}, opens questions about the extension of self-oscillation and related concepts to quantum systems (one such system, the Josephson junction, is discussed briefly in Appendix \ref{sec:Josephson}).  Section \ref{sec:business} explores the use of the concept of self-oscillation in economics and raises broader issues concerning the use of physics-inspired models to describe markets and other human organizations.

Though the presentation is intended to be largely self-contained, important results that are available in standard texts are only referenced.  The emphasis is on bringing out and organizing key concepts, especially when these are not stressed in the existing literature.  Though this is not a historically-oriented review, historical episodes and curiosities will be discussed when useful in guiding or illustrating the conceptual discussion.  Appendix \ref{sec:history} gives a brief overview of the history proper.  This might help some readers to place the subject in context and to better understand why it has not made it into physics textbooks, at least not in the form in which we approach it.  Appendix \ref{sec:sources} points out the sources most useful to a physics student wishing to learn the subject systematically.

\section{Perpetual motion}
\la{sec:perpetual}

\subsection{Energy conservation}
\la{sec:energy}

Let us begin by considering the old chimera of perpetual motion.  The state of a classical system (i.e., of an arbitrary machine) may be characterized by an $N$-dimensional, generalized-coordinate vector $\vv q$, with components $q_i$.  In the Euler-Lagrange formalism of classical mechanics, the equation of motion for the system is expressed as
\be
\frac{d}{dt} \left( \frac{\partial L}{\partial \dot q_i} \right) = \frac{\partial L}{\partial q_i}
\la{eq:EL}
\ee
where the overdot indicates the derivative of $q_i$ with respect to time $t$, and \hbox{$L \equiv T - V$} is the system's Lagrangian (where $T$ is the kinetic and $V$ the potential energy, expressed as functions of $\dot{\vv q}$, $\vv q$, and $t$).  If $L$ is not an explicit function of time, then, by \Eq{eq:EL}, the energy
\be
H \equiv \sum_i \dot q_i \frac{\partial L}{\partial \dot q_i} - L
\ee
is a constant of the motion, \hbox{$dH/dt = 0$} (see \cite{Goldstein-energy} for the conditions under which $H = T + V$).  A perpetual motion machine ``of the first kind,'' which would have greater energy every time it comes to a configuration characterized by the same $\vv q^{(0)}$, therefore requires a time-dependent Lagrangian.  The fundamental laws of Nature are believed to be time-independent and energy conservation is the reason usually given why perpetual motion of the first kind is impossible.

\subsection{Irreversibility}
\la{sec:irreversibility}

That same argument, however, implies that the machine should run forever, but no machine that runs with exactly constant energy has ever been built.  The reason is that, even though the Lagrangian of a closed system (such as the Universe as a whole) is believed to be time-independent, the Lagrangian of an open system (such as any conceivable machine that could be built by humans) will be time-dependent: mechanical energy is lost as heat leaks into the environment, causing the machine to wind down.\footnote{A device whose mechanical energy is exactly conserved is sometimes called a perpetual-motion machine ``of the third kind,'' though more commonly the term perpetual motion is reserved for machines that can do useful work, such as pulling up a weight.  On the history of the concept of ``perpetual motion,'' see \cite{Angrist}.} 

A {\it cyclic} machine that runs merely by absorbing heat from the environment and converting it into useful work is called a perpetual motion device ``of the second kind'' \cite{Fermi}.  According to Lord Kelvin's formulation of the second law of thermodynamics, such a machine is impossible, but this is just the systematic statement of an observed fact (see \cite{Fermi,Feynman-thermo}).\footnote{In the same spirit, Stevinus, the 16th century Flemish mathematician and military engineer, correctly derived the forces acting on masses rolling on inclined planes from the assumption that perpetual motion is impossible.  He was so proud of his argument that he had it inscribed on his tombstone.  Feynman jokes that ``if you get an epitaph like that on your gravestone, you are doing fine.'' \cite{Stevinus}}  The underlying reason why work can be entirely converted into heat, but heat cannot be purely converted into work, concerns the fascinating problem of the ``arrow of time,'' which still presents conceptual difficulties for theoretical physics (cf.\ \cite{Feynman-time,Wiener-time,Schrodinger-time,Carroll}).

Evidently, energy from the environment {\it may} flow into the machine and cause it to do useful work.  For instance, the water in a stream turns the wheel of a mill and heat from burning coal powers a steam engine.  This article shall focus on self-oscillation, an important type of externally-powered motion.  As we shall explain, self-oscillators are characterized by the fact that {\it their own motion} controls the phase with which the external power source drives them.  They are therefore, in a certain sense, self-driven (although not, of course, self-powered).

\subsection{Overbalanced wheels}
\la{sec:overbalanced}

A perennially popular idea for a perpetual motion machine is the ``overbalanced wheel,'' in which weights are attached to a wheel in such a way that the turning is supposed to shift the weights and keep the left and right half of the wheel persistently unbalanced.  Early examples of such proposed devices appear in the work of Indian astronomer Bh\=askara II and French draftsman Villard de Honnecourt, in the 12th and 13th centuries CE respectively \cite{Bhaskara}.  Figure \ref{fig:Worcester-wheel} shows another such a wheel, conceived in the 17th century by the Marquess of Worcester \cite{Worcester}.

That a device of this kind cannot possibly work should be obvious to a modern student, since a machine powered by gravity alone must keep lowering its center of mass in order to accelerate or to maintain its velocity in the presence of friction.\footnote{See \cite{Worcester-TPT} for an explanation of the non-operation of Worcester's wheel in terms of the explicit computation of the torques.}  Nonetheless, efforts to construct overbalanced wheels persist even today.\footnote{The documentary {\it A Machine to Die For: The Quest for Free Energy}, released in 2003 and broadcast by Australian television \cite{ToDie}, credulously showcases the work of various fringe researchers, while the few skeptics interviewed fail to adequately communicate any of the relevant physical concepts.  One of the devices featured is a large overbalanced wheel built by French retired mechanic Aldo Costa outside his home in Villiers-sur-Morin \cite{Costa}.  That wheel seems notable only for being so large that it can be turned by the wind.}

\begin{figure} [t]
\begin{center}
	\includegraphics[width=0.35 \textwidth]{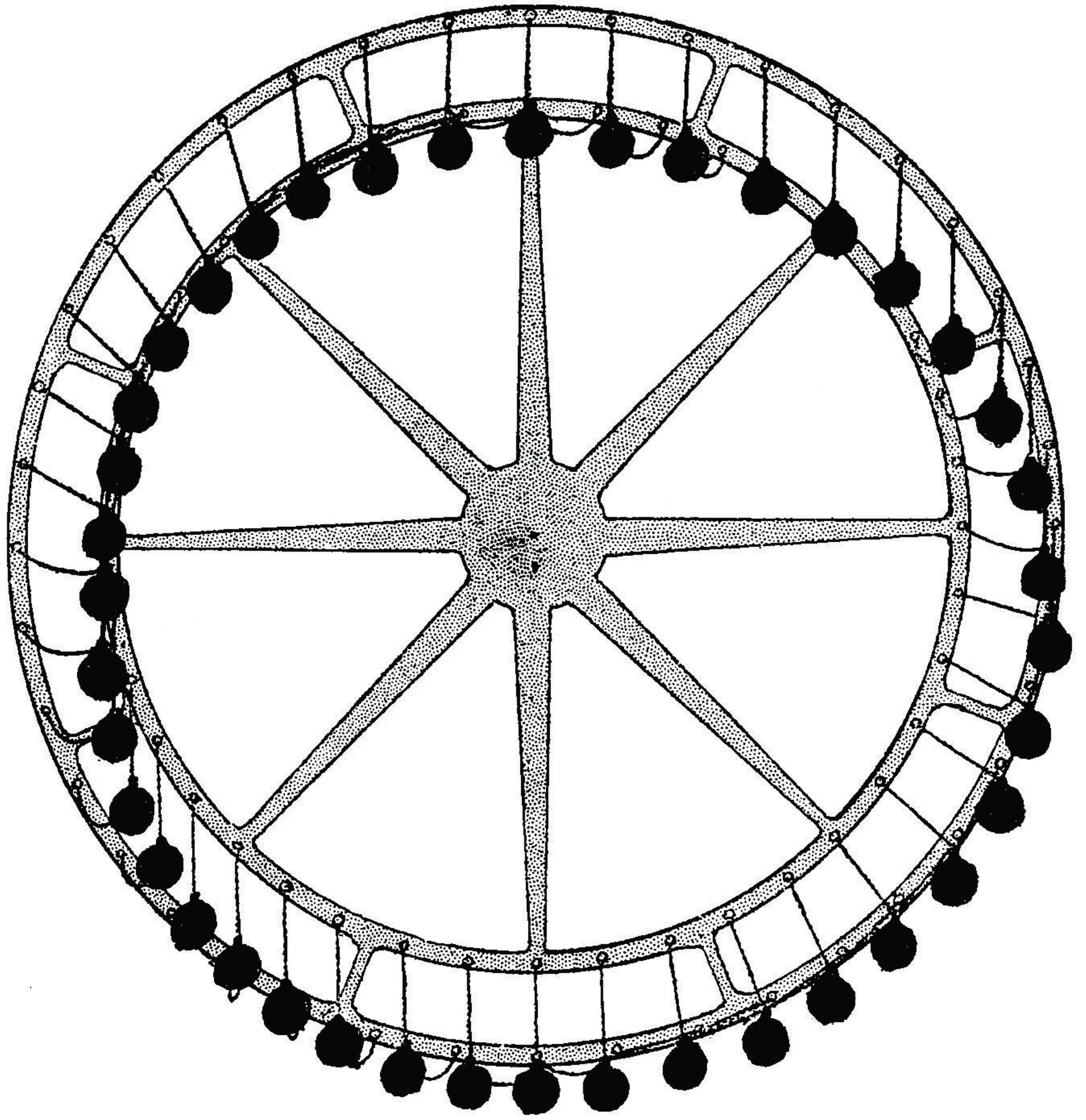}
\end{center}
\caption{\small ``Overbalanced'' wheel conceived by Edward Somerset, 2nd Marquess of Worcester, {\it circa} 1640 \cite{Worcester}.  Though the weights on the right always hang farther from the center than the weights on the left, in equilibrium two weights hang vertically in line with the wheel's center, while the twenty weights on the left balance the torque exerted by the eighteen weights on the right.  The drawing is by R. T. Gould (Fig.~14 in \cite{Gould}) and is used here with permission of his heirs. \la{fig:Worcester-wheel}}
\end{figure}

A curious episode was the exhibition of such a machine by German inventor J. E. E. Bessler, {\it alias} Orffyreus, in the early 18th century, before various political and scientific dignitaries.  Orffyreus's mechanism was hidden from view by the wheel's casing, but competent observers were unable to detect a fraud before the inventor himself destroyed the wheel in 1727, drifting thereafter into obscurity.  In an essay on the subject, Rupert T. Gould (a 20th century English naval officer and amateur scholar, best remembered for his work on John Harrison's marine chronometers\footnote{Gould is co-protagonist of the remarkable TV adaptation of Dava Sobel's {\it Longitude} \cite{Sobel}, itself a modern classic of the history of science for a general audience.  Onscreen, Gould is portrayed as a sensitive man whose life is disrupted by nervous breakdowns and his consuming obsession with restoring Harrison's historic timepieces \cite{Longitude-TV}.  The circumstances of the scandal that cost Gould his marriage and his employment in the Royal Navy ---which are inevitably more complex than what is represented in the miniseries--- are explored in detail in \cite{Betts}.  Gould's credulity about such things as Nostradamus's prophecies (see \cite{Gould}) might disappoint some admirers of the TV character.  In an essay on the novels of Charles Reade, George Orwell wrote, not without admiration, that their appeal is ``the same as one finds in R.~Austin Freeman's detective stories or Lieutenant-Commander Gould's collections of curiosities ---the charm of useless knowledge.'' \cite{Orwell}}) admits that such a device would have required ---contrary to its inventor's claims--- an external source of power other than gravity, but also deems compelling the surviving testimonies of the wheel's successful operation \cite{Gould}.

\subsection{Voice as perpetual motion}
\la{sec:Airy}

Towards the end of his discussion of Orffyreus's wheel, Gould quotes extensively from an 1830 paper by mathematician and astronomer George Biddell Airy (who succeeded a few years later to the post of Astronomer Royal), titled ``On certain Conditions under which a Perpetual Motion is possible'' \cite{Airy-perpetual}.  In fact, Airy's brief paper has nothing to do with overbalanced wheels, but is rather an early attempt to understand the operation of the human voice, motivated by Robert Willis's pioneering research on that subject \cite{Willis-larynx}.\footnote{The Rev.~Robert Willis (1800--1875) was the first English university professor to do significant research in mechanical engineering, as well as a distinguished architectural historian.  He was the grandson of Dr.~Francis Willis, the eccentric physician who attended King George III during his madness. \cite{Willis-bio}}

Consider a long tube, open at both ends, in the form of a rectangular prism, much longer than it is wide and much wider than it is tall, with a side at one of the ends replaced by a taut, flexible membrane, as shown in \Fig{fig:Willis}.  Without air flowing through the tube, the membrane sits flat and horizontal, as represented in the illustration by position $H$.  Willis found experimentally that, for a steady airflow past the membrane, there is a position $A$, slightly below $H$, at which the membrane is in equilibrium.  If it sits below that, as in $B$, the air will push the membrane out.  If the membrane sits above the equilibrium, as in $C$, it will be pulled in.\footnote{It is tempting to explain this pulling in by invoking Bernoulli's theorem, as many elementary texts do when discussing the lift on an airplane wing, but such an argument is flawed, for reasons that are clearly explained in \cite{flight}.}  Turning on the airflow may therefore cause the membrane to oscillate about the stable equilibrium at $A$.

If we picture the vocal chords as twin membranes vibrating in a steady air current, how do they draw energy in order to sustain that vibration and produce a persistent sound?  Why is the vibration about $A$ not damped out by friction and by the resistance of the air?  Both Willis and Airy noted that the answer must lie in the {\it delay} with which the stream of air exerts the restorative force that would act if the displacement were fixed.  According to Airy,
\begin{quote}
Mr.~Willis explains this [sustained vibration] by supposing that {\it time} is necessary for the air to assume the state and exert the force corresponding to any position of the [membrane]: which is nearly the same as saying that the force depends on the position of the [membrane] at some previous time. \cite{Airy-perpetual}
\end{quote}

\begin{figure} [t]
\begin{center}
	\subfigure[]{\includegraphics[width=0.5 \textwidth]{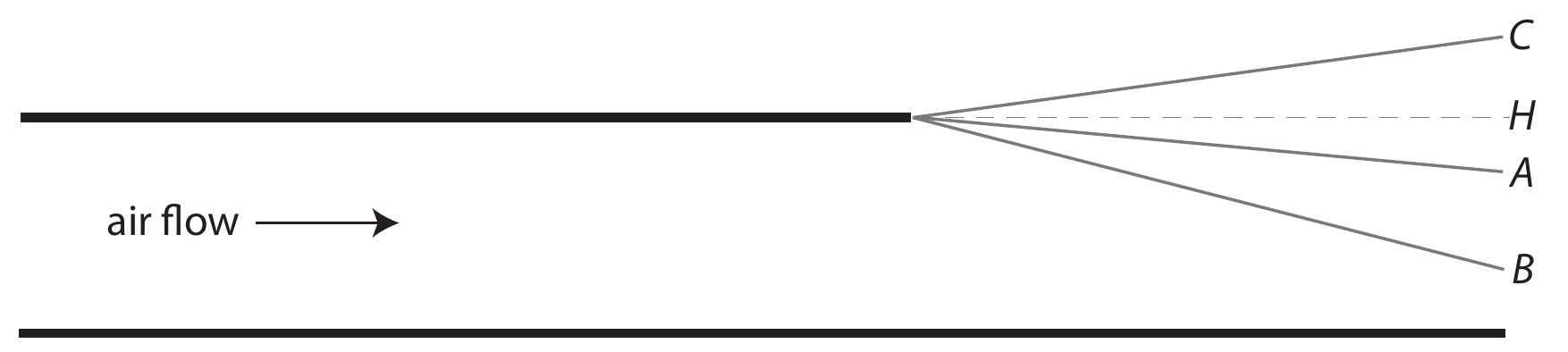}} \hskip 2 cm
	\subfigure[]{\includegraphics[width=0.3 \textwidth]{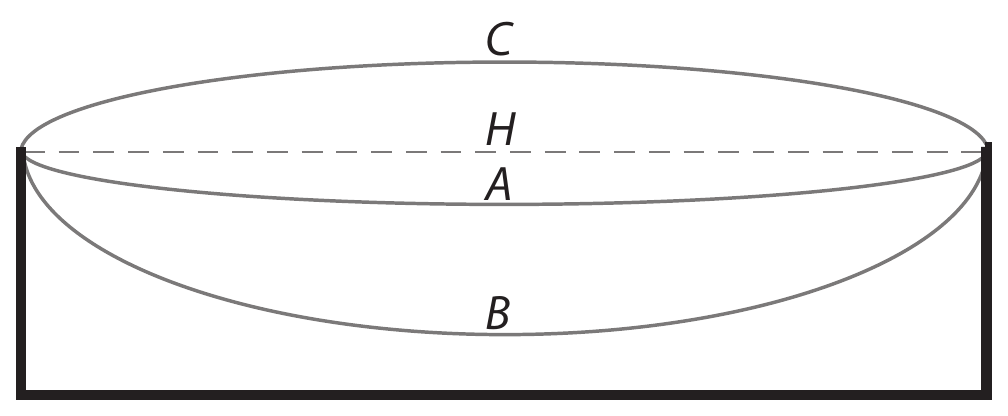}}
\end{center}
\caption{\small An open tube, shaped like a rectangular prism, with one of the sides at its end replaced by a flexible membrane (drawn in gray).  In the lateral view (a) the air flows from left to right, while in the head-on view (b) the air flows out of the page.  When the air flow starts, the membrane is initially deflected in, away from the horizontal position $H$.  For steady flow, the membrane is in equilibrium at $A$.  At $B$ the air pushes it out, while at $C$ the membrane is pulled in.  Images adapted from Fig.~22 of \cite{Willis-larynx}. \la{fig:Willis}}
\end{figure}

Airy therefore proposed modeling the vocal chords as a harmonic oscillator in which part of the restoring force depends on the displacement $q$ at an earlier time:
\be
\ddot q (t) = - a \cdot q (t) - b \cdot q (t - c) ~.
\la{eq:Airy}
\ee
He then showed, using first-order perturbation theory, that for $0 < b \ll a$ and $0 < c \sqrt{a} < \pi$, the amplitude of the oscillation grows after each period.  This is what he identified as ``perpetual motion.''  Clearly, the energy of the oscillator described by \Eq{eq:Airy} is not conserved, because the time-delayed force, $-b q (t - c)$, cannot be expressed as the derivative of any potential $V(q)$.  This is why the kinetic energy of the oscillator can be greater each successive time it passes through the equilibrium position $q=0$, as in a perpetual motion machine of the first kind.

Neither Willis nor Airy offered any detailed description of the fluid dynamics responsible for the variable force that the air exerts on the membrane.  Airy did not, for instance, justify making the delay $c$ in \Eq{eq:Airy} fixed, nor did he suggest how the parameters $a,b$ and $c$ in that equation might be related to the details of the setup shown in \Fig{fig:Willis}.  Airy merely offered an example of how the maintenance of a vibration could be modeled by a simple equation of motion that does not conserve energy, and argued for its qualitative plausibility as a model for the vocal chords.

\subsection{Delayed action}
\la{sec:delays}

If $b = 0$ in \Eq{eq:Airy} then the force always pulls the oscillator back to its equilibrium position.  But if \hbox{$0 < b \ll a$} and $0 < c \sqrt{a} < \pi$, then as the oscillator passes through $q=0$ the delayed force does not reverse its sign for a while, and therefore pushes the oscillator {\it away} from equilibrium.  The bigger the amplitude of the oscillation, the stronger this pushing grows.  Thus, the motion of the delayed oscillator may be understood as an instance of {\it positive feedback}:  the oscillator drives itself, leading to an exponentially growing amplitude.\footnote{Feedback is usually thought of as the process of taking the output of a system (in this case, the displacement $q$), subjecting it to some processing (in this case, delaying it by $c$), and then inputting the result back into the system (as represented in this instance by the term proportional to $q(t-c)$ in \Eq{eq:Airy}).  Pippard suggests that, in general, it might be better to think of feedback as a series of cross-links between the elements that compose a dynamical system (in this case, the membrane and the air flow), which forces the system to behave in the only way consistent with the relations dictated by those linkages \cite{Pippard-feedback}.  Feedback is said to be ``positive'' when it encourages the deviation of the system from some reference state or trajectory, ``negative'' when it discourages that deviation.}  Note that this positive feedback is greatest for \hbox{$c \sqrt{a} = \pi / 2$}, because then the time-delayed force always pushes in the same direction in which the oscillator is moving.

In the mid-19th century, Helmholtz invented a ``fork-interrupter,'' in which a steel tuning fork rings persistently as the movement of one of its prongs switches an electromagnet on and off \cite{Helmholtz-fork}.  Lord Rayleigh, who appears to have been unaware of the work by Willis and Airy that we have summarized in \Sec{sec:Airy}, echoes their insight early in his monumental treatise on acoustics, {\it The Theory of Sound} (first published in 1877), when he explains that Helmholtz's fork-interrupter is a ``self-acting instrument,'' whose operation is ``often imperfectly apprehended,'' and that ``any explanation which does not take account of the retardation of the [magnetic force with respect to the position of the prong] is wholly beside the mark \cite{Rayleigh-fork}.''  Doorbell buzzers work on this same principle.\footnote{An electrical circuit analogy for the self-oscillation of the vocal chords was worked out in \cite{Wegel}, though without reference to the early work of Willis and Airy.  For a more recent and detailed discussion of self-oscillating tuning forks, see, e.g., \cite{Groszkowski-fork}.}

Green and Unruh point out in \cite{Unruh} an even more elementary example of a self-oscillation associated with a time-delay: the audible tone produced by blowing air across the mouth of a bottle.  The resulting tone is sustained because of the delay in the adjustment of the airflow in the neck of the bottle to the oscillating pressure inside.  This allows more air to be drawn in when the internal pressure is high and less air to be drawn in when the internal pressure is low, thus feeding energy into the oscillation of the pressure.\footnote{In 1942, Minorsky pioneered the systematic mathematical study of the stability of dynamical systems with finite delays \cite{Minorsky}.  Bateman reviews the history of the subject in sec.\ 3.1 of \cite{Bateman}, though he is unaware of the early work of Willis and Airy.  Recently, Atiyah and Moore have speculated on the possible use of time-shifted equations of motion in relativistic field theories \cite{Atiyah}.}

\subsection{Negative damping}
\la{sec:negative-damping}

For small $c$ (i.e., $0 < c \sqrt{a} \ll 1$), \Eq{eq:Airy} can be Taylor-expanded into the equation of motion for a {\it negatively damped} linear oscillation:
\be
\ddot q - \gamma \dot q + \omega^2 q = 0~,
\la{eq:Airy-damped}
\ee
where $\omega^2 = a + b$ and $\gamma = b c$.  Negative damping corresponds to a component of the force acting in phase with the velocity $\dot q$.  The faster the oscillator moves, the more it is pushed along the direction of its motion.  The oscillator thus keeps drawing energy from its surroundings.\footnote{Note that \Eq{eq:Airy-damped} is the equation of motion for the time-dependent Lagrangian $L = (\dot q^2 - \omega^2 q^2) \cdot \exp(-\gamma t)$.}  The amplitude of the oscillation grows exponentially with time, until it becomes so large that nonlinear effects become relevant and somehow determine a limiting amplitude.  It is this which gives a regular self-oscillation.

At the end of \Sec{sec:Airy} we pointed out that the positive feedback in \Eq{eq:Airy} is maximal when $c \sqrt{a} = \pi / 2$.  In that case, even though $c$ is not small, the resulting motion may be described by \Eq{eq:Airy-damped}, with $\gamma = b$ and $\omega^2 = a$, since for sinusoidal motion $-\dot q$ has a phase of $-\pi / 2$ relative to $q$.

Self-oscillation describes not just the human voice, but also clocks, bowed and wind musical instruments, the heart, motors, and the theory of lasers, among other important kinds of mechanical, acoustic, and electromagnetic oscillations.  Surprisingly, one searches modern textbooks in theoretical physics (both elementary and advanced) largely in vain for discussion of this interesting and important phenomenon.\footnote{Most elementary physics textbooks treat only undamped, damped, and forced linear oscillations.  More advanced texts often discuss parametric resonance as well (cf.\ \cite{LL-parametric, Goldstein-parametric, Jose-parametric, Hand-parametric}), but when self-oscillation is treated at all it is usually only in the context of a mathematical discussion of the limit cycles of the van der Pol equation (cf.\ \cite{Jose-vdP,Goldstein-vdP,Pain-vdP}), a subject that we shall review in \Sec{sec:limits}.}

\section{Resonance versus self-oscillation}
\la{sec:resonance}

\subsection{Forced resonance}
\la{sec:forced-resonance}

Self-oscillation is distinct from the conceptually more familiar phenomenon of forced resonance.  In the case of a forced resonance, the damping is positive and there is a time-dependent driving term on the right-hand side of the equation of motion:
\be
\ddot q + \gamma \dot q + \omega_0^2 q = \frac{F_0}{m} \cos \left(\omega_d t \right)~.
\la{eq:forced}
\ee
The driving force produces a maximum amplitude of oscillation when the driving frequency $\omega_d$ is tuned to match the natural frequency $\omega_0$ of the undriven oscillator.  Even in the linear regime, the amplitude of a resonant oscillator diverges only if the damping vanishes.  Furthermore, the amplitude of an undamped forced resonator diverges {\it linearly} with time (see \cite{LL-resonance}), not exponentially like the amplitude of a linear self-oscillator.

A popular classroom demonstration of forced resonance is to shatter a wine glass by playing its resonant note loudly enough on a nearby speaker.  An ordinary radio tuner works by having the listener adjust the resonant frequency of an $LC$ circuit to match the frequency at which the desired radio station is being broadcast, so that the corresponding signal drives the circuit resonantly.  The radio tuner is an example of a bandpass filter (i.e., a device that allows only frequencies in a narrow band to pass through it), as represented by \Fig{fig:LC}.

\begin{figure} [t]
\begin{center}
	\includegraphics[width=0.4 \textwidth]{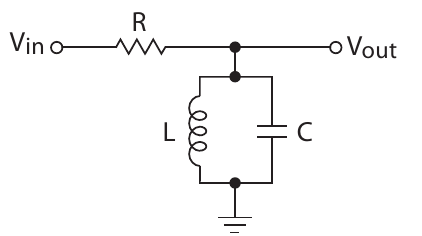}
\end{center}
\caption{\small Diagram of an $RLC$ electric circuit used as a bandpass filter.  The amplitude of $V_{\rm out}$ is maximized when the angular frequency of $V_{\rm in}$ is close to $1/\sqrt{LC}$. \cite{H&H-LC} \la{fig:LC}}
\end{figure}

\subsection{Work on oscillator}
\la{sec:work}

The net energy that an oscillator gains over a complete period $\tau$ of its motion is
\be
W_{\rm net} = \int_0^\tau dt \, \dot q F~,
\label{eq:W-net}
\ee
where $F$ is the external force.  Thus, if
\be
F = F_0 \cos \left(\omega_d t \right)
\la{eq:F-ext}
\ee
and
\be
q = A \cos \left( \omega_d t - \phi \right)
\la{eq:lag}
\ee
then
\be
W_{\rm net} = \pi A F_0 \sin \phi ~,
\la{eq:W-AF}
\ee
so that energy can steadily flow into the oscillator only if the relative phase between the external force and the oscillation is $0 < \phi < \pi$.  The most efficient transfer of power occurs when $\phi = \pi /2$, when $F$ leads $q$ by a quarter of a period.  For a forced resonator this {\it only} happens when \hbox{$\omega_d = \omega_0$}.  Moreover, if $\gamma = 0$ in \Eq{eq:forced}, then $\sin \phi =0$ for all $\omega_d \neq \omega_0$ (see \cite{Georgi-resonance}).  On the other hand, for a self-oscillating equation of motion such as \Eq{eq:Airy-damped} the phase shift is automatically $\phi = \pi /2$ by virtue of the form of the negative damping term $-\gamma \dot q$.  We will have much more to say in this article about how such a negative damping can arise in an actual physical system, without reversing the thermodynamic arrow of time.

For an undamped forced resonator, the magnitude of $F_0$ in \Eq{eq:W-AF} is fixed, as given by the inhomogeneous term in \Eq{eq:forced}.  Thus, by \Eq{eq:W-AF} and using the fact that the energy $E$ is proportional to the square of the amplitude $A$,
\be
\dot A \propto \frac{\dot E}{\sqrt{E}} \propto \frac{W_{\rm net} \, \omega_d}{A} \propto A^0~,
\ee
which implies that $A$ grows linearly with time.  On the other hand, in a self-oscillator as described by \Eq{eq:Airy-damped}, $F_0 = \gamma m \omega_d A$ scales linearly with $A$, so that
\be
\dot A \propto A~,
\ee
giving an exponential growth of the amplitude.  Thus, the exponential increase of $A$ in a linear self-oscillator reflects the fact that the motion drives itself.\footnote{Note also that the solutions to \Eq{eq:Airy-damped} must be expressible as the real part of a complex exponential, because of linearity and time-translation invariance (see \cite{Georgi-exponential}); those conditions are broken by the inhomogeneous term in \Eq{eq:forced}.}

\subsection{Flow-induced instabilities}
\label{sec:flow-induced}

Many physics texts and popular accounts attribute to forced resonance phenomena that are actually self-oscillatory.  The most notorious case is the large torsional oscillation (``galloping'') that led to the collapse of the suspension bridge over the Tacoma Narrows, in the state of Washington, in 1940 (see \Fig{fig:Tacoma}).  When it fell, the bridge was exposed to steady winds of 68 km/h (42 mph).  At that wind speed and given the dimensions of the bridge, the Strouhal frequency of turbulent vortex shedding is about 1 Hz and therefore could not have been forcing the bridge into an oscillation with the frequency observed (and documented in film) of about 0.2 Hz.\footnote{The Strouhal frequency, at which a steady flow hitting a solid obstacle sheds turbulent vortices, will be discussed in \Sec{sec:violin-aeolian}.}

\begin{figure} [t]
\begin{center}
	\includegraphics[width=0.5 \textwidth]{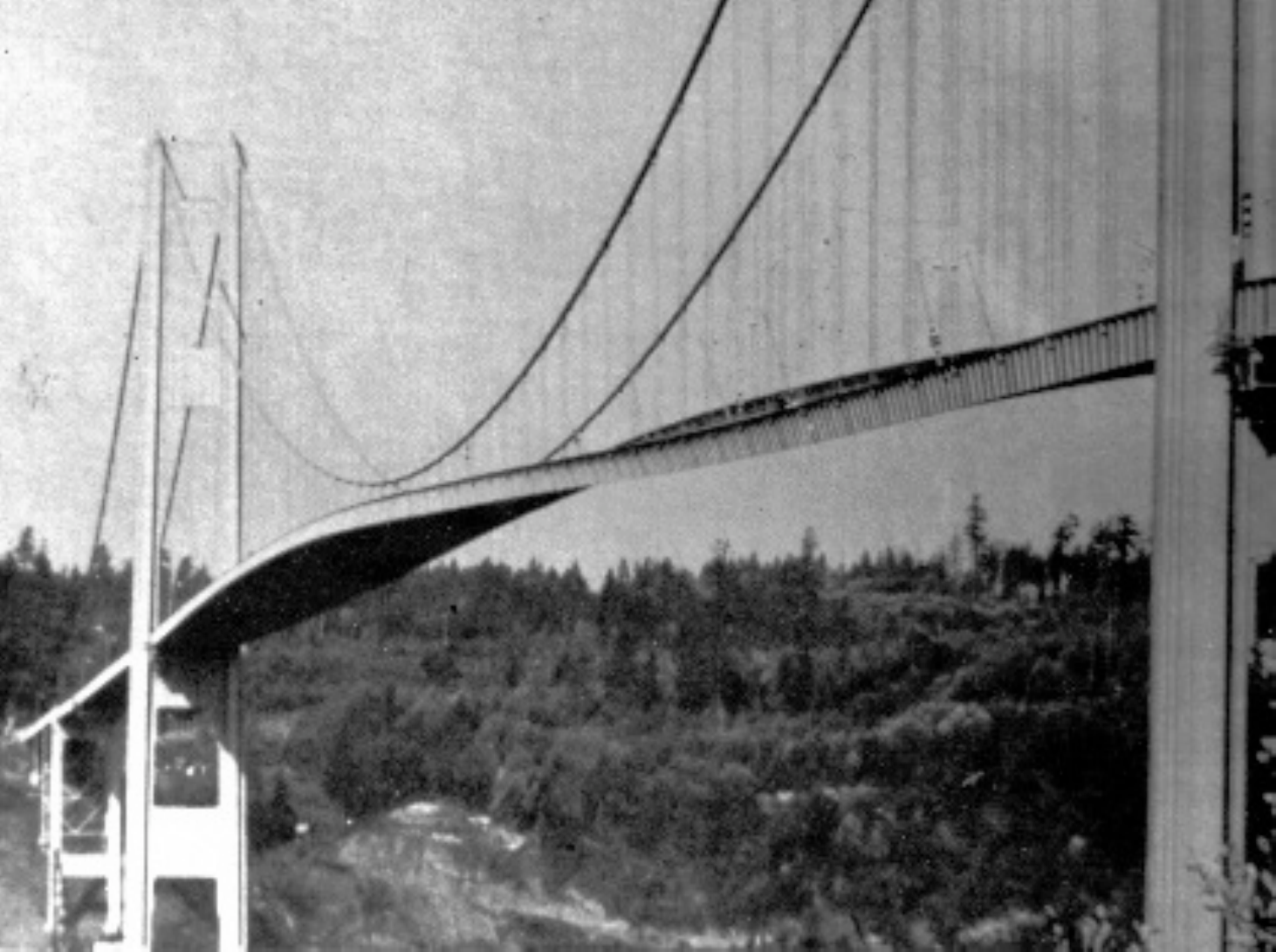}
\end{center}
\caption{\small Photograph of the first suspension bridge over the Tacoma Narrows (part of Washington state's Puget Sound) showing the large twisting motion of the bridge's central span just before it collapsed on November 7, 1940.  Even during construction, the bridge had earned the nickname ``Galloping Gertie'' because of its oscillation during the frequent high-wind conditions of the Narrows.  The picture is from \cite{Farquharson} and is used here with the permission of the University of Washington's Department of Civil and Environmental Engineering.\la{fig:Tacoma}}
\end{figure}

It has been known to engineers, since the earliest investigations of the subject by F.~B.~Farquharson\footnote{In the film footage taken on the day that the bridge fell \cite{Tacoma-movie} a man is seen walking away from an abandoned car, pipe in hand, shortly before the bridge collapses under the car.  This was Prof.~Farquharson, who had come from Seattle that morning to monitor the bridge's oscillation.  At the last moment he had attempted to rescue a black spaniel, Tubby, that had been left behind in the car when its owner fled on foot.  The dog, terrified by the violent motion, merely bit Farquharson in the finger and later perished with the bridge \cite{Tubby}.} at the University of Washington, and by T.~von K\'arm\'an and L.~G.~Dunn at Caltech \cite{Farquharson}, that the catastrophic oscillation of the bridge was a {\it flow-induced instability}, meaning that it resulted from the coupling between the solid's motion and the dynamics of the fluid driving the motion.  This is quite unlike a forced resonator, for which there would be no back-reaction of the oscillator (the bridge) on the forcing term (the wind).  This distinction, as it applies to the Tacoma Narrows bridge collapse, is lucidly made by Billah and Scanlan in \cite{Billah}, and also by Green and Unruh in \cite{Unruh}.

Like the actual fluid mechanics responsible for the vibration of the vocal chords, the dynamics of the oscillation seen in \Fig{fig:Tacoma} is rather complicated and involves turbulent flows; Pa\"idoussis, Price, and de Langre give a thorough and modern review of this subject in \cite{CrossFlow-Tacoma}.  For our purposes, it will suffice to note that the ``galloping'' resulted from a feedback between the oscillation of the bridge and the formation of turbulent vortices in the surrounding airflow.  Those vortices, in turn, drove the bridge's motion, causing a negative damping of small oscillations, like that of \Eq{eq:Airy-damped}.\footnote{In 1907, Rayleigh had reported a similar case of a tuning fork being driven at its resonant frequency by a steady cross-flow of air, even though the Strouhal frequency of vortex shedding by that flow did not match the fork's resonant frequency.  For this he could find ``no adequate mechanical explanation'' at the time.  \cite{Rayleigh-siren}}

A more recent case of large and unwanted oscillation of a bridge was the lateral swaying of the London Millennium Footbridge, after it opened in 2000.  This was also a self-oscillation: as pedestrians attempt to walk straight along a swaying bridge, they move (relative to the bridge) {\it against} the sway, thus exerting a force on the bridge that is in phase with the velocity of oscillation.  Footbridges with low-frequency, sideways modes of vibration, and a sufficiently large ratio of pedestrian load to total mass, are generally susceptible to this instability \cite{London-Structural}.  The mechanism of the back-reaction of the bridge's oscillation on the sideways motion of the pedestrians has been investigated mathematically by Strogatz et al.\ \cite{London-Sync,London-Nature}

\begin{figure} [t]
\begin{center}
	\includegraphics[width=0.5 \textwidth]{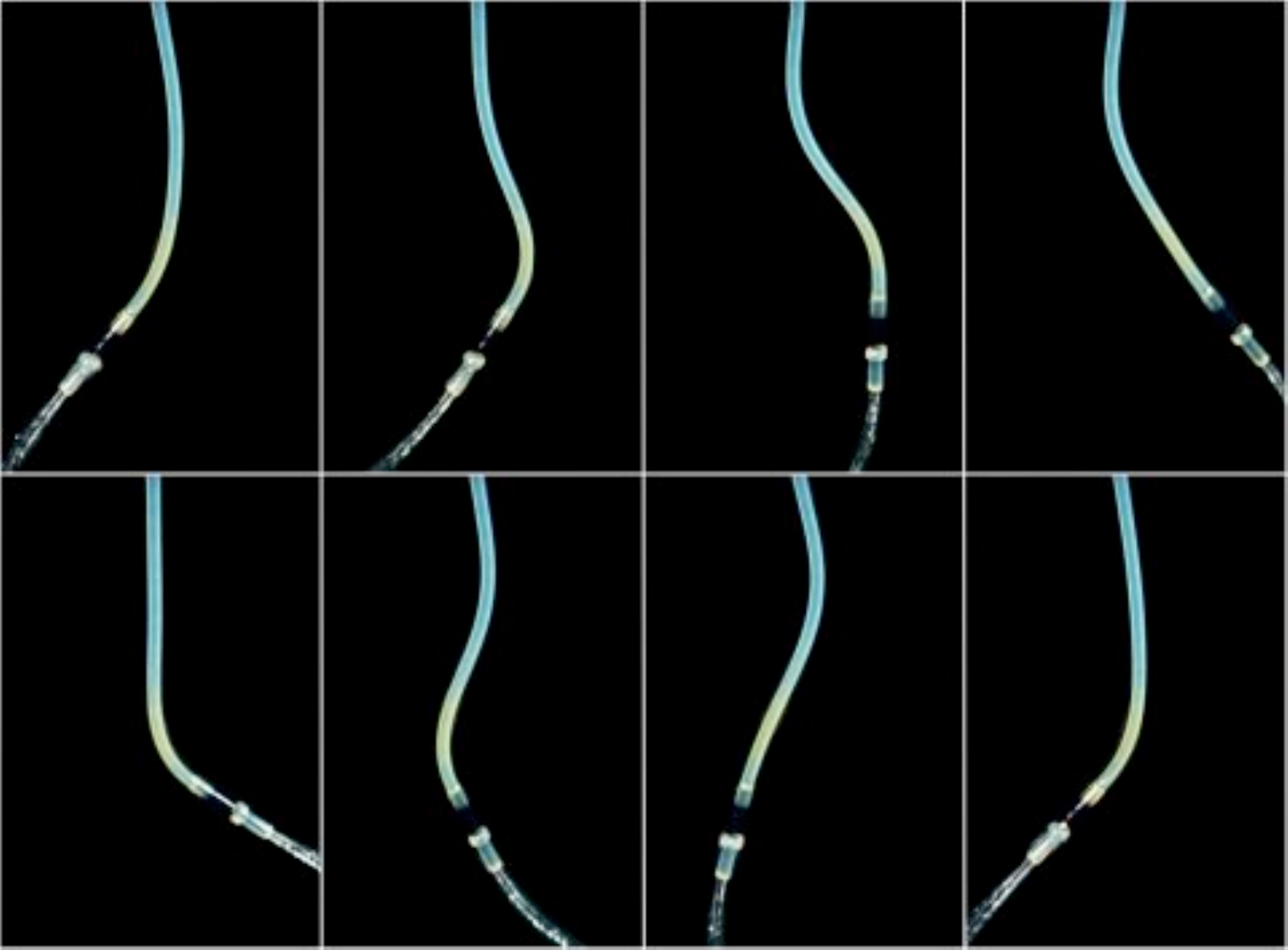}
\end{center}
\caption{\small Time-lapse pictures of a garden hose self-oscillating.  Images by Olivier Doar\'e (ENSTA) and Emmanuel de Langre (\'Ecole Polytechnique) \cite{deLangre1,deLangre2}, used here with the authors' permission.\la{fig:gardenhose}}
\end{figure}

Similar instances of flow-induced self-oscillation include the fluttering of power transmission lines and other thin solid objects in high winds \cite{CrossFlow-wake} and the vibration of an unsupported garden hose when it runs at full blast, as pictured in \Fig{fig:gardenhose} (see \cite{deLangre1,deLangre2}).  We will also see how self-oscillation explains the operation of the violin and other bowed string musical instruments, wind instruments, the human heart, and motors.

Notice that in these cases the medium must supply enough energy to sustain the oscillation, but no external rate needs to be tuned in order to produce a large periodic motion: {\it the oscillator itself sets the frequency and phase with which it is driven}.  For instance, when playing a note on the violin, there is some minimum velocity at which the bow must be drawn, but drawing it faster only makes the same note louder.  For the hose of \Fig{fig:gardenhose}, self-oscillation occurs as long as the velocity of water flow exceeds some threshold \cite{axial}.

Ocean waves are another form of self-oscillation, since they have a well-defined wavelength but are generated by the action of a steady wind.\footnote{I thank John McGreevy for bringing this point to my attention.}  Helmholtz \cite{Helmholtz-KH} and Kelvin \cite{Kelvin-KH} showed that when the relative velocity of two fluids moving in parallel directions exceeds a certain threshold, the surface of separation between them becomes linearly unstable.  Chandrasekhar treated this ``Kelvin-Helmholtz instability'' in detail in \cite{Chandra-KH}; for a brief and intuitive explanation of the relevant physics, see \cite{Tritton-KH}.  Perturbations of the surface of separation along the normal direction grow exponentially until limited by nonlinearities, as in the other negatively-damped systems that we have mentioned.  This is how the wind generates waves on the surface of a body of water. \cite{Jeffreys-waves}

\subsection{Passive versus active devices}
\la{sec:active}

In the parlance of electronics, a forced resonator is a {\it passive} device: it merely responds to the external forcing term, the response being maximal when the forcing frequency matches the resonant frequency.  A self-oscillator, on the other hand, is an {\it active} device, connected to a source from which it may draw power in order to amplify small deviations from equilibrium and maintain them in the presence of dissipation.  (The issues of power consumption and efficiency in active devices will be treated in detail in \Sec{sec:motors}.)

According to Horowitz and Hill,
\begin{quote}
devices with power gain [i.e., active devices] are distinguishable by their ability to make oscillators, by feeding some output signal back into the input. \cite{H&H-active}
\end{quote}
In other words, active devices can be identified by their ability to self-oscillate.

Passive devices (such as a mechanical resonator or an electrical transformer) can multiply the amplitude of an oscillation.  But only an active device can multiply its {\it power}, and therefore only an active device can maintain, through feedback, a steady alternating output without an alternating input \cite{H&H-active}.

In general, active devices can be thought of as involving an {\it amplification}.  For a given input signal $q_{\rm in}$, the amplifier produces an output
\be
q_{\rm out} = g \cdot q_{\rm in}~,
\la{eq:gain}
\ee
where $g > 1$ is called the ``gain.'' The power to generate $q_{\rm out}$ does not come from the signal $q_{\rm in}$, but rather from an external source.  An ideal amplifier has a constant $g$, independent of frequency or amplitude.  In practice, amplifiers always fail at large amplitudes and frequencies, causing a nonlinear distortion of the output relative to the input.

Passive devices are, for most purposes, well described by linear differential equations, whereas a practical description of active devices requires taking nonlinearities into account, as it is the nonlinearities that determine the limiting amplitude of the self-oscillation.

\subsection{Violin versus \ae olian harp}
\la{sec:violin-aeolian}

Helmholtz was the first to study systematically the physics of violin strings and his results are covered in {\it Tonempfindungen}, his groundbreaking treatise on the scientific theory of music \cite{Helmholtz-violin}.  The key point is that the friction between the bow and the violin string varies with the relative velocity between them.  When the relative velocity is zero or small, the friction is large.  As the relative velocity increases, the friction decreases.  Rosin (also called colophony or Greek pitch) is applied to the horsehairs in the violin bow to maximize this velocity dependence of the friction with the string.

The frictional force exerted by the bow on the string is therefore modulated in phase with the string's velocity, allowing the bow to do net work over a complete period of the string's oscillation (see \Eq{eq:W-net}).  Thus, as the bow is drawn over the violin string, a negative damping is obtained and the resonant vibration grows exponentially, until it reaches a limiting, nonlinear ``stick-slip'' regime.

\begin{figure} [t]
\begin{center}
	\includegraphics[width=0.6 \textwidth]{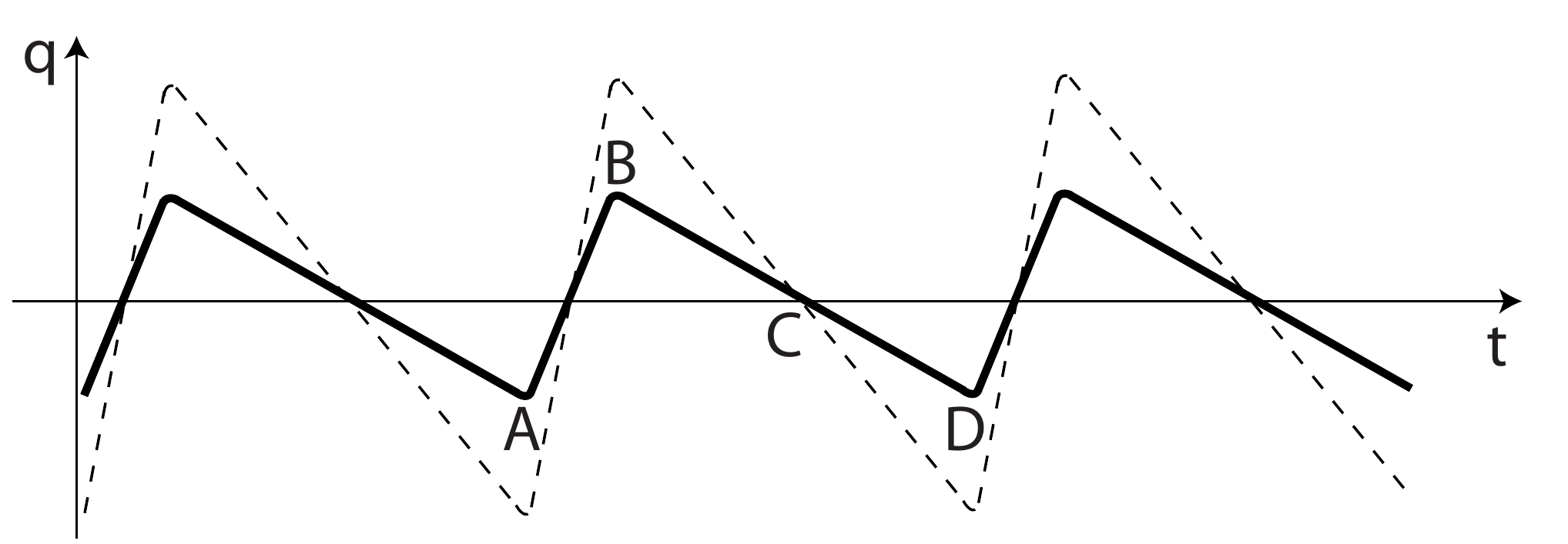}
\end{center}
\caption{\small Waveform for the displacement $q(t)$ of a violin string in the limiting ``stick-slip'' regime, as first measured by Helmholtz \cite{Helmholtz-violin}.  Between $A$ and $B$ the string moves with the bow, while between $B$ and $D$ it moves against the bow.  The dashed curve shows the waveform that results from increasing the bow's velocity.  Image adapted from Fig.~22 in \cite{Rayleigh-violin}.\la{fig:violin}}
\end{figure}

As Helmholtz observed experimentally, in this nonlinear stick-slip regime the waveform for the displacement of the violin string is triangular.  First, the violin string sticks to the bow and moves at the same velocity at which the bow is being drawn.  This is represented in \Fig{fig:violin} by the motion between $A$ and $B$.  At $B$ the string unsticks and moves back to equilibrium, which it passes at $C$.  It continues moving with nearly the same velocity until at $D$ it again becomes stuck to the bow.  Between $A$ and $B$ the frictional force exerted by the bow on the string is positive and large, while between $B$ and $D$ the frictional force is still positive but small.  Figure \ref{fig:violin} also shows that if the bow is drawn faster, the string plays the same note, but with a greater amplitude.  Within certain limits, the amplitude of the vibration is proportional to the speed of the bow.\footnote{The generation of a tone when a moist finger is dragged along the rim of a brandy glass works by the same stick-slip mechanism as the violin.  A very similar phenomenon is seen in traditional Tibetan ``singing bowls,'' which ring audibly when a leather-covered mallet is rubbed against the rim; see \cite{singing-bowls} and references therein.}

The waveform of \Fig{fig:violin} can be decomposed, by Fourier analysis, into the string's resonant frequency, called the {\it fundamental} and equal to the inverse of the period of the triangular waveform, and a series of {\it harmonic} overtones, whose frequencies are integer multiples of the fundamental.\footnote{Helmholtz measured the waveform of \Fig{fig:violin} by applying the bow very precisely at a node of a harmonic and observing the displacement at another node of that same harmonic, in order to obtain a cleaner waveform.} The generation of harmonics is a generic feature of nonlinear oscillators (see, e.g., \cite{Feynman-harmonics}).

In an ``\ae olian harp,'' one or more wires are stretched and their ends attached to an open frame, which is then placed in a location where a strong wind may pass through it.  This causes the wires to vibrate and emit audible tones.  Strouhal \cite{Strouhal} and Rayleigh \cite{Rayleigh-ToS-aeolian,Rayleigh-aeolian} demonstrated that in an \ae olian harp the wind does {\it not} act like a violin bow.  For starters, the string's motion is perpendicular to the direction of the wind.  Furthermore, Strouhal found experimentally that the frequency $f$ of the tone produced was approximately given by
\be
f \simeq 0.2 \cdot \frac{v}{d}~,
\la{eq:Strouhal}
\ee
where $d$ is the wire's diameter and $v$ the wind speed; this is the so-called Strouhal frequency of the shedding of turbulent vortices as a steady wind passes around a circular obstacle (see \Fig{fig:vortex}).\footnote{For a general flow, the number $0.2$ in \Eq{eq:Strouhal} is replaced by a power series in the inverse Reynolds number, $R^{-1} = \nu / (v d)$ (where $\nu$ is the fluid's kinematic viscosity).  The ratio \hbox{$fd/v \simeq 0.2 + 4/R + {\cal O}(1/R^2)$} is called the ``Strouhal number'' (see \cite{Rayleigh-ToS-aeolian,Batchelor-Reynolds,CrossFlow-Strouhal}).}  Evidently, the \ae olian harp {\it is} just a forced resonator, in which the frequency of the forcing term is given by \Eq{eq:Strouhal}.  Only when the wind produces an $f$ that happens to be close to the resonant frequency of the corresponding wire (as determined by its length, density, and tension), or to one of its harmonics, does the \ae olian harp ring loudly in a sustained way.

\begin{figure} [t]
\begin{center}
	\includegraphics[width=0.4 \textwidth]{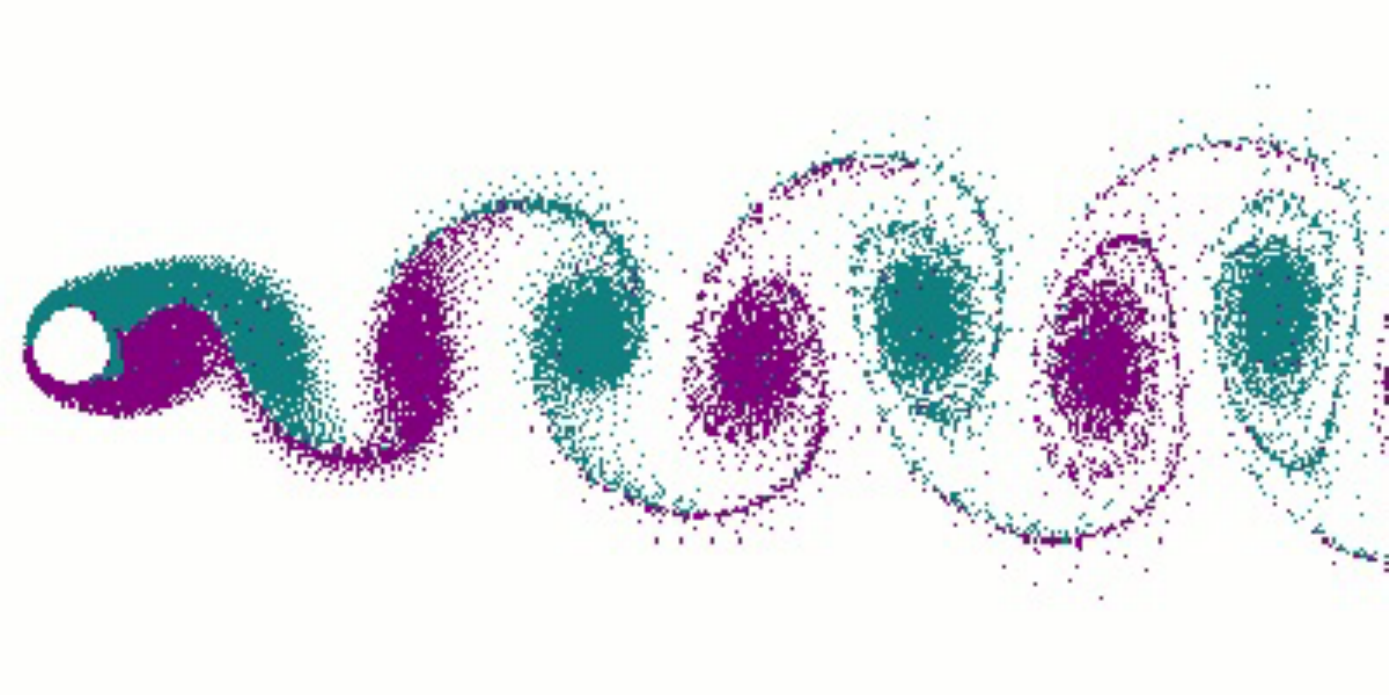}
\end{center}
\caption{\small Illustration of the turbulent vortices generated by a flow of velocity $v$ as it hits a circular obstacle of diameter $d$, on the left.  These form a ``K\'arm\'an vortex street'' \cite{Karman}.  The frequency with which the vortices are shed was experimentally found by Strouhal to obey \Eq{eq:Strouhal}.  This image is by Cesareo de la Rosa Siqueira; an animated version is available at \url{http://en.wikipedia.org/wiki/File:Vortex-street-animation.gif}. \la{fig:vortex}}
\end{figure}

\subsection{Pipe organs}
\la{sec:organs}

Unfortunately, the conceptual distinction between the \ae olian harp as a forced resonator and self-oscillating musical instruments like the flute and the flue-pipe organ is not often made clearly in the literature.  For instance, in his classic work of scientific popularization, {\it Science \& Music}, Sir James Jeans describes the operation of a flue-pipe organ as a forced resonator driven by the Strouhal vortex-shedding of the air hitting a sharp edge \cite{Jeans-organ}.  This cannot be a complete description of its operation, because the Strouhal frequency of \Eq{eq:Strouhal} depends on the velocity of the air, which would have to be tuned to the resonant frequency for the corresponding note to be sustained (recall that when the forcing and resonant frequencies are different, a forced resonator with non-zero damping vibrates only transiently at the resonant frequency, before reaching a steady state in which it moves with the forcing frequency; see \cite{Georgi-resonance}).

Jeans explains that such wind instruments work because there is a back-reaction of the oscillation of the pressure of the air within the pipe upon the Strouhal forcing term, which after some delay causes both of them to move in phase.  But this back-reaction, which is essential to the operation of the wind instruments in question, is precisely what makes them self-oscillators, rather than forced resonators.  On the details of the operation of flue-pipe organs, see \cite{Rayleigh-organ,F&R-organ}.

Note that the vortex shedding of \Fig{fig:vortex} is itself a self-oscillation, since a regular periodic motion, with frequency $f$ given by \Eq{eq:Strouhal}, is generated by the steady flow moving past the solid obstacle.  The key difference between an \ae olian harp and a flue-pipe organ is that in the former there is no appreciable back-reaction of the string's vibration on the vortex-shedding, whereas in the organ (just as in the galloping of the Tacoma Narrows Bridge), the oscillation feeds back on the vortex-shedding.  In self-oscillating ``aeroelastic flutter,'' including the motion of the bridge in \Fig{fig:Tacoma}, vortices are shed at the frequency of the fluttering, but the shedding frequency can differ significantly from the value of \Eq{eq:Strouhal} because the vortices {\it are generated by the vibration}, rather than the other way around (see \cite{Billah,CrossFlow-vortex}).\footnote{Even Pippard, who was well aware of the theory of self-oscillation, fails in \cite{Pippard-vortex} to make a sufficiently clear distinction between the solid driven by the Strouhal vortex shedding without back-reaction on the flow (as in the \ae olian harp) and aeroelastic self-oscillations driven by the feedback of the solid motion on the fluid (as in the galloping of the Tacoma Narrows Bridge).}

\subsection{Parametric resonance}
\la{sec:parametric}

Consider an equation of motion of the form
\be
\ddot q + \omega^2(t) q = 0~,
\la{eq:Hill}
\ee
where $\omega^2(t)$ is a periodic function.  Even though \Eq{eq:Hill} (``Hill's equation'' \cite{Hill}) is linear, it is evidently not time-translation invariant, and analytic solutions can be obtained only by approximation.  For $\omega^2$ of the form
\be
\omega^2(t) = \omega_0^2 \left( 1 + a \cos (\gamma t ) \right)~,
\la{eq:omega2}
\ee
with $a \ll 1$, small oscillations are seen to grow exponentially in time if the angular frequency $\gamma$ is close to $2 \omega_0 / n$, where $n$ is a positive integer.  This is the phenomenon of {\it parametric resonance}, which is strongest for $n=1$.  (Equation (\ref{eq:Hill}) with an $\omega^2$ of the form of \Eq{eq:omega2} is called the ``Mathieu equation,'' after Mathieu's investigations in \cite{Mathieu}.)

Parametric resonance resembles self-oscillation in that the growth of the amplitude of small oscillations is exponential in time, as long as there is some initial perturbation away from the unstable equilibrium at $q=0$.  On the other hand, as in the case of forced resonance, the equation of motion for parametric resonance has an explicit time-dependence.  A parametric resonator requires the tuning of $\gamma$ in \Eq{eq:omega2} to $2 \omega_0 / n$, and it fails altogether for $\gamma \to 0$.

An analysis of the flow of energy into a parametric resonator is instructive and will be useful to our discussion of the efficiency of self-oscillating motors in \Sec{sec:motors}.  As the simplest instance of parametric resonance, consider a mass $m$ hanging from a string of length $\ell_0$.  Let $\theta$ be the angle between the string and the vertical.  For small angles, the pendulum's equation of motion is
\be
\ddot \theta + \frac{g}{\ell_0} \theta = 0~,
\ee
where $g$ is the gravitational acceleration.  Suppose that the string is threaded through a hole in a solid plate above the pendulum and that one end of the string is attached to the rim of a wheel with radius $r_0 \ll \ell_0$,  turned by a motor, as shown in \Fig{fig:parametric-pendulum}.  When the motor runs, the length $\ell$ of the string varies with time, giving an equation of motion of the form of \Eq{eq:Hill}.

\begin{figure} [t]
\begin{center}
	\includegraphics[width=0.4 \textwidth]{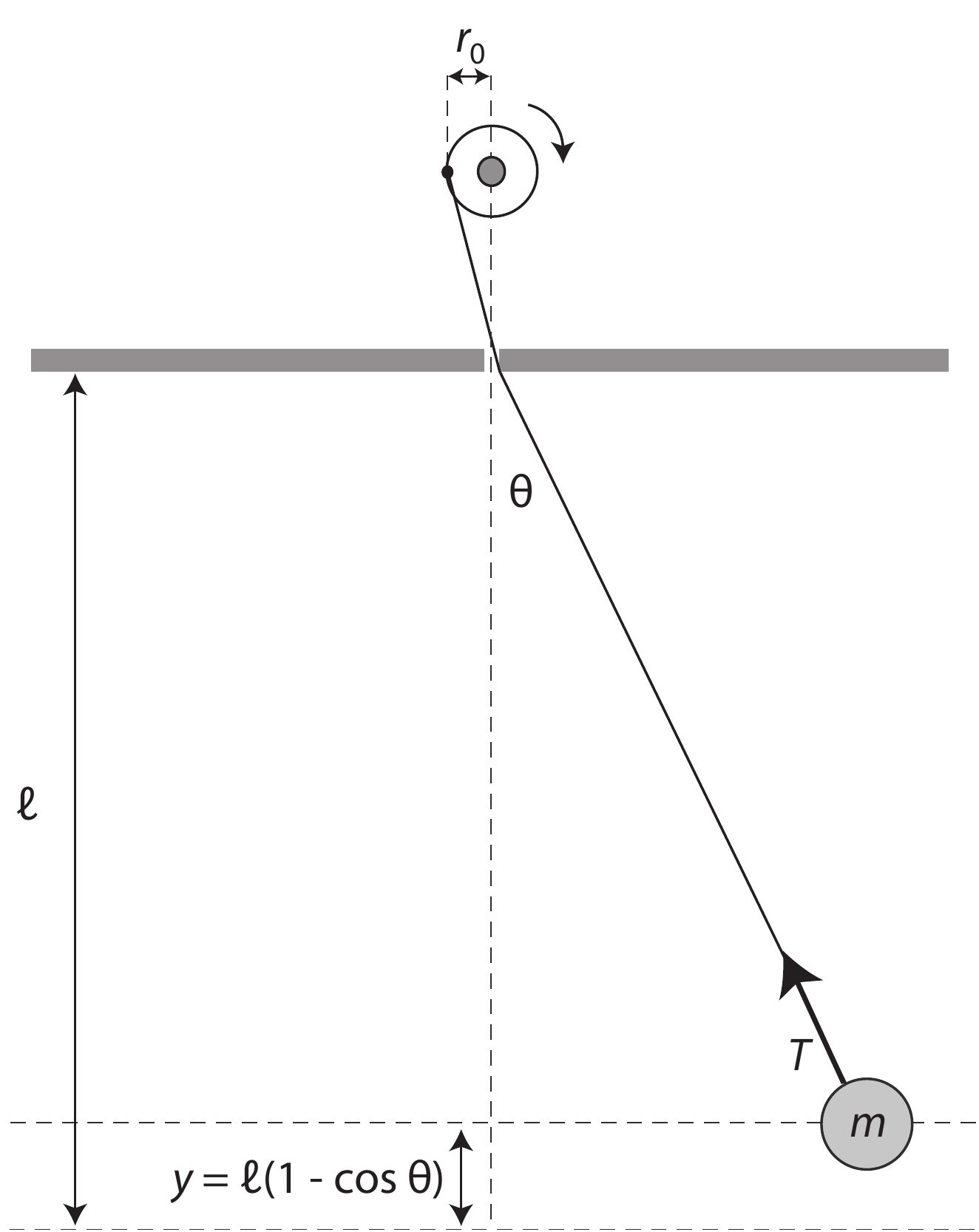}
\end{center}
\caption{\small  A string is attached to a mass $m$, while the other end is threaded through a hole in a solid plate above and attached to the rim of a small wheel driven by a motor.  The turning of the wheel causes a variation of the length $\ell$ of the pendulum, about an average value $\ell_0$.  When the wheel moves with angular velocity $2 \omega_0$, where $\omega_0 = \sqrt{g / \ell_0}$, the system achieves parametric resonance, causing the amplitude of small oscillations to grow exponentially with time.  Image adapted from Fig.~5.10 in \cite{Pippard-Response-parametric}.\la{fig:parametric-pendulum}}
\end{figure}

Why is the pendulum driven most efficiently when this motor turns with {\it twice} the angular frequency $\omega_0 = \sqrt{g/\ell_0}$ of the pendulum's free oscillation?  Note that even though the full oscillation goes as
\be
\theta(t) = \theta_0 \cos (\omega_0 t)
\la{eq:pendulum-theta}
\ee
for small angular amplitude $\theta_0$, the {\it vertical} component of the pendulum's position,
\be
y (t) = \ell \left( 1 - \cos \left( \theta_0 \cos (\omega_0 t) \right) \right)
\simeq \frac{\ell \theta_0^2}{4} \left( 1 + \cos (2 \omega_0 t) \right)~,
\la{eq:pendulum-y}
\ee
happens, simply by geometry, to move with twice the frequency $\omega_0$ in \Eq{eq:pendulum-theta}.  The reason why parametric resonance for this system is most efficient when driven with angular frequency $2\omega_0$ is that the motor causes a negative damping of $y(t)$, pulling up when the mass is going up and slackening when the mass is going down.

Note that the horizontal position of the mass, $x(t)$, moves with angular frequency $\omega_0$.  A horizontal driving force should therefore have frequency $\omega_0$, as is familiar from the pushing of a playground swing, which corresponds to an ordinary forced resonance.  It is amusing to note that Japanese children learn to use playground swings by parametric resonance: they stand up as the swing approaches its equilibrium position, at the bottom of its arc, and crouch as it reaches the top of its arc.  This makes the moment of inertia of the swing-child system oscillate with twice the frequency of the swinging and causes a faster growth of the swinging than the forced resonance method favored by Western children.\footnote{I owe this observation to Take Okui.}

Although parametric resonance is discussed mathematically in many advanced textbooks on classical mechanics (cf.\ \cite{LL-parametric, Goldstein-parametric, Jose-parametric, Hand-parametric}),\footnote{Rayleigh treated the subject in \cite{Rayleigh-parametric,Rayleigh-ToS-parametric}.} from our point of view the most instructive physical discussion is the one by Pippard in \cite{Pippard-parametric}.  Because it is not very well known, we give here an analysis ---very similar to Pippard's--- of the energy flow in a simple pendulum driven by parametric resonance.

In an ordinary pendulum, the tension $T$ of the string must obey
\be
T - mg \cos \theta = m \ell \dot \theta^2~,
\ee
so that, by \Eq{eq:pendulum-theta} and for $\theta_0 \ll 1$, 
\be
T = mg \left( 1 + \frac{\theta_0^2}{4} - \frac{3 \theta_0^2}{4} \cos (2 \omega_0 t) \right)~.
\la{eq:tension}
\ee
This tension does no work in an ordinary pendulum because it is perpendicular to the mass's velocity.  But when the motor turns with angular velocity $2 \omega_0$, it varies the length $\ell$ of the string, causing an additional velocity
\be
\dot \ell = 2 \omega_0 r_0 \cos (2 \omega_0 t)
\ee
parallel to the tension and in phase with the oscillating part of \Eq{eq:tension}.\footnote{Here we have simplified our analysis by imagining that the string is massless and cannot stretch, so that displacing one end causes an instantaneous displacement of the other end, without any additional force.  In any case, the variation of the tension $T$ induced by the action of the wheel will make no net contribution to the energy transfer of \Eq{eq:W-parametric}.} The motor therefore delivers an instantaneous power
\be
P_{\rm out} = T \dot \ell
\ee
to the mass.  The terms in $P_{\rm out}$ that are linear in $\cos (2 \omega_0 t)$ average to zero over a complete period of the motion, but the term quadratic in $\cos (2 \omega_0 t)$ gives a total energy input
\be
W_{\rm net} = \frac{3\pi \theta_0^2}{2} m g r_0  > 0
\la{eq:W-parametric}
\ee
after a full period.  Note that the fact that $W_{\rm net}$ is proportional to the {\it square} of the amplitude $\theta_0$ of the angular oscillation (as opposed to the undamped force resonator, for which $W_{\rm net}$ is linear in the amplitude) explains why the amplitude grows exponentially; see \Sec{sec:work}.  Eventually $\theta_0$ becomes large enough that the small-angle approximation fails and new terms must be added to \Eq{eq:tension}; these do not vary in phase with $\cos(2 \omega_0 t)$ and therefore reduce the efficiency with which energy is delivered by the motor to the pendulum.

A more detailed mathematical analysis is needed to account for the parametric resonance at frequencies $2 \omega_0 / n$ for integers $n > 1$; see, e.g., the treatment of the Mathieu equation in \cite{Jose-parametric}.  At those lower frequencies, parametric resonance works by a mechanism qualitatively similar to the entrainment of higher harmonics in nonlinear self-oscillators, which will be mentioned in Secs.~\ref{sec:entrainment} and \ref{sec:forced-SO}.

\section{Feedback systems}
\la{sec:feedback}

\subsection{Clocks}
\la{sec:clocks}

Mechanical and electronic timekeepers are self-oscillators, sparing the user the need to tune an external driving frequency.  Pendulum clocks and spring-driven watches, just as much as modern electronic clocks, self-oscillate by using positive feedback: the vibration of the device is amplified ---using an external source of power--- and fed back to it in order to drive it in phase with the velocity of the oscillation (see \cite{Rayleigh-clocks,Pippard-Maintained}).  In other words, clocks are {\it active devices}, as described in \Sec{sec:active}. 

This principle may be illustrated by applying feedback to the electric bandpass filter of \Fig{fig:LC}.  Let 
\be
V_{\rm in} = g V_{\rm out}~,
\la{eq:feedback}
\ee
i.e., let us feed back the output, with a gain factor of $g$.  If the terminal $V_{\rm out}$ consumes negligible current (which is possible only if the gain is effected by an active amplifier) it is easy to show that
\be
\ddot V_{\rm out} + \frac{1 - g}{RC} \dot V_{\rm out} + \frac{1}{LC} V_{\rm out} = 0~.
\la{eq:RLCg-eom}
\ee
Thus, for $g > 1$ (i.e., for {\it positive} feedback) the damping term is negative and the circuit will self-oscillate with angular frequency $\omega = 1 / \sqrt{LC}$.  The limiting amplitude of the oscillation is determined by the nonlinear saturation of the amplifier, which reduces the effective $g$ for large voltages.  This is the principle on which all electronic clocks operate \cite{Pippard-Maintained}.\footnote{See \cite{H&H-clocks} for a thorough review of how electronic oscillators are implemented in practice.}

A perplexing philosophical question about time is what {\it defines} the notion of regularity by which we evaluate physical clocks in order to establish how accurate they are.  Why do we time the rotation of the Earth with an atomic clock\footnote{The second is now defined as ``the duration of 9,192,631,770 periods of the radiation corresponding to the transition between the two hyperfine levels of the ground state of the caesium 133 atom'' \cite{SI}.  In practice, this is implemented in atomic clocks by the self-oscillation of a microwave cavity whose resonant frequency (i.e., the analog of the value of $1/\sqrt{LC}$ in \Eq{eq:RLCg-eom}) is adjusted to maximize the rate at which caesium atoms passing through the cavity undergo the hyperfine transition in question \cite{Time}.  Note that this is {\it not} an adjustment of the frequency of a driving force to make it match the cavity's resonant frequency: rather, it is the value of the resonant frequency that is adjusted to ensure its constancy (see also \cite{Pippard-standard}).} and not the other way around?  We submit that any reasonable answer must depend on the theoretical concept of self-oscillation, as represented by \Eq{eq:RLCg-eom}, when the value of $LC$ can be tied to a quantity that is presumed fixed in our accepted description of the fundamental laws of physics, and in the {\it weakly nonlinear} regime in which the effective gain approaches unity.  We shall characterize this weakly nonlinear regime in more detail in \Sec{sec:weakly-nonlinear}.\footnote{For a modern discussion of this problem in the context of mechanical clock escapements, see \cite{AJP-escapement}.}  

\subsection{Relaxation oscillations}
\la{sec:relaxation}

The simplest electronic oscillators are $RC$ circuits in which the driving voltage switches between two fixed levels when the output reaches an upper and lower threshold.  Conceptually, this is akin to a sandglass, turned over as soon as the upper chamber becomes empty.  Such devices are known as ``relaxation oscillators,'' because the output relaxes to a fixed value ---with a time constant given by the value of $RC$--- before the driving voltage switches.  The switching is done by a ``Schmitt trigger,'' which has $g \gg 1$, but whose output voltage is confined between fixed upper and lower limits \cite{Schmitt, H&H-Schmitt}.

An even simpler relaxation oscillator is the Pearson-Anson flasher, which charges a capacitor until its voltage reaches a threshold, thereupon causing a neon lamp to discharge the capacitor quickly with an accompanying flash of light \cite{Pearson-Anson}.  This is illustrated in \Fig{fig:Pearson}.  Relaxation oscillators are useful because of their simplicity, but they are not good for precision timekeeping.\footnote{Until the development of practical pendulum clocks by Huygens in the late 17th century \cite{Huygens-pendulum}, all mechanical timekeepers were relaxation oscillators (see \cite{LeC-review-Fr}).  The most sophisticated mechanism of this sort was the ``verge and foliot escapement,'' which became common in Europe in the 14th century \cite{verge}.} 

\begin{figure} [t]
\begin{center}
	\subfigure[]{\includegraphics[width=0.45 \textwidth]{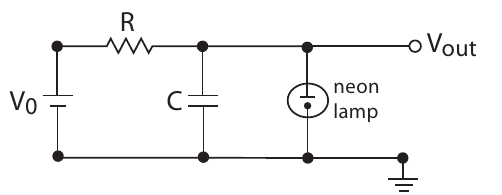}} \hskip 1 cm
	\subfigure[]{\includegraphics[width=0.45 \textwidth]{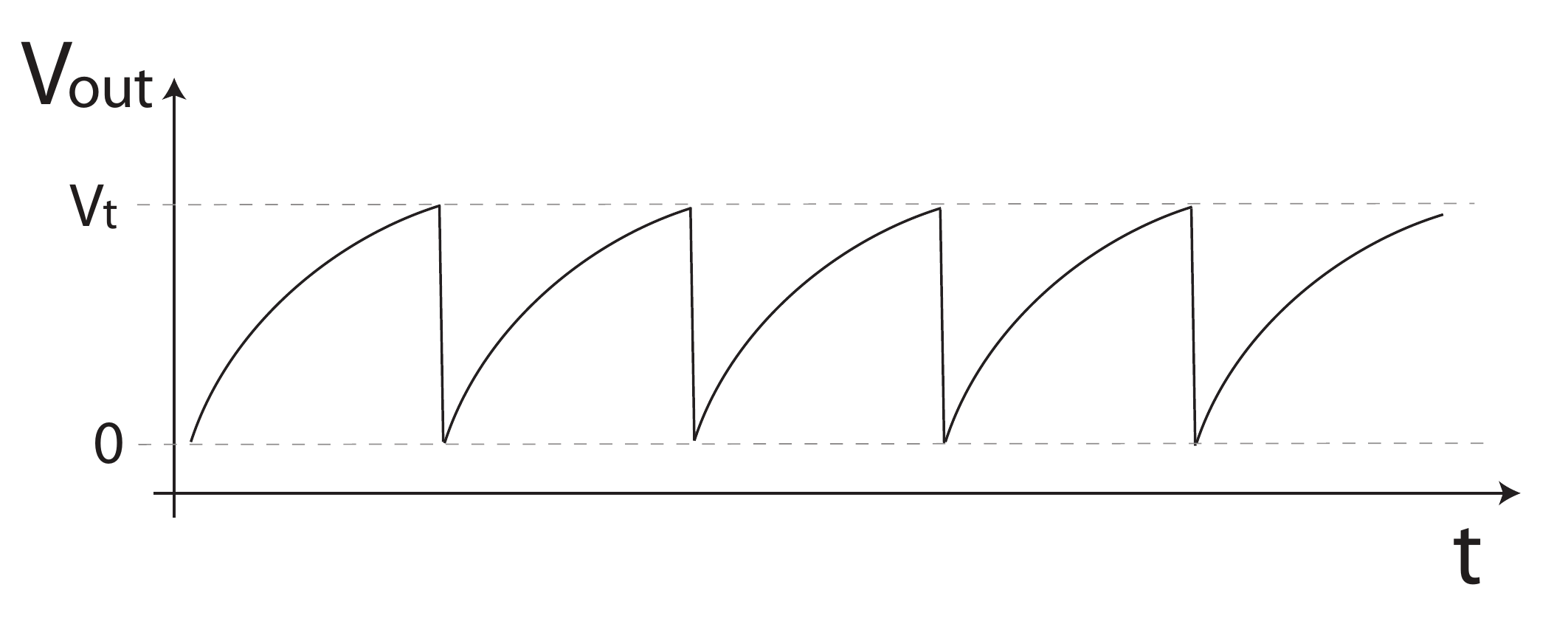}}
\end{center}
\caption{\small A Pearson-Anson flasher, which charges the capacitor $C$ until it reaches a threshold voltage \hbox{$V_t < V_0$}, whereupon the neon lamp discharges rapidly, producing a flash of light.  The circuit diagram is shown in (a), while the resulting waveform $V_{\rm out} (t)$ is shown in (b). \la{fig:Pearson}}
\end{figure}

A relaxation oscillator like the one shown in \Fig{fig:Pearson} has no resonant frequency, since $L \to \infty$, implying \hbox{$\omega \to 0$}; the actual period of oscillation depends on the switching at the thresholds, which fix the amplitude.  Relaxation oscillators with finite $L$ and $C \to \infty$ are also common: for example, an automobile's turn signal relies on an $RL$ circuit connected to a steady voltage.  A bimetallic conducting strip (called a ``thermal flasher'') is connected in series with $L$, so that when the current reaches some threshold the strip heats to the point that it bends and opens the circuit.  The strip then quickly cools and bends back, closing the circuit and starting the cycle again.

We shall see in \Sec{sec:vdP} that relaxation oscillations can be understood as a particular regime of self-oscillation.  This will require incorporating into the equation of motion the nonlinearity associated with the switching.

\subsection{Van der Pol oscillator}
\la{sec:vdP}

In the 1920s, Dutch physicist Balthasar van der Pol and his collaborators developed a model of electrical self-oscillation based on the nonlinear, ordinary differential equation
\be
\ddot V - \left( \alpha - \beta V^2 \right) \dot V + \omega^2 V = 0~,
\la{eq:vdP}
\ee
where $\alpha$, $\beta$, and $\omega^2$ are positive constants \cite{vdP,vdP-Appleton,vdP-relax}.  This oscillator has a negative linear damping $-\alpha \dot V$ and its amplitude is limited by the nonlinear, positive damping term $\beta V^2 \dot V$.\footnote{Rayleigh had earlier proposed an equation of the form \hbox{$\ddot q - \alpha \dot q + \beta \dot q^3 / 3 + \omega^2 q = 0$} to describe self-oscillators such as clocks, violin strings, and clarinet reeds \cite{Rayleigh-clocks,Rayleigh-vdP}.  Note that \Eq{eq:vdP} can be obtained from Rayleigh's equation by substituting $V = \dot q$ and differentiating.}

Van der Pol's self-oscillator may be implemented in the laboratory by using a tunnel diode to apply nonlinear feedback to an $RLC$ bandpass.\footnote{In van der Pol's original implementation, the active element in the circuit was a triode vacuum tube.  Such devices are now largely obsolete, though they are sometimes still used in high-power radio frequency (RF) amplifiers and in some audio amplifiers.  For a review of triode circuits with positive feedback, see, e.g., \cite{Groszkowski-triode}.}  Assuming that a negligible amount of electrical current flows out of the terminal $V_{\rm out}$ in the circuit of \Fig{fig:vdP}(a), 
\be
C \ddot V_{\rm out} + \frac{1}{L} V_{\rm out} = - \dot I_{\rm diode}~.
\la{eq:diode-eom}
\ee
If one can contrive to get
\be
I_{\rm diode} = - \frac{\alpha C}{g} V_{\rm in} + \frac{\beta C}{3 g^3} V_{\rm in} ^3 + \hbox{const.}
\la{eq:diode-IV}
\ee
and
\be
V_{\rm in} = g V_{\rm out}~,
\la{eq:vdP-g}
\ee
the equation of motion for $V_{\rm out}$ will have the form of \Eq{eq:vdP}, with $\omega = 1 / \sqrt{LC}$.

\begin{figure} [t]
\begin{center}
	\subfigure[]{\includegraphics[width=0.45 \textwidth]{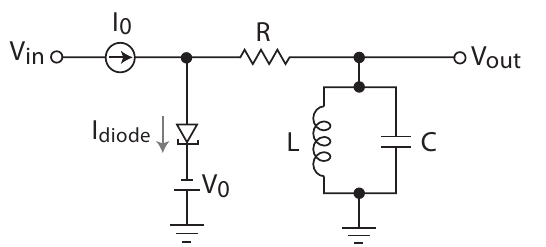}} \hskip 1 cm
	\subfigure[]{\includegraphics[width=0.22 \textwidth]{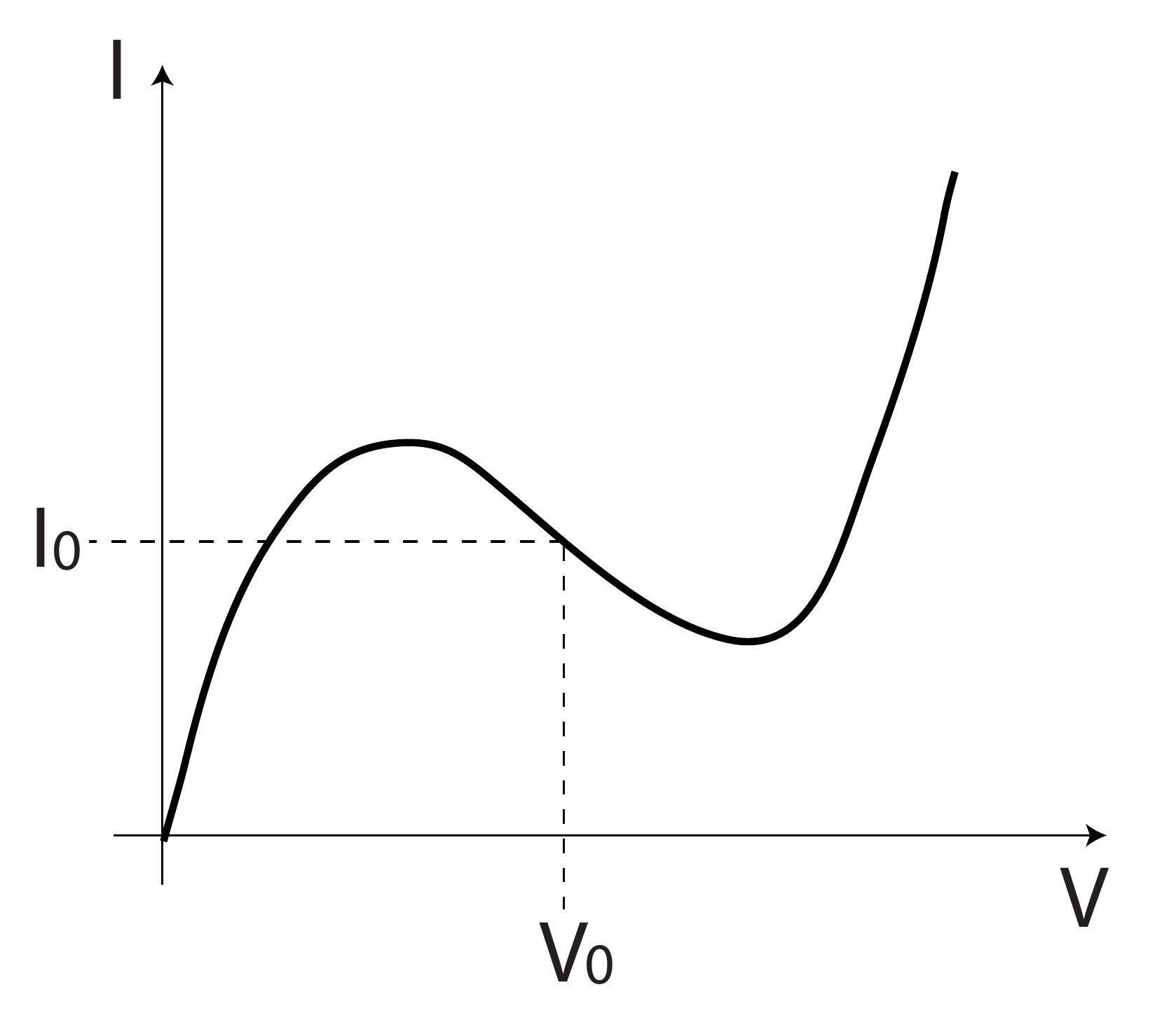}}
\end{center}
\caption{\small (a): Circuit diagram for the electronic van der Pol self-oscillator.  To obtain the necessary nonlinear properties of the diode current, a tunnel diode is biased to a point $(V_0, I_0)$ along its characteristic $I$-$V$ curve where the slope is negative and the concavity $I''(V_0)$ is negligible, as shown schematically in (b).  Images adapted from \cite{vdP-Scholarpedia}. \la{fig:vdP}}
\end{figure}

In practice, implementing \Eq{eq:diode-IV} requires biasing the tunnel diode with a voltage source $V_0$ and a current source $I_0$, corresponding to an inflection point along the diode's characteristic $I$-$V$ curve with negative slope, as shown in \Fig{fig:vdP}(b) (see \cite{vdP-Scholarpedia}).   Meanwhile, \Eq{eq:vdP-g} can be enforced by a simple op-amp multiplier or follower (see \cite{multiplier}).

\begin{figure} [t]
\begin{center}
	\subfigure[]{\includegraphics[width=0.4 \textwidth]{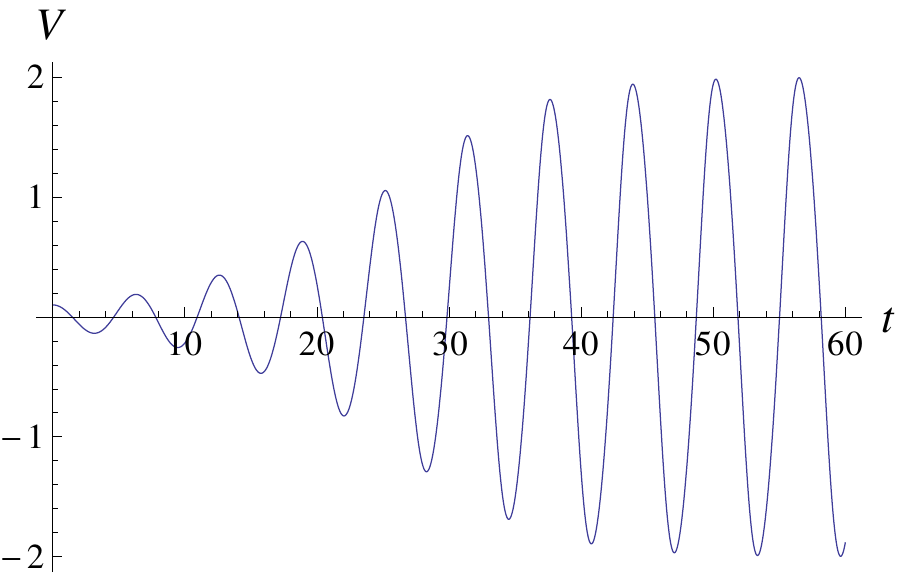}} \hskip 1 cm
	\subfigure[]{\includegraphics[width=0.4 \textwidth]{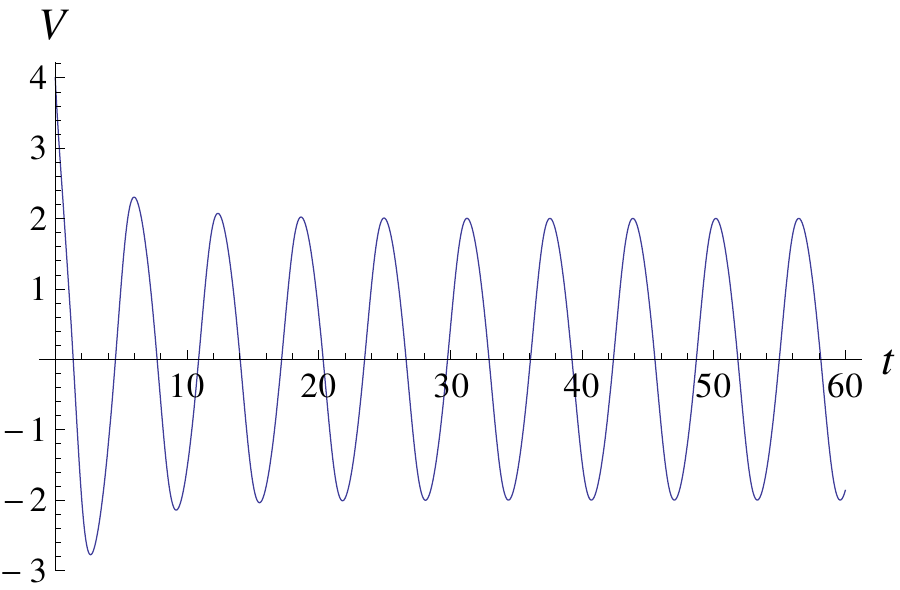}}
\end{center}
\caption{\small Numerical solutions to the van der Pol equation \hbox{$\ddot V - 0.2 \left( 1 - V^2 \right) \dot V + V = 0$} for initial conditions: (a) $V(0) = 0.1$, $\dot V(0) = 0$; and (b) $V(0) = 4$, $\dot V(0) = -4$.\la{fig:vdP-sinusoidal}}
\end{figure}

The steady-state amplitude of oscillation for \Eq{eq:vdP} is
\be
V_0 = 2 \sqrt{\frac{\alpha}{\beta}}~.
\la{eq:V0}
\ee
For that fixed amplitude, the average damping vanishes over a complete period of the oscillation, so that the oscillation can be maintained at its natural frequency without net energy either being gained or lost (see \cite{Lasers-vdP}).  For $\alpha \ll \omega$, small oscillations build up to amplitude $V_0$, while large oscillations decay down to it, as shown respectively by the waveforms of Figs.~\ref{fig:vdP-sinusoidal}(a) and (b).

On the other hand, for $\alpha \gg \omega$, small displacements grow very quickly, causing them to overshoot $V_0$, whereupon the nonlinear term $\beta V^2 \dot V$ causes the amplitude to decay back down, until it eventually shoots off in the other direction, as shown by the waveform of \Fig{fig:vdP-relax}(a). This produces a cycle of rapid buildup and slower decay that van der Pol identified as a relaxation oscillation \cite{vdP-relax,vdP-review} (see also \cite{Friedlander}).  Such an oscillation is periodic but not sinusoidal.  The period is not $2\pi/\omega$, but instead is approximately proportional to $\alpha/\omega^2$.  As shown in \Fig{fig:vdP-relax}(b), large, energetic oscillations decay down to the same limit cycle to which small oscillations build up.

In electronics, one might be used to thinking of a relaxation oscillation, such as the waveform of \Fig{fig:Pearson}(b), as a linear evolution periodically reset by an external intervention.  The most interesting conceptual feature of the van der Pol equation in the $\alpha \gg \omega$ regime is that it incorporates the rapid nonlinear switching of a relaxation oscillator {\it into the solution to the equation of motion}.\footnote{In the parlance of theoretical economics, the van der Pol model ``endogenizes'' the switching; see \cite{Goodwin-endogenous} and \Sec{sec:business}.}  A close-up of this switching is shown in \Fig{fig:vdP-relax}(b).  In \Sec{sec:limits} we shall return to the van der Pol equation and characterize its limit cycles in greater detail.

The term ``relaxation oscillation'' was coined by van der Pol in \cite{vdP-relax}, though the practical use of relaxation oscillators is very ancient (see, e.g., \cite{Gradstein}).  Van der Pol chose the name on account of the period of being determined by the relaxation time ($RC$ or $L/R$ for non-resonant linear circuits).  Friedl\"ander called the same concept {\it Kippschwingungen} (``tipping oscillation'') \cite{Friedlander}, a term still used in Germany.

\begin{figure} [t]
\begin{center}
	\subfigure[]{\includegraphics[width=0.4 \textwidth]{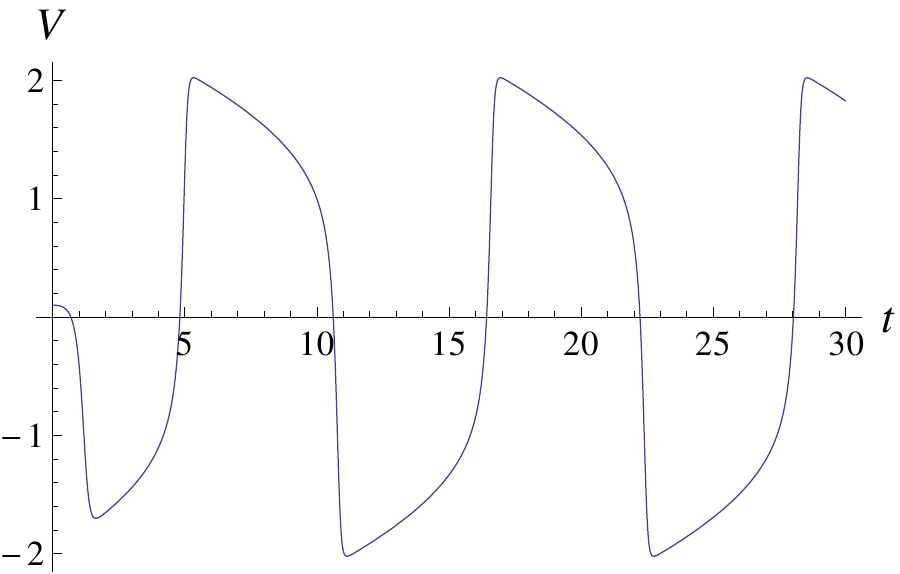}} \hskip 1 cm
	\subfigure[]{\includegraphics[width=0.4 \textwidth]{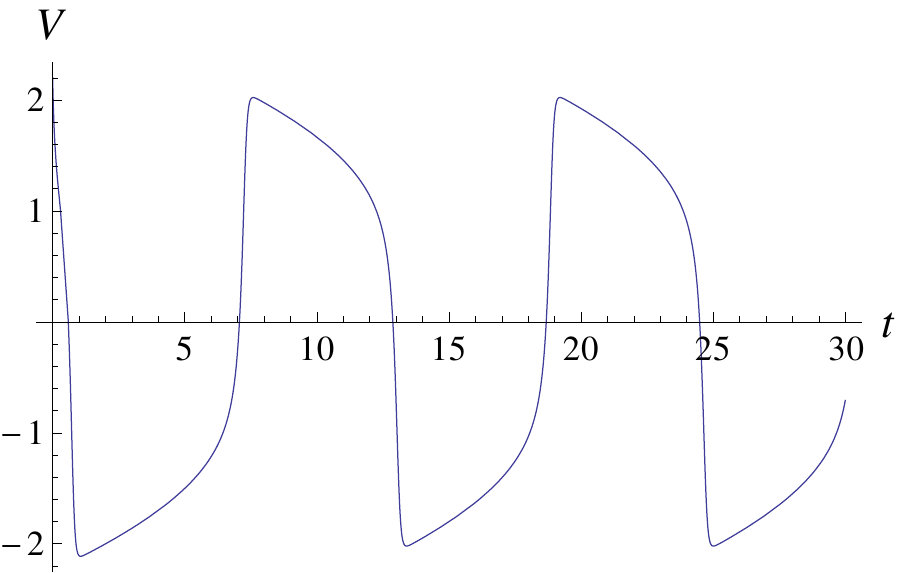}}
	\subfigure[]{\includegraphics[width=0.4 \textwidth]{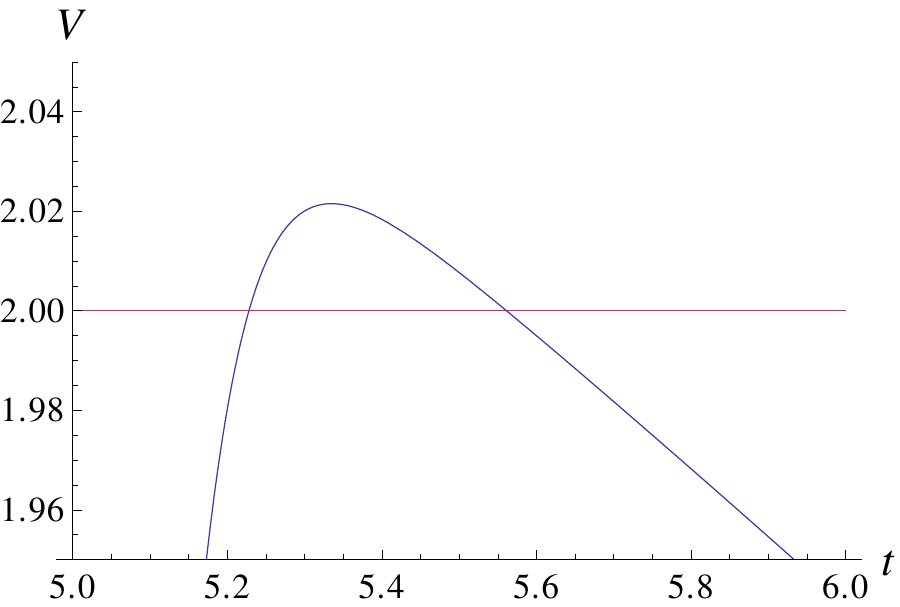}}
\end{center}
\caption{\small Numerical solution to the van der Pol equation \hbox{$\ddot V - 5 \left( 1 - V^2 \right) \dot V + V = 0$}, for: (a) $V(0) = 0.1$ and $\dot V(0) = 0$; and (b) $V(0) = 2.2$ and $\dot V(0) = -12$.  These show the sequence of fast buildups and slow decays that van der Pol identified as a relaxation oscillation \cite{vdP-relax,vdP-review}. Plot (c) gives a close-up of the waveform (a) as it overshoots $V_0 = 2$ and starts to decay. \la{fig:vdP-relax}}
\end{figure}

The vortex shedding illustrated in \Fig{fig:vortex} is a relaxation oscillation of the point of separation of the viscous boundary layer of the flow around the solid obstacle \cite{CrossFlow-Strouhal}.  The fact that it is a relaxation oscillation explains why the frequency in \Eq{eq:Strouhal} depends of the velocity of the flow.

\subsection{The heart is a self-oscillator}
\la{sec:heart}

Already in the second century CE, Galen noted that
\begin{quote}
the heart, removed from the thorax, can be seen to move for a considerable time [...] a definite indication that it does not need the nerves to perform its own function. \cite{Galen}
\end{quote}
In the 15th century, Leonardo da Vinci captured the same observation succinctly and poetically:
\begin{quote}
{\it Del core.  Questo si move da s\`e, e non si ferma, se non eternal mente.}  (``As to the heart, it moves itself, and doth never stop, except it be for eternity.'') \cite{DaVinci}
\end{quote}
This corresponds, in a modern language, to the observation that the heartbeat is a self-oscillation.

The heartbeat is controlled by the electric potential in the sino-atrial node (SAN), a bundle of specialized cells that act as the heart's natural pacemaker and are located in the upper part of the right atrium.  Van der Pol and his collaborator, Johannes van der Mark, developed a model of the SAN electrical potential as a relaxation oscillation \cite{vdP-heart}.  By coupling three relaxation oscillators they succeeded in reproducing many of the features of electrocardiograms for both healthy and diseased hearts.  The model of \Eq{eq:vdP} is still relevant to the theory of the heart's electrophysiology (see, e.g., \cite{Pacemakers} and references therein).

The form of the SAN potential is shown schematically in \Fig{fig:SAN}.  Note that, unlike the waveforms in \Fig{fig:vdP-relax}, the SAN potential is not symmetric: the negative decay phase after the switching at the lower threshold (segment $AB$ in the solid curve in \Fig{fig:SAN}) is slower than the positive decay phase after the switching at the upper threshold (segment $CD$ in the same curve).\footnote{Throughout this article, we use the word ``threshold'' to refer to the level at which nonlinearity causes a rapid switching in a relaxation oscillation.  This is the standard usage in electronics (see, e.g., \cite{H&H-Schmitt}).  In electrophysiology the word is usually reserved for a different concept: the level to which the cellular membrane must be depolarized to trigger the firing of an action potential (see, e.g., \cite{Kandel}).  In \Fig{fig:SAN} the ``thresholds'' in the electronic sense are the levels at $A$ and $C$, while the ``threshold'' in the electrophysiological sense is the level at $B$.}  We will see how to accommodate this asymmetry mathematically in \Sec{sec:asymmetric}.

The rate of the heartbeat is now believed to be regulated primarily by the strength of the ``funny current'' $I_f$ (also called, rather less bizarrely, the ``hyperpolarization-activated'' or ``pacemaker'' current) \cite{DiFrancesco1, DiFrancesco2}.  This $I_f$ is given by an inward flow of positively-charged ions across the cellular membrane.  It occurs during the negative decay phase (segment $AB$ in \Fig{fig:SAN}) of the relaxation oscillation of the membrane potential: this phase is known in cardiology as ``pacemaker'' or ``diastolic'' depolarization.\footnote{Experiments show that the activation of $I_f$ is fairly gradual, making the switching at $A$ less sharp than shown in \Fig{fig:SAN} (see \cite{DiFrancesco1,DiFrancesco2}).  We avoid this complication in the interest of conceptual clarity.}

\begin{figure} [t]
\begin{center}
	\includegraphics[width=0.5 \textwidth]{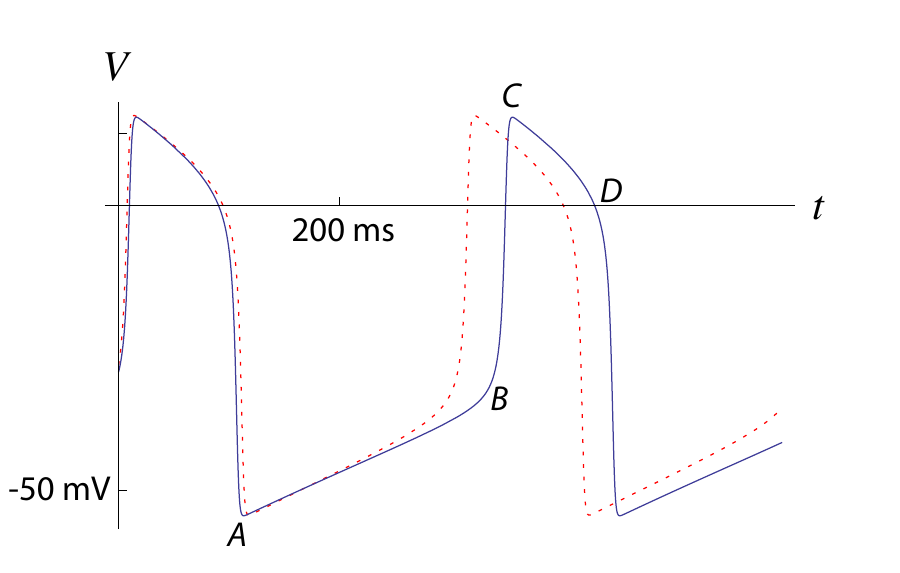}
\end{center}
\caption{\small The solid curve schematically represents the waveform for the heart's SAN potential.  The potential switches nonlinearly at $A$ and $C$.  This switching is followed by slow ``decay'' phases (corresponding to segments $AB$ and $CD$).   The rest of the waveform is composed of rapid ``buildup'' phases.  The dotted curve shows the speeding up of the negative decay phase that results from the delivery of adrenaline to the SAN.  Image adapted from \cite{DiFrancesco1}.\la{fig:SAN}}
\end{figure}

The increase in $I_f$ caused by a greater concentration of adrenaline is believed to account for the speeding up of the heartbeat when adrenaline is delivered to the SAN \cite{adrenaline}. The dotted curve in \Fig{fig:SAN} illustrates how this speeding up appears in the waveform for the SAN potential.  Conversely, $I_f$ may be suppressed, and the heartbeat therefore slowed down, by the use of drugs such as ivabradine \cite{DiFrancesco2}.

A generalization of van der Pol's equation was proposed by FitzHugh \cite{FitzHugh} as a general model of the oscillation of the membrane potential of excitable cells (neurons and muscle fibers).  Nagumo et al.\ then implemented FitzHugh's equations as an electrical circuit with a tunnel diode \cite{Nagumo}.\footnote{FitzHugh called it the ``Bonhoeffer-van der Pol model'' \cite{FitzHugh} which was also the name then used by Nagumo et al.\ \cite{Nagumo}, but it is universally known today as ``FitzHugh-Nagumo.''}  For a brief review of this model and its applications, see \cite{FHN-Scholarpedia}.  See \cite{Izhikevich} for a thorough treatment of the uses of the FitzHugh-Nagumo model in modern neuroscience.  We will characterize this model in more detail in \Sec{sec:FHN}.

\subsection{Entrainment}
\la{sec:entrainment}

Two or more coupled sinusoidal self-oscillators with different $\omega$'s will end up oscillating synchronously, as long as the coupling is sufficient to overcome the difference of the $\omega$'s.  This ``entrainment'' of self-oscillators (also called frequency, phase, or mode locking) was first reported by Huygens in 1665 for a pair of adjacent pendulum clocks mounted on the same vertical support \cite{Huygens-entrain}.  The same phenomenon was investigated experimentally by Lord Rayleigh in the early 20th century, using a pair of weakly coupled ``fork interrupters'' (see \Sec{sec:delays}) with slightly different resonant frequencies \cite{Rayleigh-siren}.  We have already mentioned, at the end of \Sec{sec:violin-aeolian}, how entrainment is important to the operation of wind instruments without reeds, such as flue-pipe organs and flutes.

Entrainment is possible because the frequency of a nonlinear vibration may be adjusted by varying its amplitude (we shall have more to say on this point in \Sec{sec:forced-SO}).  Van der Pol gave a sophisticated mathematical treatment of the subject in \cite{vdP-locking}.  Among modern physics textbooks, \cite{Lasers-vdP} and \cite{Pippard-locking} treat the subject in some detail.

Entrainment can also cause an oscillator to move with an integer multiple (a ``harmonic'') of the frequency of the other.  This simply reflects the fact that the response of a nonlinear oscillator produces harmonics (see \cite{Feynman-harmonics,LL-harmonics}).  Self-oscillating musical instruments ---like pipe organs, clarinets, or the human voice--- maintain {\it only} harmonic overtones \cite{F&R-locking}.\footnote{Musical instruments that are not self-oscillators are played either by striking or plucking.  Of these, string instruments have ---to a good approximation--- only harmonic overtones because the ends of the string are fixed and must therefore be nodes of any standing wave (see \cite{Jeans-strings}).  Other percussion instruments (e.g., bells, xylophones, drums, cymbals, etc.) give some non-harmonic overtones.  The more pronounced those overtones are, the less clear the pitch is.}  Higher, non-harmonic modes of the free oscillator are usually either not excited (as in the case of a bottle, which rings at an almost pure frequency when air is blown across the mouth) or entrained so that they become harmonic (as in pipe organs) \cite{Rayleigh-harmonics}.\footnote{There do exist certain acoustic self-oscillators, like the air horn and the vuvuzela, in which the higher overtones are not entrained, giving noisy, unmusical sounds.}  Entrainment also explains why vowel sounds in human speech are relatively easy to identify and produce, since they correspond to timbres given by a few harmonics (called, in this context, ``formants'') produced by the vibration of air in the mouth and nasal cavities \cite{Rayleigh-vowels,Vowels}.\footnote{The earliest work on understanding the production of vowel sounds was by the same Robert Willis mentioned in \Sec{sec:Airy}; see \cite{Willis-vowels}.  Rayleigh quotes Willis at length in \cite{Rayleigh-vowels}, describing his work as ``remarkable'' and as leaving ``little to be effected by his successors,'' though Helmholtz and other German scientists who investigated the problem in the mid-19th century were unaware of it.  Even more unfortunately, Rayleigh's point about the entrainment of the formants was ignored by many acousticians in the 20th century.  The resulting confusion is discussed in \cite{Vowels-Rayleigh}.}

Relaxation oscillators ---whose period is not governed by a resonant frequency--- are easier to entrain than sinusoidal oscillators.  They also exhibit a unique variant of entrainment, in that they can be locked into a {\it subharmonic} of the driving frequency.  This phenomenon was discovered by van der Pol and van der Mark, who called it ``frequency demultiplication'' \cite{demultiplication}.  Relaxation oscillators can also, in some circumstances, be weakly entrained at a rational fraction of the driving frequency \cite{Pippard-locking}, or even at an irrational multiple (this last phenomenon, called {\it quasiperiodicity}, will be relevant to the discussion in \Sec{sec:forced-SO}).

The entrainment of relaxation oscillators is immensely important in theoretical biology.  It explains, for instance, why all the potentials of the individual cells in the heart's SAN oscillate in unison, how thousands of fireflies can flash synchronously, and how the daily rhythm that governs the human body is established: see \cite{synchronization} and references therein.

Van der Pol also pointed out in \cite{vdP-review} the similarity between the coupling of two of his oscillators (a problem he had first treated in \cite{vdP-hysteresis}) and the system of nonlinear equations proposed by Lotka \cite{Lotka} and Volterra \cite{Volterra} as a model of predator-prey populations; see also \cite{Lasers-vdP,Arnold-LV,AJP-relax,Pop-review,Pop-Scholarpedia}.  For recent popular discussions of the role of entrainment in the determination of biological rhythms and other phenomena, see \cite{Sync,chronobiology}.  For a thorough mathematical treatment of entrainment and its modern applications, see \cite{Pikovsky}.

\subsection{Chaos}
\la{sec:chaos}

Another interesting feature of \Eq{eq:vdP} in the relaxation regime ($\alpha \gg \omega$) is that if a forcing term $F_0 \cos (\omega_d t)$ is added to the right-hand side, then the solutions may, for certain values of $\alpha$, be chaotic.  That is, solutions may show sensitive dependence on initial conditions, making the precise behavior of the oscillator effectively unpredictable, even though it is governed by a deterministic equation.

During their experiments with frequency demultiplication in electrical circuits, van der Pol and van der Mark were the first to observe the onset of chaos in a simple nonlinear system \cite{demultiplication}.  At the time, however, the did not fully appreciate the significance of this phenomenon.\footnote{Edward Lorenz's work on the ``Lorenz attractor'' \cite{Lorenz}, often cited as marking the beginning of modern chaos theory, appeared more than thirty years later.}  Inspired by van der Pol's work, Mary Cartwright and J.~E.~Littlewood proceeded to investigate the problem mathematically \cite{Cartwright-Littlewood,Cartwright,Cartwright-Littlewood-rev}.  For a modern review of this subject, see \cite{Thompson-Stewart}.

In \Sec{sec:forced-SO} we shall discuss entrainment and chaos of self-oscillators again.  First, however, we will introduce mathematical tools to characterize them more precisely.  In the process, the close connection between control theory and the study of self-oscillation will become evident.

\section{Control theory}
\la{sec:control}

\subsection{General linear systems}
\la{sec:linear}

Let an $N$-dimensional system be in equilibrium for \hbox{$\vv q = 0$}.  For small perturbations, the equation of motion may usually be approximated by the linear relation
\be
{\MM M} \ddot {\vv q} + {\MM C} \dot {\vv q} + {\MM K} {\vv q} = 0~,
\la{eq:linear}
\ee
where ${\MM M}$, ${\MM C}$, and ${\MM K}$ are real-valued, constant $N \times N$ matrices describing, respectively, the masses (or generalized inertias), dampings, and elasticities.  By linearity and time-translation invariance, the corresponding motion may be expressed as
\be
\vv q (t) = \Re{ \sum_{i = 1}^N w_i e^{\alpha_i t} \vv A_i}~.
\la{eq:linear-soln}
\ee
The complex numbers $\alpha_i$ are the roots of the $2N$th-degree polynomial
\be
\det \left( \alpha^2 {\MM M} + \alpha {\MM C} + {\MM K} \right)~.
\la{eq:frequencies}
\ee
The corresponding time-independent vectors $\vv A_i$ may be found by solving for
\be
\left(\alpha_i ^2 {\MM M} + \alpha_i {\MM C} + {\MM K} \right) \vv A_i = 0
\la{eq:modes}
\ee
and are called the ``normal modes.''  The complex weights $w_i$ in \Eq{eq:linear-soln} are determined by the initial conditions.\footnote{In the modern theoretical physics literature, it is conventional to write $-i \omega$ instead of $\alpha$ (see, e.g., \cite{Georgi-resonance}).} 

If $\alpha_i$ is a root of \Eq{eq:frequencies} then its conjugate $\alpha_i^\ast$ will be a root as well (the physical reason for this is that the imaginary part of $\alpha_i$ corresponds to a periodic motion, which will appear reversed after time-translating by the corresponding half-period).  When ${\MM C} = 0$, conjugate $\alpha_i$'s correspond to the same normal mode $\vv A_i$, which can be chosen to be real-valued \cite{Georgi-modes}.  Otherwise, complex-conjugating \Eq{eq:modes} tells us that if \hbox{$\alpha_j = \alpha_i^\ast$} then \hbox{$\vv A_j = \vv A_i^\ast$}.

Whenever the real part of one of the $\alpha_i$'s is positive, the amplitude of the corresponding mode grows exponentially with time.  This means that the system can self-oscillate in that mode.

\subsubsection{Stability criterion}
\la{sec:stability}

Thomson (the future Lord Kelvin) and Tait drew attention to the signs of the real parts of the roots of \Eq{eq:frequencies}.  In the first edition of their {\it Treatise on Natural Philosophy}, published in 1867, they noted that a root with a positive real part would correspond to ``a motion returning again and again with continually increasing energy through the configuration of equilibrium,'' which they dismissed as unphysical perpetual motion \cite{T&T-perpetual}.  Maxwell proposed studying the stability of mechanical systems by examining the conditions for the real parts of all the roots of to be non-positive.  His interest in this question grew out of the practical concern of understanding the possible instabilities of machines whose rate of operation is controlled by a ``governor.''  His 1868 paper on the subject, ``On Governors'' \cite{Maxwell}, is now commonly cited as a founding document of modern control theory.\footnote{Huygens had proposed a centrifugal governor for clocks in \cite{Huygens-governor}.  His design was later adapted for use in windmills and water wheels (see \cite{Bateman}), but the interest of engineers and scientists in mechanical governors and their stability was sparked primarily by the work of James Watt, who in 1788 introduced a centrifugal governor into his steam engine design \cite{Watt-governor}.  Mathematician Norbert Wiener coined the term {\it cybernetics} after Maxwell's 1868 paper:  $\kappa \upsilon \beta \varepsilon \rho \nu \acute{\eta} \tau \eta \varsigma$ means ``steersman'' in ancient Greek and is the source of the English word ``governor'' \cite{Wiener-governor}.}

The general stability criterion for linear systems was worked out by Edward Routh in 1877 \cite{Routh} and, independently, by Adolf Hurwitz in 1895 \cite{Hurwitz}.  For a single degree of freedom with a second-order equation of motion, the Routh-Hurwitz criterion simply states that a linear system is stable if the elastic and the damping term are both non-negative.  For more complicated linear systems, the criterion is expressed in terms of the non-negativity of a series of determinants built from the coefficients in the polynomial in \Eq{eq:frequencies}.  Bateman reviews the history and the mathematics of this problem in \cite{Bateman}.  In the physics literature, this subject is discussed by Rayleigh in \cite{Rayleigh-stability} and, more modernly, by Pippard in \cite{Pippard-stability}.

A consequence of the Routh-Hurwitz stability criterion is that a linear system cannot self-oscillate if the matrices ${\MM M}$, ${\MM C}$, and ${\MM K}$ in \Eq{eq:linear} are all symmetric (i.e., ${\MM M}^\T = {\MM M}$, etc., where the superscript $\T$ indicates matrix transposition), unless ${\MM C}$ has negative eigenvalues.  To prove this, let us define bilinear forms
\bea
T(\vv v, \vv u) &\equiv& \vv v^\T {\MM M} \vv u \nn
F(\vv v, \vv u) &\equiv& \vv v^\T {\MM C} \vv u \nn
V(\vv v, \vv u) &\equiv& \vv v^\T {\MM K} \vv u ~.
\la{eq:forms}
\eea
By \Eq{eq:modes}, for any pair of normal modes $\vv A_{i,j}$,
\bea
\alpha_i^2 T(\vv A_j, \vv A_i) + \alpha_i F(\vv A_j, \vv A_i) + V(\vv A_j, \vv A_i) = 0 \nn
\alpha_j^2 T(\vv A_i, \vv A_j) + \alpha_j F(\vv A_i, \vv A_j) + V(\vv A_i, \vv A_j) = 0~.
\la{eq:quadratics}
\eea
For symmetric ${\MM M}$, ${\MM C}$, and ${\MM K}$, the bilinear forms of \Eq{eq:forms} are also symmetric (i.e., $T(\vv v, \vv u) = T(\vv u, \vv v)$, etc.) and therefore, by \Eq{eq:quadratics}, $\alpha_{i,j}$ are the two roots of the same quadratic polynomial, so that
\be
\alpha_i + \alpha_j = -\frac{F(\vv A_j, \vv A_i)}{T(\vv A_j, \vv A_i)}~.
\la{eq:quadratic-sum}
\ee

As long as $\alpha_i$ is not purely real, we may choose \hbox{$\alpha_j = \alpha_i^\ast$}, \hbox{$\vv A_j = \vv A_i^\ast$}.  Let \hbox{$\vv A_i = \vv a_i + i \vv b_i$}, for real-valued vectors $\vv a_i, \vv b_i$.  The symmetry of the bilinear forms implies that
\be
2 \Re{\alpha_i} = - \frac{F(\vv A_i^\ast, \vv A_i)}{T(\vv A_i^\ast, \vv A_i)}
= - \frac{F(\vv a_i ,\vv a_i)+F(\vv b_i, \vv b_i)}{T(\vv a_i ,\vv a_i)+T(\vv b_i, \vv b_i)}~.
\la{eq:realpart}
\ee
By the positivity of the kinetic energy, \hbox{$T(\vv v, \vv v) > 0$} for any real-valued $\vv v$.  It is easy to show that the minimum value of $F(\vv v, \vv v)$ for unit $\vv v$ is given by the smallest eigenvalue of ${\MM C}$ (see, e.g., \cite{B&S-bilinear}).

By \Eq{eq:realpart}, a symmetric linear system can therefore only self-oscillate if ${\MM C}$ has a negative eigenvalue, which is not normally the case in simple mechanical systems.  Evidently, an undamped or positively damped linear system will be unstable if $\MM K$ in \Eq{eq:linear} has negative eigenvalues.  In that case, the corresponding roots of \Eq{eq:frequencies} are real-valued and the exponentially-growing solutions are not oscillatory.

For instance, M.~Stone worked out in detail the conditions of instability on a generator-governor system and stressed that the possibility of self-oscillation depends on the asymmetry of the matrices
\be
{\MM C} = \left( \begin{array}{c c} c_1 & 0 \\ \gamma & c_2 \end{array} \right) ; ~~
{\MM K} = \left( \begin{array}{c c} k_1 & - \kappa \\ 0 & k_2 \end{array} \right) ~,
\la{eq:CK-governor}
\ee
where $c_{1,2}$, $k_{1,2}$, $\gamma$, and $\kappa$ are all positive \cite{Stone,Baker}.  Stone also simulated such asymmetric linear systems using electrical circuits with active components \cite{Stone}.

\subsubsection{Gyroscopic systems}
\la{sec:gyroscopic}

Note that asymmetric couplings in \Eq{eq:linear} cannot be obtained directly from a time-independent Lagrangian.  Such asymmetry is possible only if $\vv q$ describes perturbations about a time-dependent state (e.g., about the steady rotation of the generator considered by Stone in \cite{Stone}).  Asymmetric systems can self-oscillate by absorbing energy via the underlying motion at $\vv q = 0$.

\begin{figure} [t]
\begin{center}
	\includegraphics[width=0.35 \textwidth]{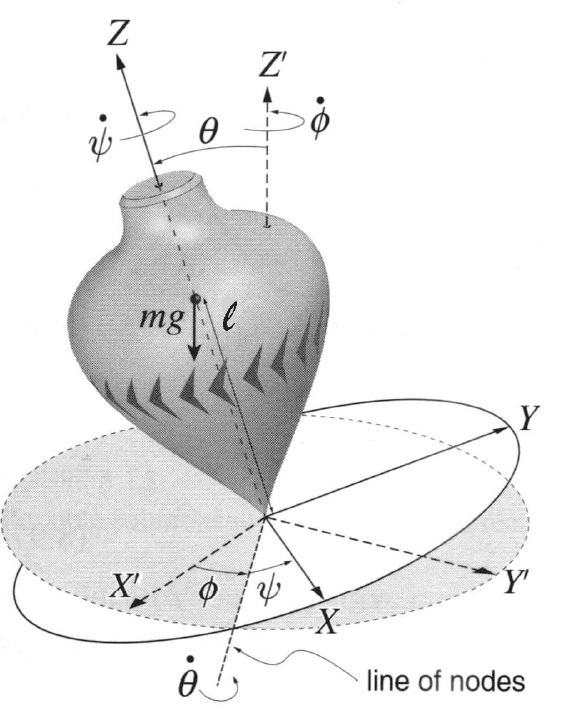}
\end{center}
\caption{\small The motion of a rigid body can be characterized by the time-dependent rotation between a non-inertial body frame, $XYZ$, attached to the rotating rigid body, and an inertial space frame, $X'Y'Z'$.  This rotation can be parametrized by the Euler angles, $(\theta, \phi, \psi)$.  The ``line of nodes'' is the intersection of the $XY$ and $X'Y'$ planes.  For the symmetric top, the angle $\theta$ gives the tilt with respect to the vertical, $\psi$ gives the top's rotation about its axis, and $\phi$ gives the precession of the axis about the vertical.  The top has mass $m$ and its center of mass is a distance $\ell$ from the tip that touches the floor.  This image is taken from figure 8.14 in \cite{Hand-top} and is used here with the authors' permission.\la{fig:top}}
\end{figure}

A way of obtaining an anti-symmetric contribution to ${\MM C}$ is to eliminate a cyclic variable with non-zero momentum.  As a simple instance of this, consider the case of the heavy symmetric top.  In terms of the Euler angles $(\theta, \phi, \psi)$ ---illustrated in \Fig{fig:top}--- the Lagrangian for the spinning top is
\be
L = \frac{I_1}{2} \left( \dot \theta^2 + \dot \phi^2 \sin^2 \theta \right) + \frac{I_3}{2} \left( \dot \psi + \dot \phi \cos \theta \right)^2 - m g \ell \cos \theta ~,
\ee
where $I_{1,3}$ are moments of inertia, $m$ the mass of the top, $g$ the gravitational acceleration, and $\ell$ the distance from the top's point of contact with the ground to its center of mass.  The cyclic variable $\psi$ is associated with a conserved momentum
\be
\frac{\partial L}{\partial \dot \psi} = I_3 \left( \dot \psi + \dot \phi \cos \theta \right) = b~.
\ee
By Legendre-transforming in $\psi$ (see \cite{LL-Routhian}) we obtain the reduced Lagrangian (or ``Routhian'')
\be
R = L - b \dot \psi = \frac{I_1}{2} \left( \dot \theta^2 + \dot \phi^2 \sin^2 \theta \right) + b \dot \phi \cos \theta - m g \ell \cos \theta - \frac{b^2}{2 I_3^2}~.
\la{eq:R-top}
\ee

Suppose that the top starts out spinning vertically (\hbox{$\theta = 0$}).  Changing variables to
\be
\xi \equiv \sin \theta \cos \phi ~, ~~~ \eta \equiv \sin \theta \sin \phi~,
\ee
and linearizing, we obtain, up to total derivatives
\be
R = \frac{I_1}{2} \left( \dot \xi^2 + \dot \eta^2 \right) + \frac{mg\ell}{2} \left( \xi^2 + \eta^2 \right) - \frac{b}{2} \left(\xi \dot \eta - \eta \dot \xi \right)~.
\label{eq:R-linear}
\ee
Thus, eliminating $\psi$ introduced terms of the form
\be
\gamma_{i j} q_i \dot q_j ~,~\hbox{with~} \gamma_{i j} \neq \gamma_{j i}
\la{eq:gyroscopic}
\ee
into the effective Lagrangian for linear perturbations about the motion with $\theta = 0$ and $b^2 > 0$.  Such terms are called ``gyroscopic'' and make an antisymmetric contribution to the damping matrix ${\MM C}$.  Whittaker describes the corresponding theory of vibrations about steady motion in \cite{Whittaker-gyroscopic}.\footnote{Older sources label terms of the form of \Eq{eq:gyroscopic} as ``gyrostatic,'' following the usage introduced by Kelvin and Tait in \cite{T&T-gyrostatic}.  See \cite{Nichols} for an early discussion of asymmetric couplings in electrical circuits.}

The resulting equations of motion for small tilt are
\be
\left\{
\begin{array}{c c} 
I_1 \ddot \xi + b \dot \eta - m g \ell \xi = 0 \\
I_1 \ddot \eta - b \dot \xi - m g \ell \eta = 0
\end{array} ~.
\right.
\la{eq:top-eom}
\ee
The determinant of \Eq{eq:frequencies} therefore has roots
\be
\pm i \alpha =  \frac{1}{2 I_1} \left( b \pm \sqrt{b^2 - 4 I_1 m g \ell} \right)~.
\ee
Thus, if the top spins rapidly ($b^2 > 4 I_1 mg\ell$), the vertical axis is stable and the top ``sleeps.''  For $b^2 < 4 I_1 mg\ell$, the axis spirals away from the vertical, until nonlinearities limit the tilt, or the top falls down.  For the nonlinear dynamics of the frictionless top see, e.g., \cite{LL-top}.

The oscillation of the top's tilt, called {\it nutation}, is powered by the top's gravitational potential $mg\ell$.  The kinetic energy $b^2 / 2 I_3$ associated with that spinning is conserved (unless friction is taken into account) and does not power the nutation.  This is qualitatively different from the other instances of self-oscillation that we have discussed because there is no negative damping, and therefore also no feedback driving the oscillator.  This can be seen more clearly by re-writing \Eq{eq:R-top} in terms of a single complex variable
\be
\Phi \equiv \sqrt{\frac{I_1}{2}} e^{i \phi} \sin \theta~,
\ee
which, for small tilt $\sin^2 \theta \ll 1$, gives a linearized Routhian
\be
R = | \dot \Phi |^2 + \mu^2 \left|  \Phi \right|^2 + i B \left( \Phi^\dagger \dot \Phi - \Phi \dot \Phi^\dagger \right)~,
\label{eq:R-linear-c}
\ee
where the superscript $\dagger$ indicates complex conjugation and
\be
\mu \equiv \sqrt{\frac{m g \ell}{I_1}} ~~ \hbox{and} ~~ B \equiv \frac{b}{2I_1}~.
\ee
The gyroscopic term in \Eq{eq:R-linear-c} can now be eliminated by the transformation
\be
\Phi \to e^{i B t} \Phi ~,
\la{eq:phase}
\ee
under which the expression for the Routhian becomes simply
\be
R = | \dot \Phi |^2 + \left( \mu^2 - B^2 \right) \left|  \Phi \right|^2~.
\la{eq:hybrid}
\ee
This is a quadratic action for $\Phi$, with an unstable potential when $\mu^2 > B^2$.

Thus, if the top's spinning is absorbed into the phase of the complex variable $\Phi$ by the transformation \Eq{eq:phase}, nutation looks simply like a ``rolling away'' from an unstable equilibrium at $| \Phi | = 0$.  The full nonlinear action for the top may give a turning point at large $| \Phi |$, making a repetitive nutation possible (see \cite{LL-top}).  This is qualitatively different from a self-oscillation generated and maintained by feedback, as in the systems discussed in \Sec{sec:feedback}.  If we ignored the precession variable $\phi$, then we would not describe the variation of the tilt $\theta$ as self-oscillatory, but simply as an undamped motion in a potential well with a local maximum at the origin.\footnote{I owe this insight to Take Okui.}

\subsection{Limit cycles}
\la{sec:limits}

Linear stability analysis determines whether small perturbation about equilibrium will decay away or grow exponentially with time.  But physical oscillations cannot grow forever.  In order to understand the {\it limiting} oscillatory regime of unstable systems it is necessary to take nonlinearities into account, as we shall do in this section for the simplest cases.\footnote{This section is the only one whose contents may be found in somewhat similar form in standard physics textbooks (cf.\ \cite{Jose-vdP,Goldstein-vdP,Pain-vdP}).  But the points that we wish to emphasize and to use in subsequent sections are enough unlike the ones covered in those texts that we have judged it advisable to review the subject here in some detail.}

Consider an equation of motion of the form
\be
\ddot x + f(x) \dot x + g(x) = 0~.
\la{eq:Lienard-fg}
\ee
It will be convenient to define a new variable \hbox{$y \equiv \dot x + F(x)$}, with $dF/dx = f(x)$, so that \Eq{eq:Lienard-fg} may be re-expressed as a system of two first-order differential equations
\be
\left\{ \begin{array}{l} \dot x = y - F(x) \\ \dot y = -g(x) \end{array} \right. ~.
\la{eq:Lienard-xy}
\ee
This is the Li\'enard transformation \cite{Lienard}, which is useful for understanding the limit cycles of self-oscillators.  It is a variation of the phase-space method pioneered by Poincar\'e for the study of nonlinear differential equations \cite{Poincare}; see also \cite{Lyapunov}.  It may also be used to show rigorously that the van der Pol oscillator always has a unique limit cycle to which all solutions tend, regardless of initial conditions (see \cite{Ye}).

The vertical isocline curve, defined by
\be
\frac{dy}{dx} = \frac{\dot y}{\dot x} = \infty~,
\ee
is given, for the system described by \Eq{eq:Lienard-xy}, by
\be
y = F(x)~.
\ee
Note that, for the circuit of \Fig{fig:vdP}, the function $F$ corresponds to the characteristic $I$-$V$ curve of the diode, with the inflection point $(V_0,I_0)$ displaced to the origin.

Let us consider, in particular, the van der Pol oscillator of \Eq{eq:vdP}.  For mathematical convenience, we rescale by $\omega$ to work with dimensionless quantities ($t \to t \omega$, \hbox{$\alpha \to \alpha / \omega$}, and $x \equiv V \sqrt{\beta/\alpha}$) so that the equation becomes
\be
\ddot x - \alpha \left(1 - x^2 \right) \dot x + x = 0 ~.
\la{eq:vdP-unit}
\ee
By \Eq{eq:Lienard-xy}, for this equation of motion the Li\'enard transformation gives
\be
\left\{ \begin{array}{l} \dot x = y + \alpha \left(x - x^3/3 \right) \\ \dot y = - x \end{array} \right.~.
\la{eq:Lienard-vdP}
\ee
The rate of change of the radius $r$ of the trajectory in the Li\'enard plane is
\be
\dot r = \frac{d}{dt} \sqrt{x^2 + y^2} = \frac{x \dot x + y \dot y}{r}~.
\la{eq:rdot-general}
\ee

Using first-order perturbation theory in $\alpha \ll 1$ and averaging over a complete period of the unperturbed system, we obtain that
\be
\left \langle \dot r \right \rangle = \frac{\alpha r}{8} \left(4 - r^2 \right)~,
\la{eq:rdot-vdP}
\ee
which proves that for $\alpha \ll 1$ the steady-state amplitude is $r = 2$ (see \Eq{eq:V0}).  Note that \Eq{eq:rdot-vdP} also establishes that for small $\alpha$ this limit cycle is stable and that all trajectories tend towards it.

\subsubsection{Weakly nonlinear regime}
\la{sec:weakly-nonlinear}

Figures \ref{fig:Lienard-sinusoidal}(a) and (b) show the Li\'enard trajectories corresponding to the solutions plotted in Figs.~\ref{fig:vdP-sinusoidal}(a) and (b), respectively.  Shown along with the trajectories is the vertical isocline curve $y = \alpha (-x + x^3/3)$, for $\alpha = 0.2$.  Figure \ref{fig:Lienard-sinusoidal}(a) illustrates how small oscillations spiral out towards $r=2$, while \Fig{fig:Lienard-sinusoidal}(b) illustrates how large oscillations spiral in to that same limiting radius.  In either case the system approaches a nearly circular trajectory, corresponding to an approximately sinusoidal limit cycle for $x(t)$.

\begin{figure} [t]
\begin{center}
	\subfigure[]{\includegraphics[width=0.3 \textwidth]{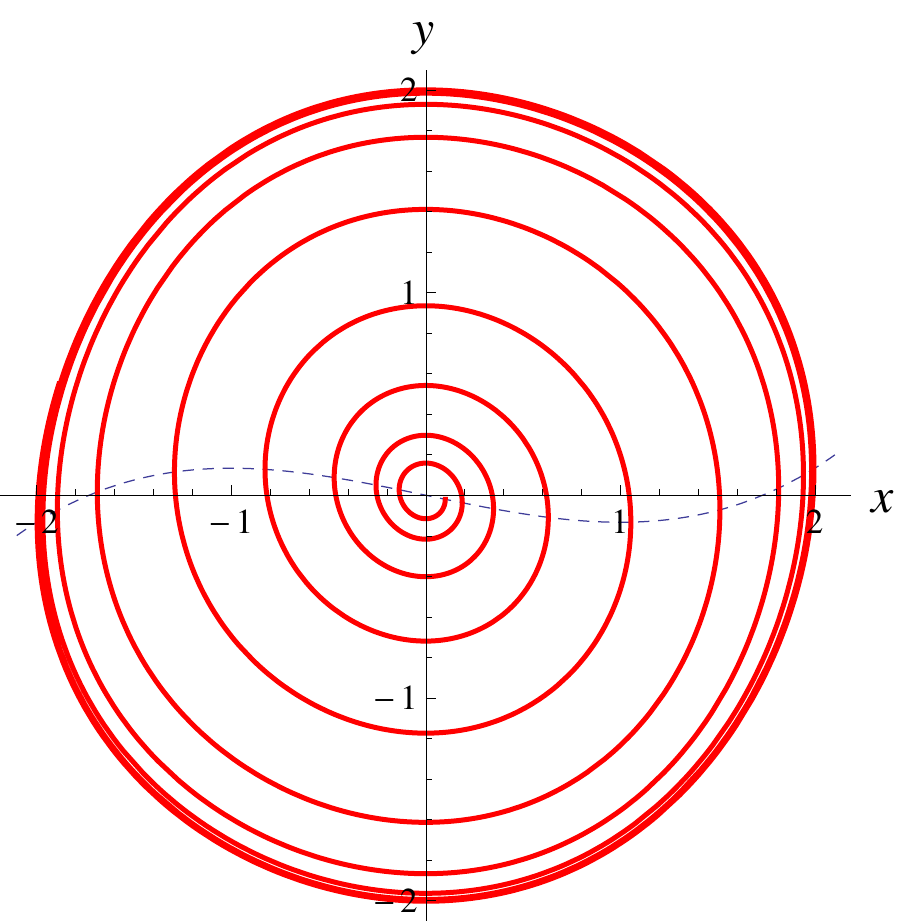}} \hskip 2 cm
	\subfigure[]{\includegraphics[width=0.35 \textwidth]{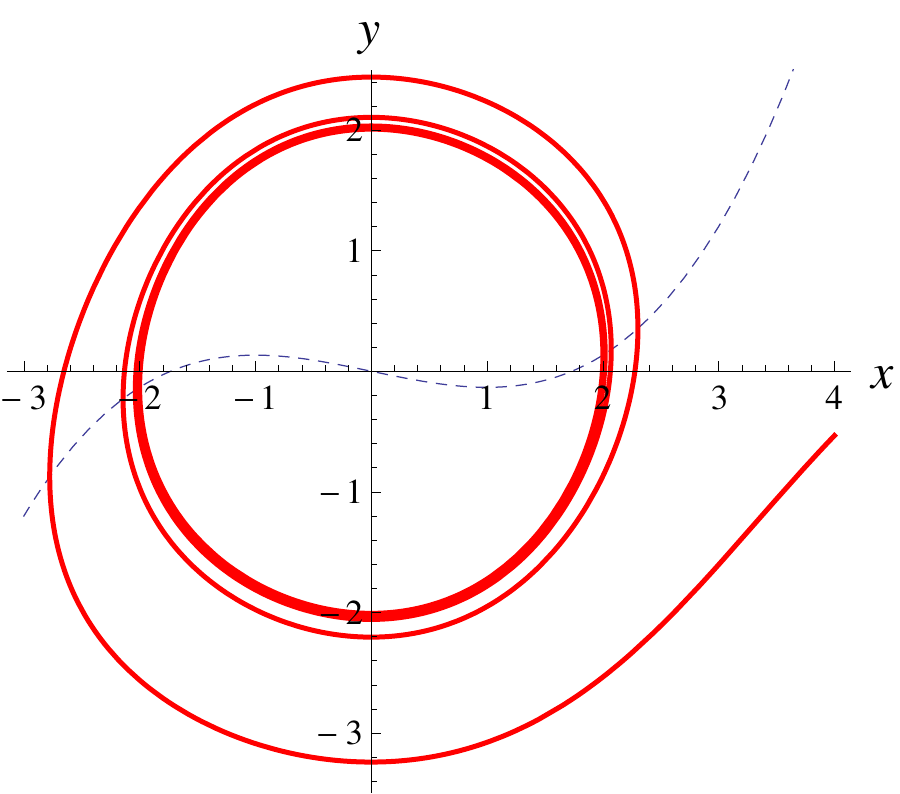}}
\end{center}
\caption{\small Solutions to the van der Pol equation \hbox{$\ddot x - \alpha \left( 1 - x^2 \right) \dot x + x = 0$}, for $\alpha = 0.2$, represented in the Li\'enard plane defined by the transformation of \Eq{eq:Lienard-vdP}, for initial conditions: (a) $x(0) = 0.1$, $y(0) = -0.02$ (i.e., $\dot x(0) = 0$); and (b) $x(0) = 4$, $y(0) = -0.53$ (i.e., $\dot x(0) = -4$).  The dashed blue line is the vertical isocline $dy/dx = \infty$.\la{fig:Lienard-sinusoidal}}
\end{figure}

Treating $\alpha$ in \Eq{eq:vdP-unit} as a perturbation and expanding the steady-state solution as a Fourier series, we find that the limit cycle gives
\be
x(t) = 2 \cos t - \frac{\alpha}{4} \sin 3 t + {\cal O}(\alpha^2)~.
\la{eq:vdP-Fourier}
\ee
(see \cite{Rayleigh-clocks,LeC-review-Eng}).  Thus, in the limit $\alpha \to 0^+$, the steady self-oscillation gives a fixed-amplitude waveform with the same shape and frequency as the corresponding linear, undamped oscillation.  This is the weakly nonlinear limit of self-oscillation pursued in the design of accurate clocks, as we mentioned in \Sec{sec:clocks}.

As $\alpha$ is increased, the harmonic terms in \Eq{eq:vdP-Fourier} grow larger, making the waveform less sinusoidal.  The highly-nonlinear regime $\alpha \gg 1$ gives non-sinusoidal relaxation oscillations, which we introduced in \Sec{sec:relaxation}.

\subsubsection{Relaxation limit cycle}
\la{sec:relax-limit}
 
For $\alpha \gg 1$, it is convenient to rescale $y \to \alpha y$, so that \Eq{eq:Lienard-vdP} becomes
\be
\left\{ \begin{array}{l} \dot x = \alpha \left( y + x - x^3/3 \right) \\ \dot y = - x / \alpha \end{array} \right.~.
\la{eq:Lienard-vdPr}
\ee
For most values of $x$ and $y$, we have that $|\dot x | \gg |\dot y|$, so that the flow is almost horizontal.  Only for $\left| y - x + x^3 / 3 \right| \sim {\cal O} (|x| / \alpha^2)$ (i.e., very close to the vertical isocline) are $|\dot x |$ and $|\dot y|$ of the same order.

\begin{figure} [t]
\begin{center}
	\subfigure[]{\includegraphics[width=0.4 \textwidth]{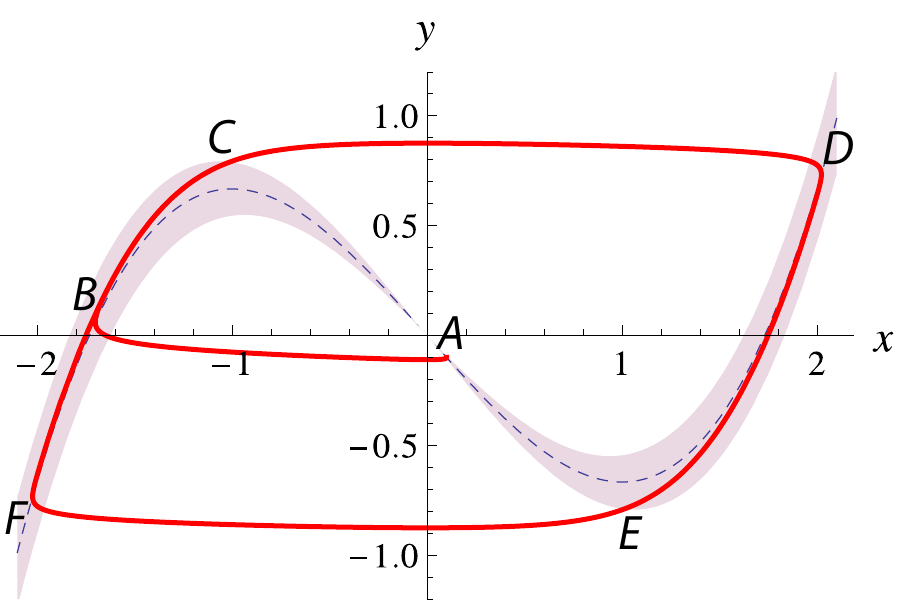}}
	\subfigure[]{\includegraphics[width=0.4 \textwidth]{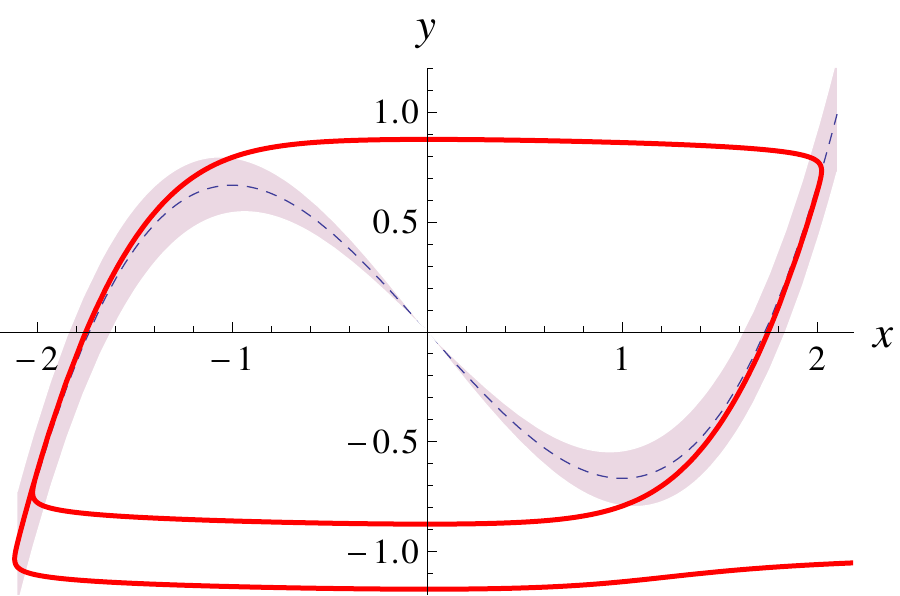}}
	\subfigure[]{\includegraphics[width=0.4 \textwidth]{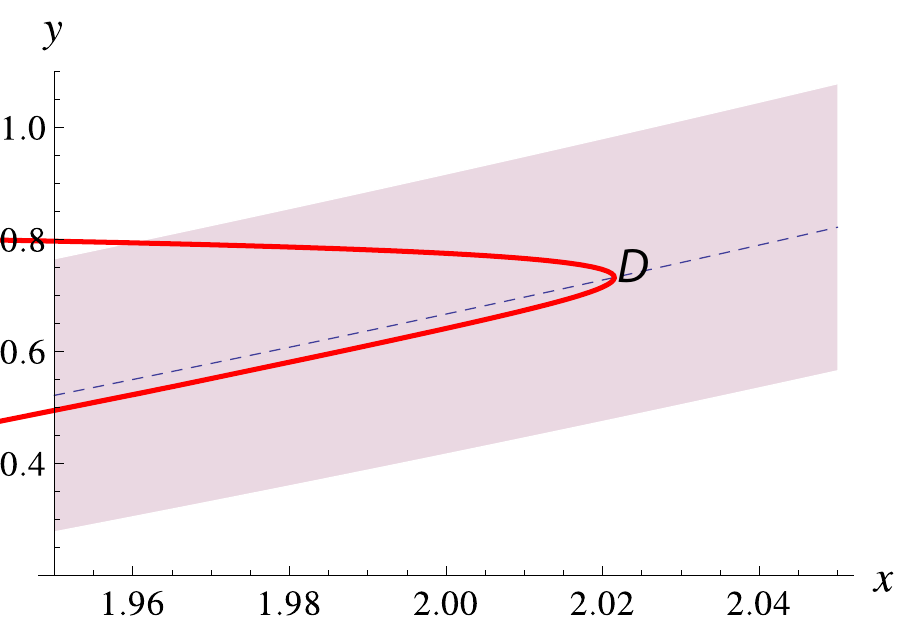}}
\end{center}
\caption{\small Solutions to the van der Pol equation \hbox{$\ddot x - \alpha \left( 1 - x^2 \right) \dot x + x = 0$}, for $\alpha = 5$, represented in the Li\'enard plane, defined by the transformation of \Eq{eq:Lienard-vdPr}, for initial conditions: (a) $x(0) = 0.1$, $y(0) = -0.1$ (i.e., \hbox{$\dot x(0) = 0$}); and (b) $x(0) = 2.2$, $y(0) = -1.05$ (i.e., \hbox{$\dot x(0) = -12$}). The vertical isocline $y = - x + x^3/3$ is indicated by the dashed blue curve.  The shaded region around it corresponds to values of $(x,y)$ for which $|\dot x|$ and $|\dot y|$ are of similar order.  Plot (c) is a close-up of (a) around the maximum of the oscillation, where the Li\'enard trajectory crosses the vertical isocline.\la{fig:Lienard-relax}}
\end{figure}

For $\alpha = 5$, the Li\'enard trajectory is shown in \Fig{fig:Lienard-relax}(a):  The system starts at a point $A$, given by $x(0) = 0.1$ and $y(0) = -0.1 + 0.1^3 / 3 \simeq -0.1$ (i.e., $\dot x (0) = 0$).  It soon develops a large negative $\dot x$ and shoots off almost horizontally until it crosses the vertical isocline at point $B$.  The trajectory then approximately follows that isocline curve ($y = - x + x^3/3$) for a while, but the isocline has a maximum in $y$ at $x = - 1$, whereas $\dot y$ in \Eq{eq:Lienard-vdPr} cannot change sign while $x < 0$.  At $C$ the system therefore re-enters the region of rapid horizontal flow and shoots off in the positive $x$ direction, until at $D$ it  crosses the vertical isocline on the $x > 0$ side.  The system will continue in a limit cycle $DEFC$: this corresponds to the succession of buildups and decays of the waveform that we had plotted in \Fig{fig:vdP-relax}(a).

Figure \ref{fig:Lienard-relax}(b) shows the trajectory of the same dynamical system, but now starting at $x(0) = 2.2$ and $y(0) = -12/5 - 2.2 + 2.2^3 / 3 \simeq -1.05$ (i.e., $\dot x(0) = -12$).  This corresponds to the waveform that we had plotted in \Fig{fig:vdP-relax}(b).  The switching in \Fig{fig:vdP-relax}(c) is seen here as the crossing of the vertical isocline, shown in close-up in \Fig{fig:Lienard-relax}(c).

We can now justify the result, quoted in \Sec{sec:vdP}, that for the van der Pol equation \Eq{eq:vdP} with $\alpha \ll \omega$ the period of the relaxation oscillation is approximately proportional to $\alpha / \omega^2$.  The reason is that the Li\'enard trajectory for \Eq{eq:Lienard-vdPr} spends most of its time near the vertical isocline (corresponding to what we have called the ``decay'' phases of the oscillation, represented by segments $DE$ and $FC$ in \Fig{fig:Lienard-relax}(a)), where $|\dot x|$ and $|\dot y|$ are ${\cal O} (1 / \alpha)$.   The length of one of those segments of the trajectory is ${\cal O}(1)$.  In units in which $\omega = 1$, the time that it takes the system of execute this part of the trajectory is therefore $\sim \alpha$.

As $\alpha \to \infty$ the region around the vertical isocline where the trajectory is not horizontal shrinks to zero width.  Therefore the limit cycle must be composed of two straight horizontal segments (the infinitely fast buildups, indicated by segments $CD$ and $EF$ in \Fig{fig:sharp}(a)) and two segments that run infinitesimally close to the vertical isocline between its local extrema at $y = \pm 2/3$ (i.e., the decays, which determine the period and are indicated in \Fig{fig:sharp}(a) by segments $DE$ and $FC$).  The overshooting of the thresholds $x= \pm 2$ (around $D$ and $F$) thus shrinks to zero.

\begin{figure} [t]
\begin{center}
	\subfigure[]{\includegraphics[width=0.4 \textwidth]{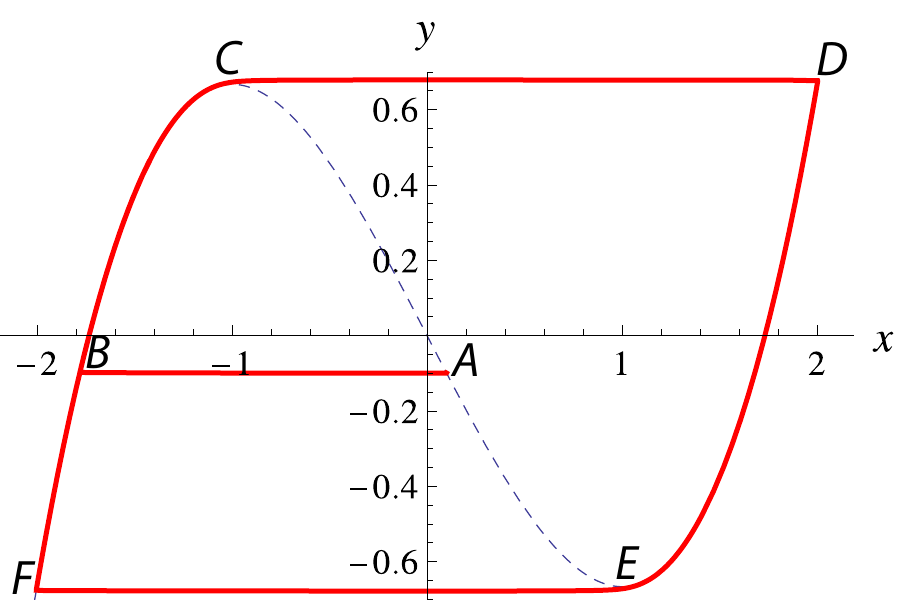}} \hskip 1 cm
	\subfigure[]{\includegraphics[width=0.4\textwidth]{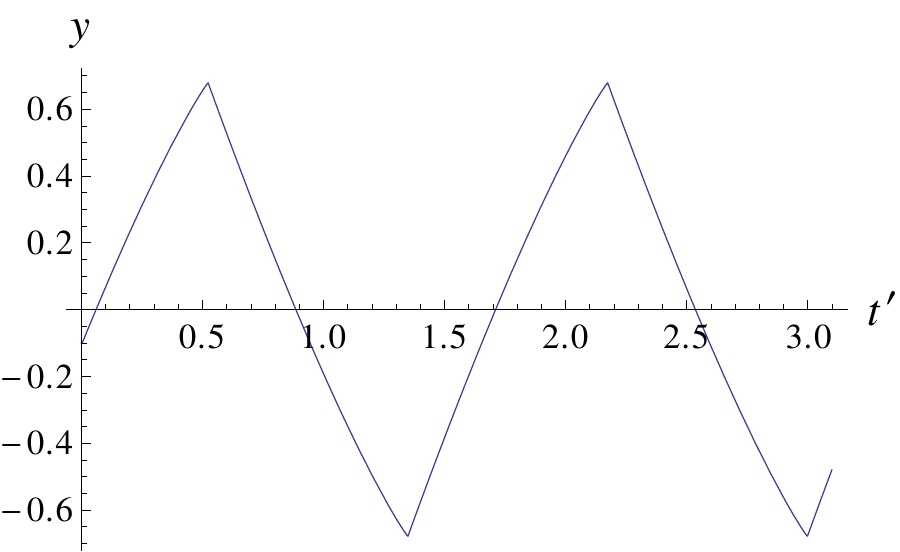}}
\end{center}
\caption{\small (a) Li\'enard trajectory for the equation \hbox{$\ddot x - \alpha \left( 1 - x^2 \right) \dot x + x = 0$}, for $x(0) = 0.1$, $y(0) = -0.1$ (i.e., $\dot x(0) = 0$), as $\alpha \to \infty$.  In the limit cycle $DEFC$, the segments $EF$ and $CD$ have constant $y = \pm 2/3$ respectively, while segments $DE$ and $FC$ follow the vertical isocline curve, $y = - x + x^3/3$, indicated by the dashed blue curve.  Plot (b) shows the value of $y \equiv \dot x / \alpha - x + x^3/3$, as a function of $t' \equiv t / \alpha$.\la{fig:sharp}}
\end{figure}

Figure \ref{fig:sharp}(b) gives the waveform for $y$, which shows the steady decays interrupted by discontinuous switching, familiar from $RC$ relaxation oscillators in electronics (see \Sec{sec:relaxation}).  Note that, if the variable $x$ is identified by the voltage $V_{\rm out}$ in \Fig{fig:vdP}(a) then, by \Eq{eq:Lienard-vdPr},  $y$ is proportional to the current through the inductor.  The behavior of $y$ is therefore that of a simple $RL$ relaxation oscillation in which the capacitance has become negligible and the resonant frequency has therefore effectively vanished.  (This is similar to the ``thermal flasher'' mentioned in \Sec{sec:relaxation}.)

The period of the relaxation oscillation in the $\alpha \to \infty$ limit can be easily calculated by using \Eq{eq:Lienard-vdPr}, which gives
\be
dt = \frac{dy}{\dot y} = - \alpha \frac{\left( \frac{dy}{dx} \right) dx}{x}~.
\ee
For the segments $DE$ and $FC$ of the limit cycle in \Fig{fig:sharp}(a), $dy/dx = -1 + x^2$, while for segments $EF$ and $CD$ $dy/dx = 0$.  The corresponding period is therefore
\be
\tau = 2 \alpha \int_2^1 dx \, \frac{1 - x^2}{x} = \alpha \left( 3 - 2 \log 2 \right) \simeq 1.6 \alpha~,
\la{eq:relax-period}
\ee
(in units in which the resonant frequency $\omega$ is 1).  Note that \Eq{eq:relax-period} implies that as $\alpha$ (the coefficient of negative damping) increases, the period $\tau$ of the oscillation {\it grows}, which might appear counterintuitive at first.  A larger $\alpha$ speeds up the buildup phases ($CD$ and $EF$ in \Fig{fig:sharp}(a)) but these are already very brief in the relaxation regime.  Meanwhile, the decay phases ($DE$ and $FC$) become longer if $\alpha$ is increased. 

In general, self-oscillators give a regular output {\it robustly}: they approach the same limit cycle regardless of initial conditions and transient perturbations.  It is this, above all else, which accounts for their technological value, such as their use as motors and clocks.

\subsubsection{Asymmetric oscillations}
\la{sec:asymmetric}

\begin{figure} [t]
\begin{center}
	\subfigure[]{\includegraphics[width=0.4 \textwidth]{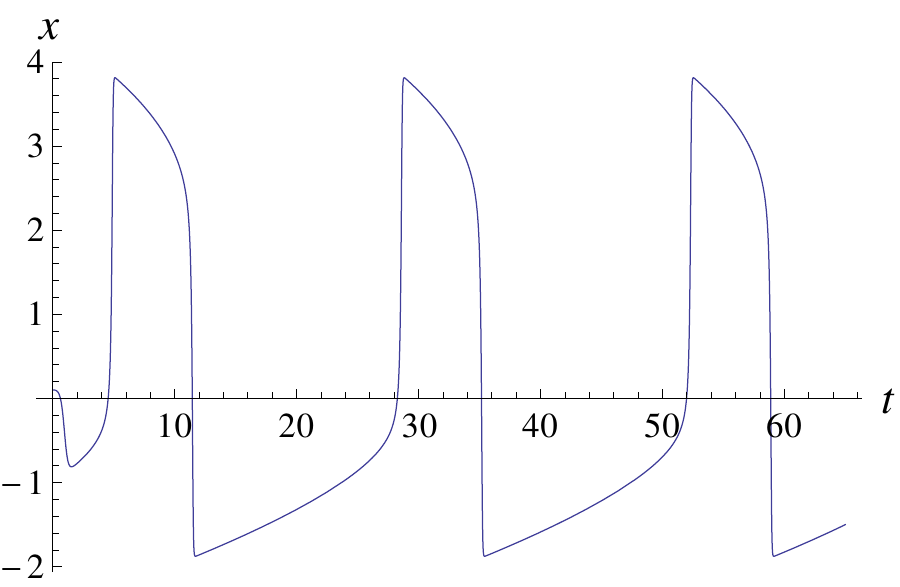}} \hskip 1 cm
	\subfigure[]{\includegraphics[width=0.4 \textwidth]{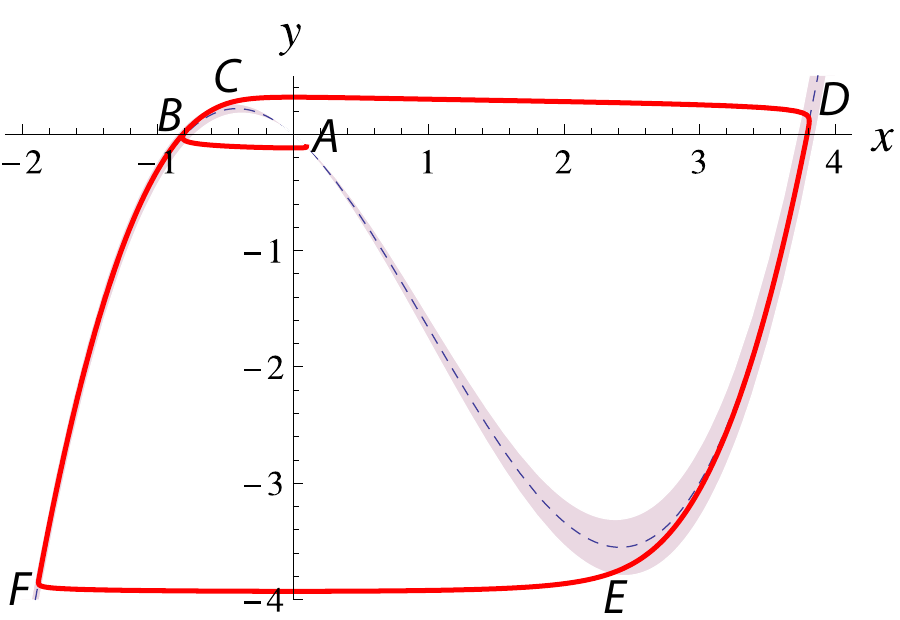}}
\end{center}
\caption{\small Solution to \hbox{$\ddot x - 5 \left( 1 + 2 x - x^2 \right) \dot x + x = 0$}, for $x(0) = 0.1$, $\dot x(0) = 0$, represented: (a) as a function of time and (b) in the Li\'enard plane defined by \Eq{eq:Lienard-xy}, with the rescaling $y \to 5y$.\la{fig:asymmetric}}
\end{figure}

Relaxation oscillations in nature are often asymmetric: for instance, the heart's SAN potential spends more time near its lower threshold than near its upper one, as discussed in \Sec{sec:heart}.  We can accommodate this asymmetry by adding a linear term to $f(x)$ in \Eq{eq:Lienard-fg}.  Figure \ref{fig:asymmetric} shows the waveform and Li\'enard trajectory for such a system.  The asymmetry reflects the displacement of the vertical isocline in the positive $x$ direction.  Since $\dot y \propto x$ (see \Eq{eq:Lienard-vdPr}), the non-horizontal part of the Li\'enard trajectory is faster just after the switching at the upper threshold (when $|x|$ is large) than just after the switching at the lower threshold (when $|x|$ is small).  In \Fig{fig:asymmetric}(b), these are segments $DE$ and $FC$ respectively.

So far we have only considered cases in which the function $F(x)$ in \Eq{eq:Lienard-xy} is symmetric about its inflection point.  In the corresponding oscillators, the mechanical energy moves up and down twice during each period (for instance, in \Fig{fig:sharp}(a) it spikes up sharply during the jumps at $CD$ and $EF$).  There is a class of self-oscillators in which $F(x)$ is asymmetric and the mechanical energy goes up and down only once during each period.  These are called ``two-stroke'' oscillations and were first described in detail in \cite{two-stroke}.

An instance of a two-stroke oscillation is given by the limit cycle of \Eq{eq:Lienard-xy} with $g(x) = x$ and a characteristic function
\be
F(x) = \rho \left( e^x - 2 x \right)~,
\ee
for \hbox{$0 < \rho < 1$}.  The nonlinear positive damping $\exp(x) \dot x$ is only important for $x > 0$, while for $x < 0$ the oscillation is almost linear.  For $\rho \ll 1$, the limit cycle is sinusoidal, but for $\rho \gtrsim 0.4$, the asymmetry becomes clear: the oscillator undergoes rapid switching only once per period, at a positive threshold, and this is the only time when the oscillator's energy peaks.  Figure \ref{fig:2stroke} illustrates this behavior for $\rho = 0.75$.

\begin{figure} [t]
\begin{center}
	\subfigure[]{\includegraphics[width=0.4 \textwidth]{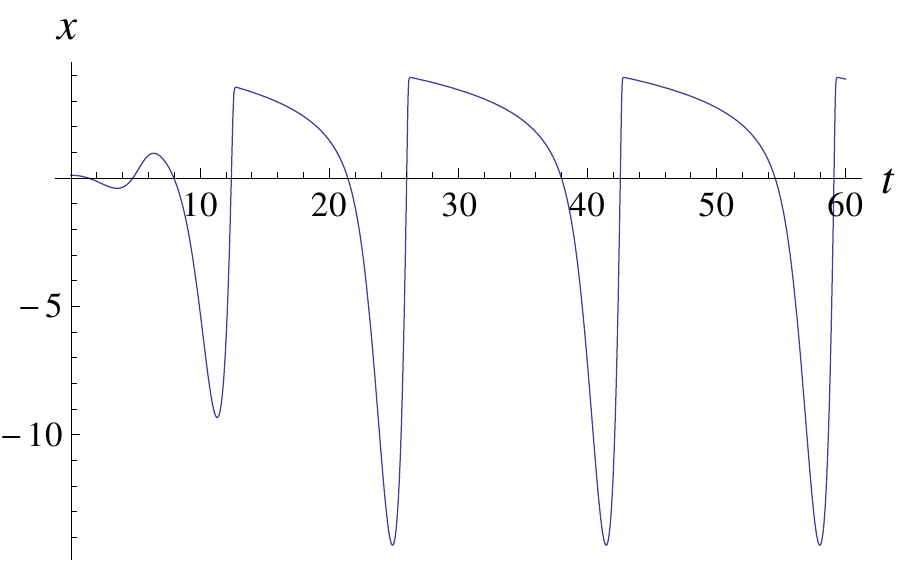}} \hskip 1 cm
	\subfigure[]{\includegraphics[width=0.4 \textwidth]{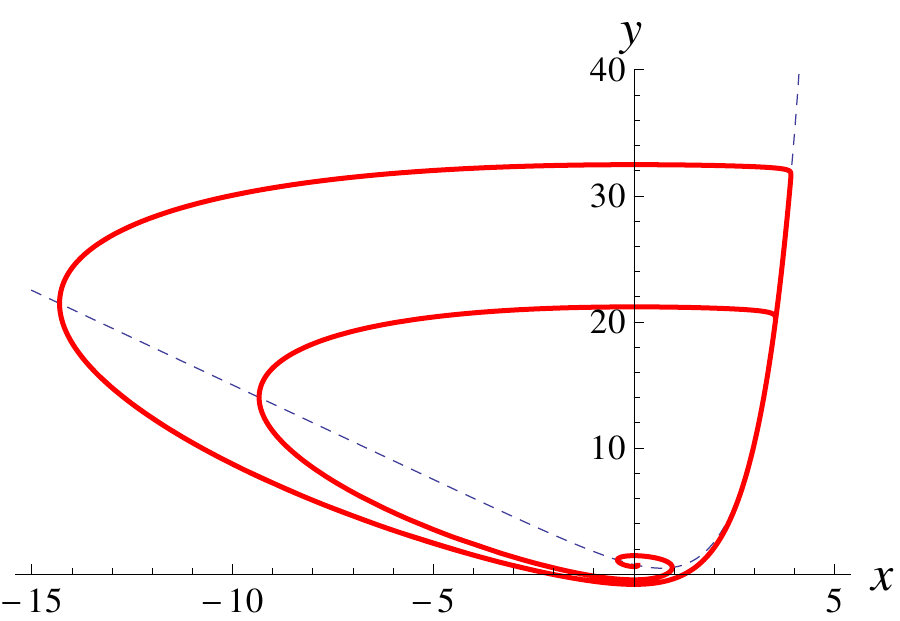}}
\end{center}
\caption{\small Solution to \hbox{$\ddot x + 0.75 \left( \exp(x) - 2 \right) \dot x + x = 0$}, for $x(0) = 0.1$, $\dot x(0) = 0$, represented: (a) as a function of time and (b) in the Li\'enard plane defined by \Eq{eq:Lienard-xy}.  The vertical isocline $y = 0.75( \exp(x) - 2 x )$ is shown as the dashed blue curve.\la{fig:2stroke}}
\end{figure}

\subsubsection{Forced self-oscillators}
\la{sec:forced-SO}

The Li\'enard transformation can also be applied to forced oscillations obeying an inhomogeneous equation of the form
\be
\ddot x + f(x) \dot x + g(x) = \phi(t)~,
\la{eq:Lienard-forced}
\ee
giving 
\be
\left\{ \begin{array}{l} \dot x = y - F(x) \\ \dot y = -g(x) + \phi(t) \end{array} \right.~,
\la{eq:Lienard-forced-xy}
\ee
where $dF/dx = f(x)$.

\begin{figure} [t]
\begin{center}
	\subfigure[]{\includegraphics[width=0.4 \textwidth]{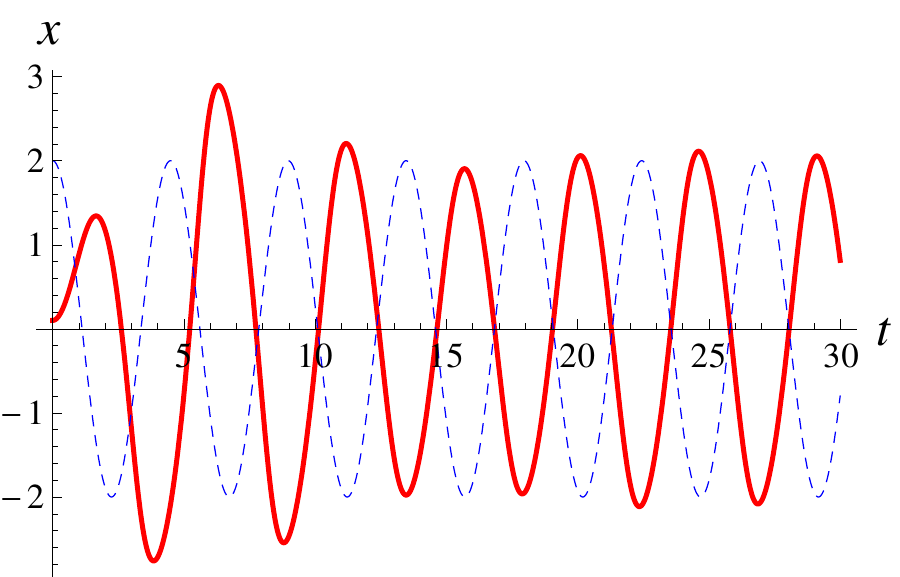}} \hskip 1 cm
	\subfigure[]{\includegraphics[width=0.22 \textwidth]{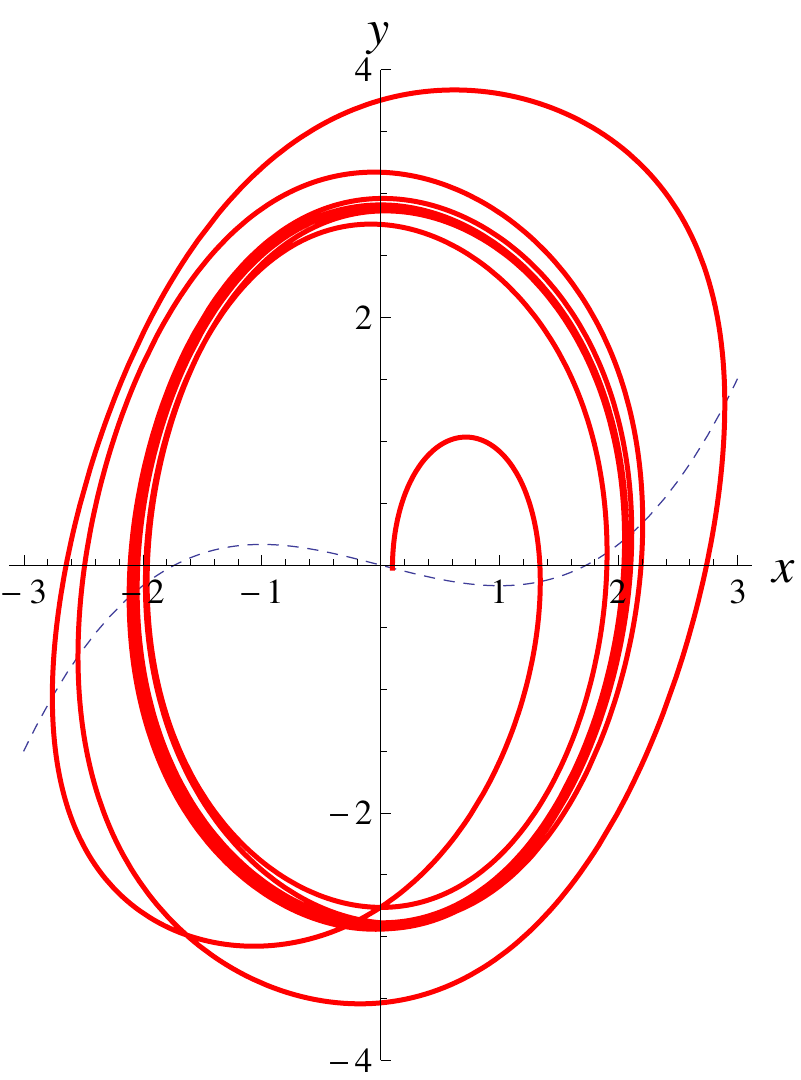}}
\end{center}
\caption{\small Solution to \hbox{$\ddot x - 0.25 \left( 1 - x^2 \right) \dot x + x = 2 \cos (1.4 t)$}, for $x(0) = 0.1$, $\dot x(0) = 0$, represented: (a) as a function of time (in red), with the forcing term $2 \cos (1.4 t)$ shown as a dashed blue curve, and (b) in the Li\'enard plane defined by \Eq{eq:Lienard-forced-xy}, with the vertical isocline indicated by the dashed blue curve.  This solution shows entrainment between the oscillator (with a resonant angular frequency of 1) and the forcing term (with angular frequency of 1.4).\la{fig:entrainment}}
\end{figure}

As an illustration of the entrainment behavior discussed in \Sec{sec:entrainment}, consider
\be
\ddot x - \alpha \left( 1 - x^2 \right) \dot x + x = F_0 \cos \left(\omega_d t \right)
\la{eq:vdP-forced}
\ee
for $\alpha = 0.25$, $\omega_d = 1.4$, and $F_0 = 2$.  The plots in \Fig{fig:entrainment} show that the oscillation is entrained by the forcing term after two cycles, and that thereafter it moves precisely in antiphase with it, much as Huygens originally reported for the pair of pendulum clocks \cite{Huygens-entrain}.  A self-oscillator can also be entrained at an integer multiple (``harmonic'') of the driving frequency, as was mentioned in \Sec{sec:entrainment}.

Though the quantitative characterization of entrainment can be cumbersome (see \cite{Lasers-vdP,Pippard-locking,F&R-locking,Pikovsky}), the qualitative explanation of the kind of entrainment shown in \Fig{fig:entrainment} is straightforward: when the relative phase $\phi$ between the (nearly) sinusoidal motion of the oscillator and the forcing term in \Eq{eq:vdP-forced} is such that $\sin \phi \neq 0$, the external force is doing net work on the oscillator (see \Sec{sec:work}) and therefore causing the amplitude to vary.  Because of the nonlinearity, a change in the amplitude also changes its frequency, and for some range of parameters this means that the oscillation can adjust itself in response to the external driving force, until it reaches a steady, phase-locked motion.

In the relaxation regime ($\alpha \gg 1$) for \Eq{eq:vdP-forced}, $x(t)$ can also be locked into a submultiple $\omega_d / n$ of the driving frequency (``frequency demultiplication'') \cite{demultiplication}.  An instance of demultiplication is shown in \Fig{fig:demultiply}.  An interesting demonstration, originally conducted by van der Pol and van der Mark, was to have a Pearson-Anson relaxation oscillator (see \Fig{fig:Pearson}) drive a loudspeaker, and in turn to drive that oscillator at a constant frequency $\omega_d$ \cite{vdP-vdM}.  As the capacitance in the Pearson-Anson circuit is dialed up, increasing its natural period, the output is locked at successive submultiples of $\omega_d$, producing a series of discrete tones that sound much like a descending musical scale played on the bagpipes (see also \cite{LeC-review-Fr}).

\begin{figure} [t]
\begin{center}
	\includegraphics[width=0.45 \textwidth]{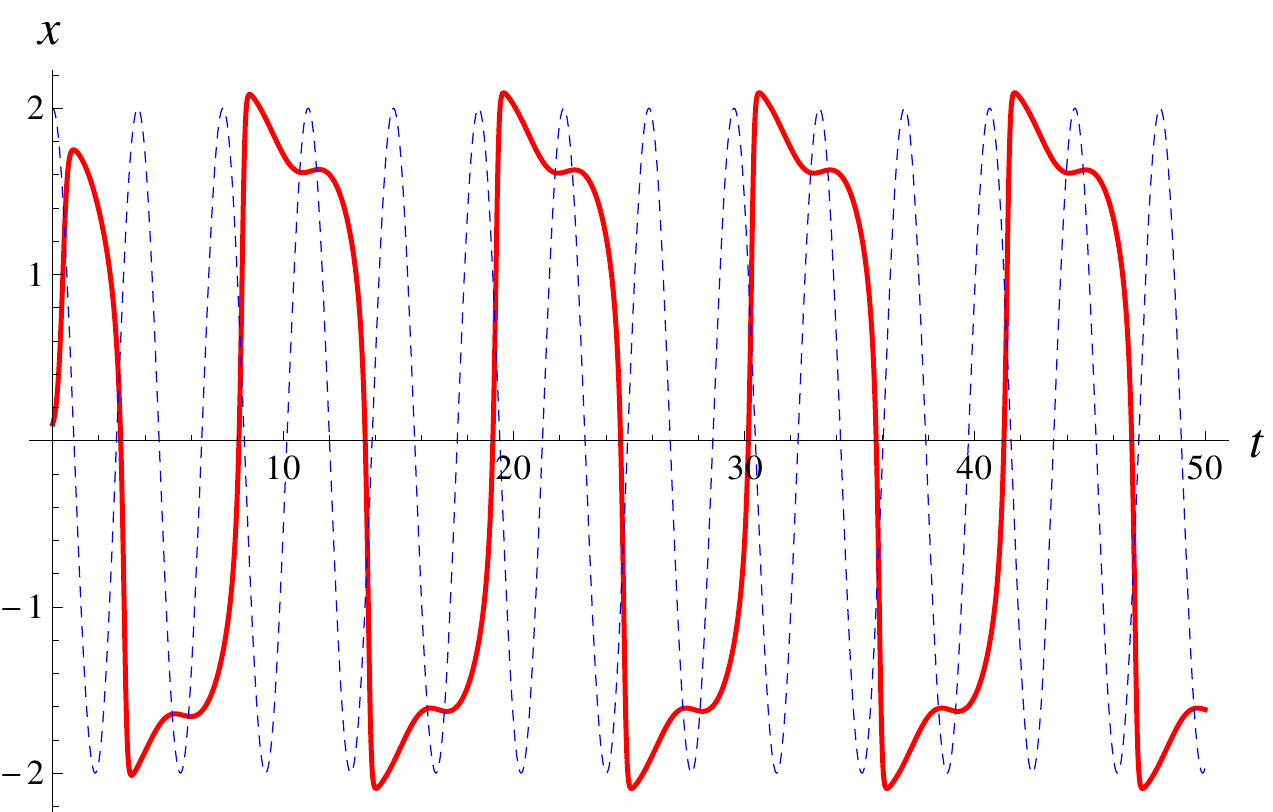}
\end{center}
\caption{\small The solution to \hbox{$\ddot x - 5 \left( 1 - x^2 \right) \dot x + x = 2 \cos (1.6 t)$}, for $x(0) = 0.1$, $\dot x(0)=0$, is shown in red.  The forcing term $2 \cos (1.6 t)$ is shown in dashed blue.  The waveform $x(t)$ is locked at a third of the forcing frequency.\la{fig:demultiply}}
\end{figure}

During their experiments on frequency demultiplication, van der Pol and van der Mark also noticed that under certain conditions the output of the loudspeaker became very noisy.  The reason is that, for some values of the parameters in \Eq{eq:vdP-forced}, the ratio of the period of the entrained motion to the period of the forcing term becomes irrational: this is called the {\it quasiperiodic} regime.  For large $\alpha$, the quasiperiodic and the periodic regimes may overlap, leading to solutions to \Eq{eq:vdP-forced} that are chaotic, as we mentioned at the end of \Sec{sec:entrainment}.  This is the case, e.g., for $\alpha = 3$, $\omega_d = 1.788$, and $F_0 = 5$, as illustrated in \Fig{fig:chaos}.

The subject of the onset of chaos in driven nonlinear systems is treated in detail in modern textbooks on dynamical systems.  See, e.g., the discussion of the Kolmogorov-Arnol'd-Moser (KAM) theorem in \cite{Jose-KAM}.  For an interesting discussion of such systems, with emphasis on their possible use in macroeconomics, see \cite{Goodwin-chaos}.

\begin{figure} [t]
\begin{center}
	\subfigure[]{\includegraphics[width=0.45 \textwidth]{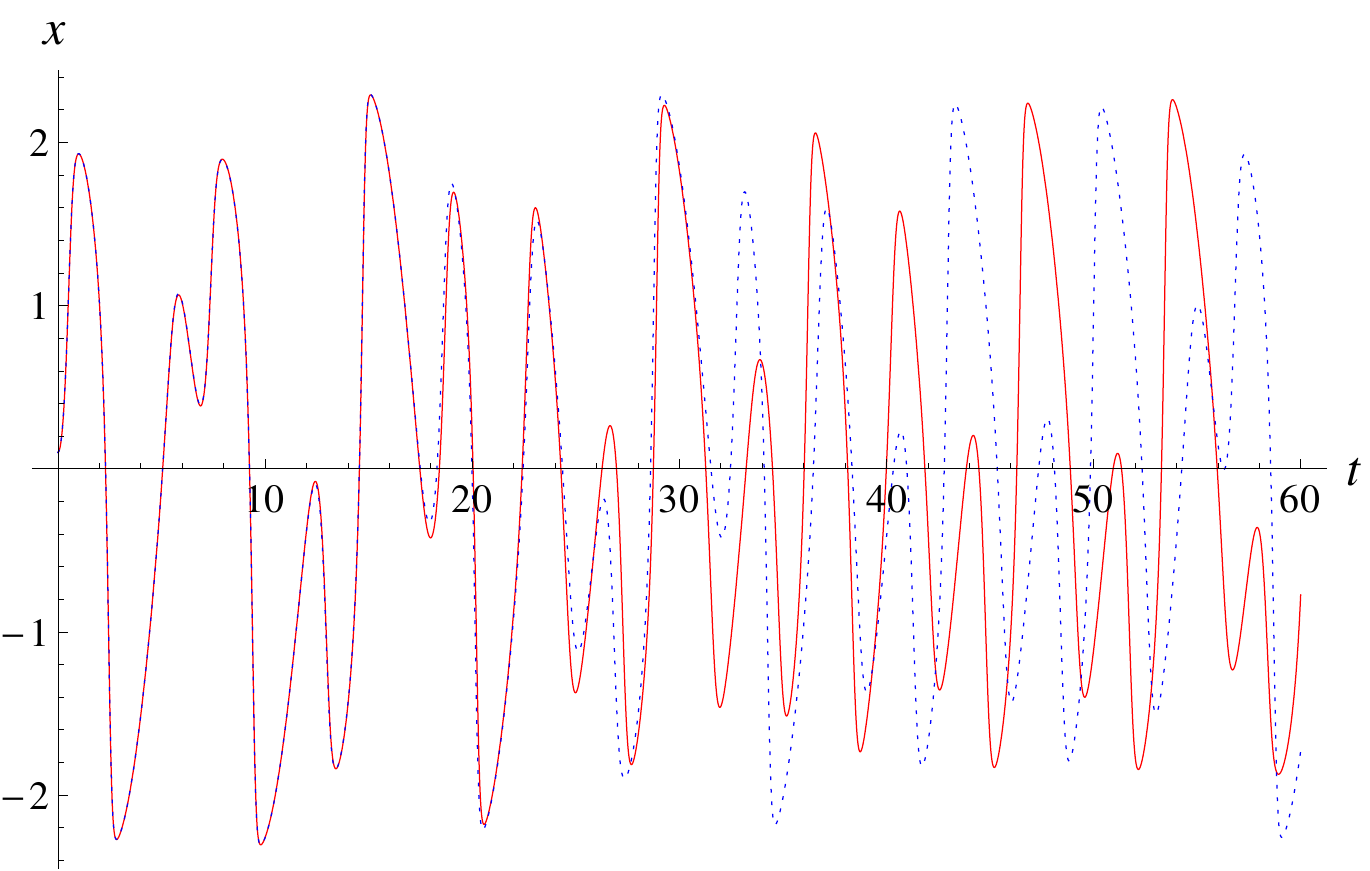}} \hskip 1 cm
	\subfigure[]{\includegraphics[width=0.4 \textwidth]{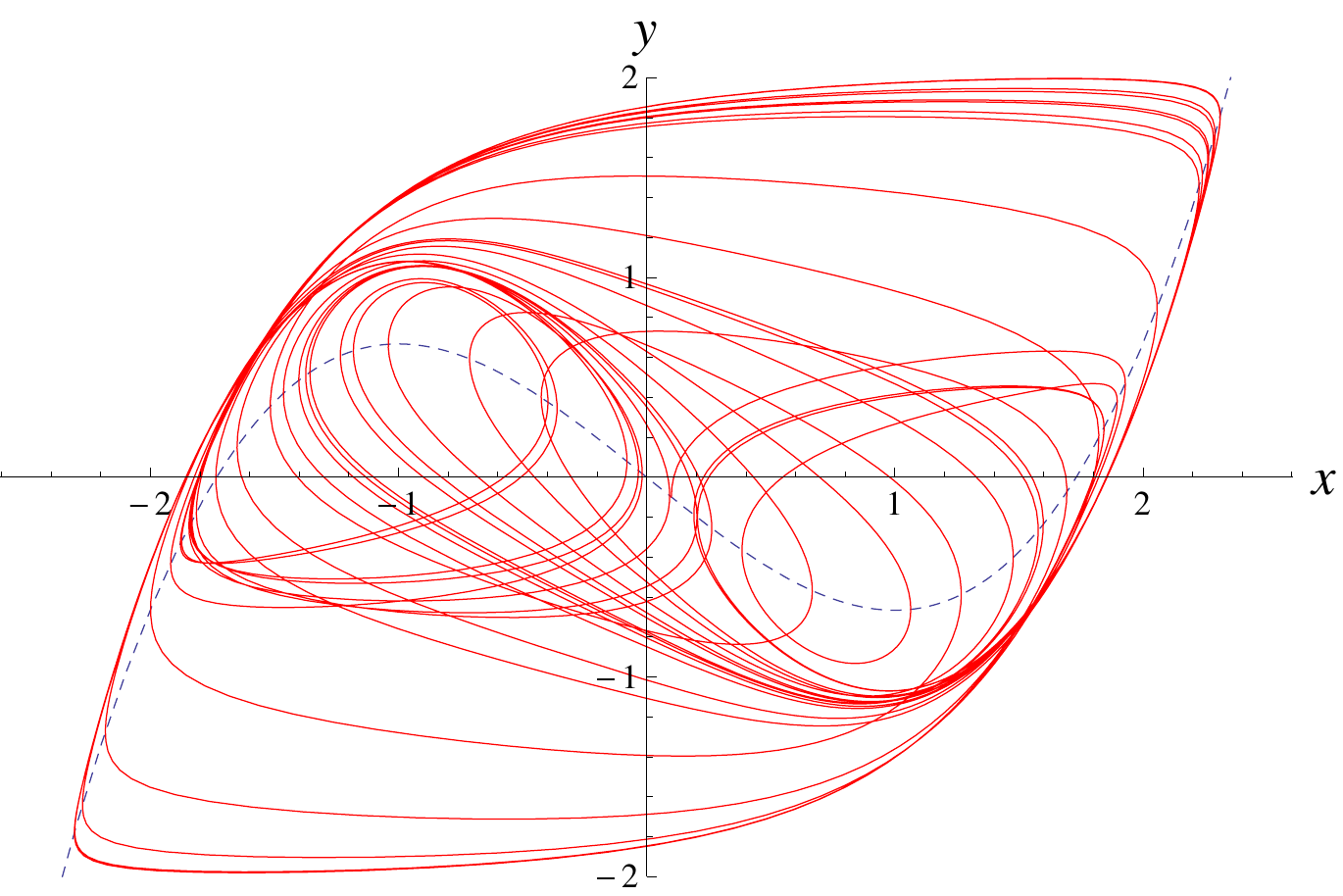}}
\end{center}
\caption{\small Solutions to \hbox{$\ddot x - 3 \left( 1 - x^2 \right) \dot x + x = 5 \cos (1.788 t)$} show sensitive dependence to initial conditions.  In (a) the waveform in solid red starts at $x(0) = 0.1$, $\dot x(0) = 0$ and the one in dotted blue at $x(0) = 0.1$, $\dot x(0) = 0.01$.  The chaotic Li\'enard trajectory (with $y \equiv \dot x / 3 - x + x^3 / 3$) for the first of those solutions is shown in (b), for $0 \leq t \leq 100$.  The vertical isocline is indicated by the dashed blue curve.\la{fig:chaos}}
\end{figure}

\subsubsection{FitzHugh-Nagumo model}
\la{sec:FHN}

As we mentioned in \Sec{sec:heart}, the FitzHugh-Nagumo model of neuronal action potentials \cite{FitzHugh,Nagumo} is of considerable significance in modern neuroscience (see \cite{Izhikevich}).  It was introduced by FitzHugh as a simplification of a four-variable model proposed earlier by Alan Hodgkin and Andrew Huxley to account for their observations of the behavior of the squid giant axon \cite{Hodgkin-Huxley}.  (For that work, Hodgkin and Huxley shared the 1963 Nobel Prize in physiology or medicine with Sir John Eccles \cite{Nobel-Med63}.)

Mathematically, the FitzHugh-Nagumo model is characterized by
\be
\left\{ \begin{array}{l} \dot x = y - F(x) + I(t) \\
\dot y = - x - c y \end{array} \right. ~,
\la{eq:FHN-xy}
\ee
which is equivalent to the second-order, inhomogeneous equation
\be
\ddot x + \left( f(x) + c \right) \dot x + \left( x + c F(x) \right) = c I (t) + \dot I (t)~,
\la{eq:FHN-fg}
\ee
where $f(x) = F'(x)$.  This can be easily implemented electrically by adding a resistor in parallel with the inductance $L$ in the circuit of \Fig{fig:vdP}(a) and supplying a time-varying current $I_0 + I(t)$.  Then $x$ corresponds to the voltage $V_{\rm out}$ and $y$ corresponds to the current flowing from ground through the inductor.  As before, $F(x)$ is given by the characteristic $I$-$V$ curve of the tunnel diode.  Many authors use the same $F(x)$ as in the van der Pol oscillator:
\be
F(x) = \alpha (-x + x^3/3)~.
\la{eq:FHN-F}
\ee

In modeling neuronal action potentials, $I(t)$ is usually interpreted as a slowly-varying stimulus current, so that $\dot I$ is negligible and the driving is not a significant effect.  Rather, what is important is that a steady $I \neq 0$ displaces the point along the characteristic $I$-$V$ curve of \Fig{fig:vdP}(b) about which the circuit operates.

In \Eq{eq:FHN-xy}, the equilibrium point \hbox{$\dot x = \dot y =0$} is given by the intersection of the vertical isocline \hbox{$y = F(x) - I$} with the horizontal isocline $y = - x / c$.  For $c > 0$, the stimulus current $I$ displaces that equilibrium away from the inflection point (authors usually choose a $c$ small enough so that the horizontal isocline can intersect the vertical isocline only once, always giving a single equilibrium).  This displacement makes the oscillation asymmetric, for the same reason described in \Sec{sec:asymmetric}.  For sufficiently large $| I |$, the equilibrium is displaced to a point along the characteristic curve where the slope is positive and self-oscillation therefore ceases, since the linear damping term is no longer negative and the corresponding equilibrium is therefore stable.

\begin{figure} [t]
\begin{center}
	\subfigure[]{\includegraphics[width=0.5 \textwidth]{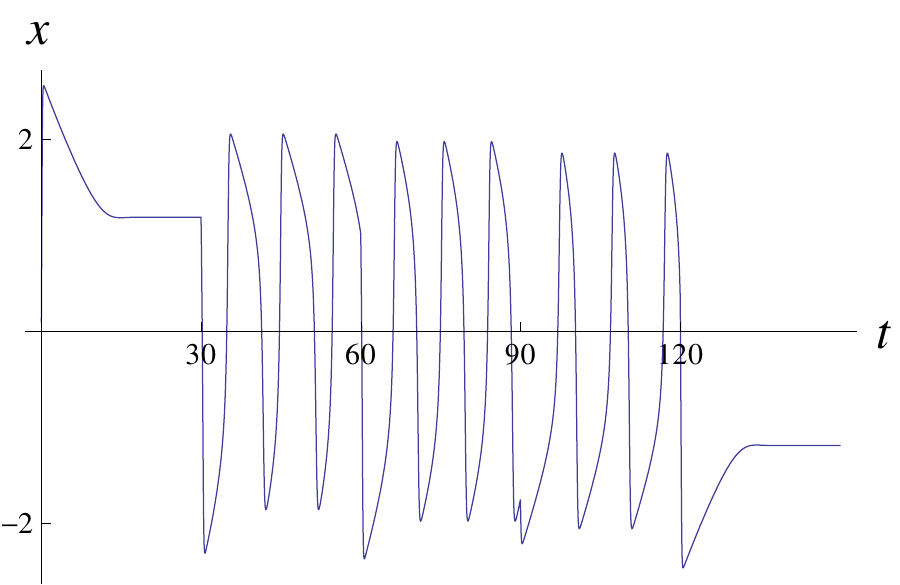}} \hskip 2 cm
	\subfigure[]{\includegraphics[width=0.25 \textwidth]{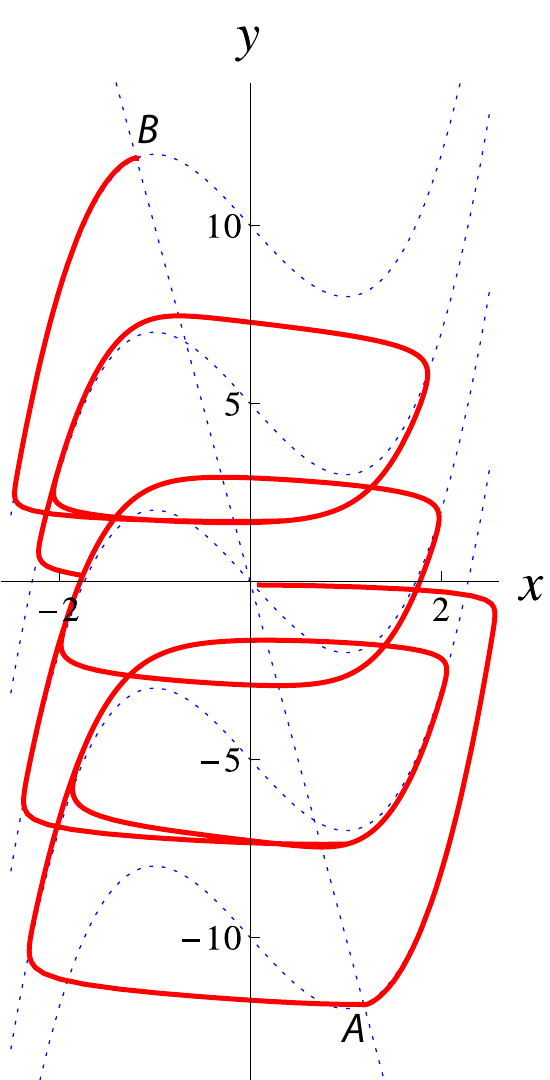}}
\end{center}
\caption{\small Solution to the FitzHugh-Nagumo equation (\Eq{eq:FHN-xy}) for $F(x) = x^3 - 3x$ and $c = 0.1$, as $I$ is reduced by steps for size 5, starting from $I=10$, at time intervals of 30.  Plot (a) shows the waveform for $x(t)$.  Plot (b) shows the Li\'enard trajectory, with the horizontal isocline and the successive vertical isoclines shown in dashed blue. \la{fig:FHN}}
\end{figure}

In this way, the model captures the ability of neuronal firing to be turned on and off.  Note that the introduction of the term $-cy$ in the expression for $\dot y$ in \Eq{eq:FHN-xy}, which corresponds to adding a resistor in parallel with the inductor in the circuit of \Fig{fig:vdP}(a), is what allows one to adjust the point of operation along the diode's characteristic curve by varying the current $I_0$ without having also to adjust the voltage $V_0$ of the battery.

Figure \ref{fig:FHN} illustrates how the relaxation oscillation turns on and off, and how the asymmetry of the oscillation changes, as the value of the stimulus current $I$ is varied.  In \Fig{fig:FHN}(a), the waveform for $x(t)$ is plotted as $I$ is successively stepped down from 10 to 5 (at $t = 30$), from 5 to 0 (at $t=60$), from 0 to -5 (at $t=90$), and from -5 to -10 (at $t=120$).  The same oscillation is shown as a trajectory in the Li\'enard plane in \Fig{fig:FHN}(b), where the straight horizontal isocline $y = -10 x $ and the successive vertical isoclines $y = x^3 - 3x - I$ are plotted as dashed lines.  The points marked $A$ and $B$ are stable equilibria, corresponding to the non-oscillatory phases of the waveform for $x(t)$, when $I = \pm 10$ respectively.

\subsection{Poincar\'e-Lyapunov stability analysis}
\la{sec:Lyapunov}

Poincar\'e \cite{Poincare-stability} and Lyapunov \cite{Lyapunov-stability} formulated stability analysis ---including the stability at a point and the stability of time-dependent trajectories such as limit cycles--- in a more sophisticated mathematical language, based on Poincar\'e's phase-space method for the study of nonlinear differential equations.\footnote{Poincar\'e's work on this subject, which he called the ``qualitative theory of differential equations,'' marks the birth of modern differential topology; see \cite{Dieudonne}.  It is also the source of most of the modern mathematical theory of dynamical systems, including chaos theory. \cite{history-chaos}}  It is this formulation which is most familiar to scientists today.  See \cite{Lyapunov} for a brief summary of this work.  Arnol'd gives a particularly readable introduction to this subject in his remarkable textbook on ordinary differential equations \cite{Arnold-Lyapunov}.  Other treatments, written for physicists, include \cite{Andronov-Lyapunov,Jose-Lyapunov}.

\subsubsection{Stability in phase-space}
\la{sec:phasespace}

Consider, for instance, a one-dimensional system with a nonlinear equation of motion
\be
\ddot x + \left( \mu + x^2 \right) \dot x + x = 0 ~,
\la{eq:Hopf}
\ee
where $\mu$ is now thought of as a parameter which can be varied externally.  The Li\'enard transformation\footnote{In the mathematical literature, it is more common to use Poincar\'e's phase-space transformation, with $y \equiv \dot x$.  We prefer to invoke the closely-related Li\'enard transformation here, since we used it extensively in \Sec{sec:limits}.} for \Eq{eq:Hopf} gives
\be
\left\{ \begin{array}{l} \dot x = y - \mu x - x^3/3 \\ \dot y = -x \end{array} \right. ~.
\la{eq:Hopf-Lienard}
\ee
The point $(x,y) = 0$ is an equilibrium.  The linearized equation of motion about it can be written in matrix form as
\be
\left( \begin{array}{l} \dot x \\ \dot y \end{array} \right) = 
\left( \begin{array}{c c} -\mu & 1 \\ -1 & 0 \end{array} \right)
\left( \begin{array}{l} x \\ y \end{array} \right) ~,
\la{eq:Hopf-matrix}
\ee
whose solution is
\be
\left( \begin{array}{l} x(t) \\ y(t) \end{array} \right) = 
\exp \left[ t \left( \begin{array}{c c} -\mu & 1 \\ -1 & 0 \end{array} \right) \right]
\left( \begin{array}{l} x_0 \\ y_0 \end{array} \right) ~,
\la{eq:Hopf-matrix-soln}
\ee
for initial values $x(t=0) = x_0$, $y(t=0) = y_0$.

By diagonalizing the matrix in \Eq{eq:Hopf-matrix}, this solution can be expressed as
\be
\left( \begin{array}{l} x(t) \\ y(t) \end{array} \right) =
{\MM B} \left( \begin{array}{c c} e^{\lambda_1 t} & 0 \\ 0 & e^{\lambda_2 t} \end{array} \right) {\MM B}^{-1}
\left( \begin{array}{l} x_0 \\ y_0 \end{array} \right) ~,
\ee
where
\be
\lambda_{1,2} = \frac{1}{2} \left(-\mu \pm i \sqrt{4 - \mu^2} \right)
\la{eq:Hopf-eigenvalues}
\ee
and
\be
{\MM B} =  \left( \begin{array}{c c} \lambda_1 & \lambda_2 \\ -1 & -1 \end{array} \right) ~.
\ee
Thus, for $0 < \mu < 2$ the real part of $\lambda_{1,2}$ is negative and trajectories will spiral in towards the origin.  On the other hand, for $-2 < \mu < 0$, trajectories that start near the origin will spiral out, until the cubic term in \Eq{eq:Hopf-Lienard} can no longer be neglected.  (For a thorough treatment of linear differential equations in the plane, see \cite{B&S-plane}.)

In other words, the origin is a stable equilibrium for $\mu > 0$ and an unstable equilibrium for $\mu < 0$.  The physical reason for this is obvious: as $\mu$ crosses 0 the linear damping term changes sign.  A negative damping leads to self-oscillation, as we discussed in \Sec{sec:negative-damping}.

\subsubsection{Hopf bifurcation}
\la{sec:Hopf}

\begin{figure} [t]
\begin{center}
	\includegraphics[width=0.4 \textwidth]{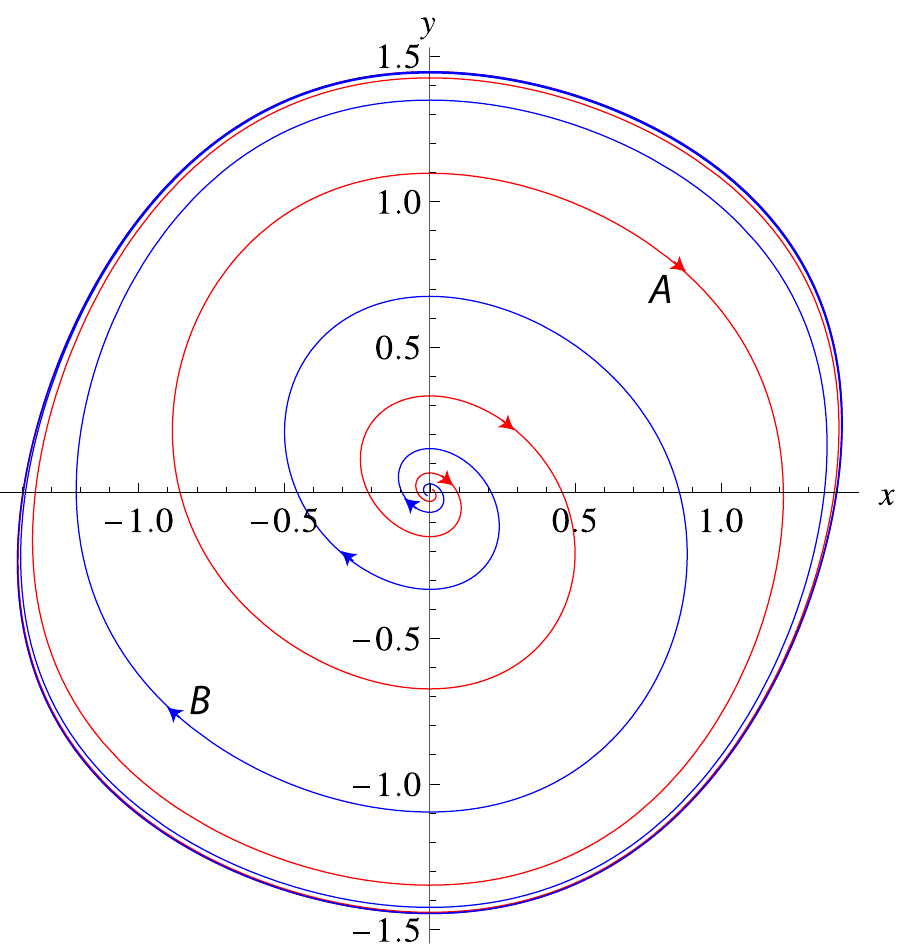}
\end{center}
\caption{\small Solutions to \Eq{eq:Hopf-Lienard} for negative damping $\mu = -0.5$.  The orbits labeled $A$ (shown in red) and $B$ (in blue) are related by $(x,y) \to (-x,-y)$ and $(\dot x, \dot y) \to (-\dot x, -\dot y)$.  They bifurcate at the origin and converge to the same nonlinear limit cycle.\la{fig:Hopf}}
\end{figure}

A system such as that of \Eq{eq:Hopf} has a ``Hopf bifurcation'' at $\mu = 0$.  In the mathematical theory of dynamical systems, the term {\it bifurcation} (or {\it branching}) is applied to any qualitative change in the solutions to the equations of motion that occurs as a parameter is varied past a critical value; see \cite{bifurcation-Scholarpedia} and references therein.  In this particular case, ``bifurcation'' is appropriate because, as the parameter $\mu$ changes sign from positive to negative, the origin is destabilized, which causes pairs of orbits related by the invariance of \Eq{eq:Hopf} under $x(t) \to -x(t)$ (or, equivalently, by the symmetry of \Eq{eq:Hopf-Lienard} under $(x,y) \to (-x,-y)$; $(\dot x, \dot y) \to (-\dot x, -\dot y)$) to bifurcate from the origin.  These orbits spiral out and approach the same nonlinear limit cycle, as illustrated in \Fig{fig:Hopf}.

This picture of the transition between stability and instability upon variation of a parameter past a critical point was introduced by Poincar\'e in \cite{Poincare-stability}.  The case of two-dimensional phase-space was worked out in detail by Andronov and his collaborators in the 1930s; see \cite{Andronov-Hopf}.  The corresponding analysis in an arbitrary (but finite) number of dimensions was published by Hopf in 1942 \cite{Hopf}.  The same phenomenon is therefore also called Andronov-Hopf or Poincar\'e-Andronov-Hopf bifurcation.  For a full mathematical treatment of this subject, including the generalization to infinite-dimensional phase-spaces (such as are used to describe equations with finite delays and certain partial differential equations, including the Navier-Stokes equations for fluids), see \cite{Marsden-Hopf}.

The physics involved in Hopf bifurcation is the same that was described by the older Routh-Hurwitz stability analysis.  It is even possible that the greater mathematical elegance and sophistication of the treatment due to Poincar\'e et al.\ may have helped divert attention from the energetic considerations that we stress in this article.  Note, for instance, that the mathematical treatment of \Eq{eq:Hopf} in terms of varying $\mu$ past its critical point at zero obscures the fact that $\mu > 0$ occurs quite naturally because of dissipation, whereas $\mu < 0$ requires a feedback between the oscillator and an external source of power, as we discussed in \Sec{sec:feedback}.  In our view at least, it is the mechanism of this feedback, rather than the characterization of the solutions of \Eq{eq:Hopf} and similar nonlinear equations of motion, that is most useful and interesting to a physicist.

Hopf bifurcation is well-covered in the current literature on the theory of dynamical systems; see, e.g., \cite{Strogatz-Hopf}.  We will therefore not comment further on this particular formulation of stability analysis.

\section{Motors}
\la{sec:motors}

Andronov, Vitt, and Kha\u{\i}kin, defined self-oscillation as the production of ``a periodic process at the expense of a non-periodic source of energy'' \cite{Andronov-defn}.  Only a regular periodic motion is technologically useful, but most of the sources of power provided by nature are steady or slowly varying with respect to the time scales relevant to human technology.  Therefore self-oscillators, which turn a steady input into an alternating output, are essential to engineering.

\subsection{Turbines}
\la{sec:turbines}

The oldest motors are turbines, which use a steady flow to drive a circular motion.  This circular motion can then be transformed into the one-dimensional ``reciprocating'' action of a piston,\footnote{The two principal mechanisms for converting between circular motion and reciprocating linear motion are scotch yokes and cranks.  A scotch yoke converts a uniform circular motion into a strictly sinusoidal motion of the same period (see \cite{yoke}) but lubrication is needed to avoid significant power loss and wear from friction as the pin that is attached to the wheel moves inside the yoke.  A crank is usually more practical, but it will introduce distortion, so that a uniform rotation of the wheel will not give perfectly sinusoidal reciprocation.} or if the turbine drives a generator, into an alternating electrical current.  In this sense, turbines are self-oscillators, since they can produce a periodic motion without a periodic input.

The reciprocating motion generated by a turbine is similar to a relaxation oscillation in that the amplitude is fixed but the frequency may vary depending on the speed of the flow and the friction on the turbine (the turbine's resonant frequency is zero, because there is no restoring force to move it back towards an equilibrium).  However, unlike the relaxation oscillators described in \Sec{sec:relaxation}, a turbine can produce a {\it sinusoidal} reciprocation.  Rather than a sudden switching at the thresholds, for a piston connected to a turbine the nonlinear switching is gradual, being given by the turbine's circular turning.

\subsection{Heat engines}
\la{sec:heat}

In practical steam engines, the feedback of the flow of steam on the turbine's motion is effected by a {\it valve}, which modulates the pressure of the steam in phase with the linear velocity of the piston attached to the turbine.  The operation of such a steam engine is characterized mathematically as a self-oscillation in \cite{Andronov-steam}.  Note that the action of the valve in such engines is not, strictly speaking, thermodynamical.  For instance, the setup analyzed in \cite{Andronov-steam} would work equally well if the working substance were a fluid moved by a mechanical pump, instead of steam generated in a boiler.

It is also possible to achieve the feedback necessary for self-oscillation by controlling the release of heat, rather than the mechanical motion of the working substance.  This is the case of a Diesel engine, in which the compression of the working gas by the piston causes the fuel to ignite, injecting a large amount of heat into the gas.  In an Otto engine (such as the one in a gasoline-powered automobile), the injection of heat into the working gas at the phase of maximum compression is triggered by a spark-plug, whose action is timed accordingly.  (On the Diesel and Otto thermodynamic cycles, see, e.g., \cite{Moran}.)

There is also a class of ``thermoacoustic self-oscillators,'' which we shall describe in detail in \Sec{sec:thermoacoustic}, in which a positive feedback is naturally established by the flow of heat between the working substance and its surroundings.  A particularly fascinating and scientifically relevant instance of such a heat engine is seen in Cepheid variable stars, which we shall describe in \Sec{sec:cepheids}.

\subsection{Limit efficiency}
\la{sec:efficiency}

Philippe Le Corbeiller\footnote{Philippe Emmanuel Le Corbeiller (1891--1980), was a French-American electrical engineer, mathematician, and physicist, whose intellectual interests were remarkably broad.  He was educated at the \'Ecole Polytechnique, worked on telegraph and radio systems, and earned a doctorate in mathematics from the Sorbonne.  Le Corbeiller emigrated to the United States during World War II and became a professor of applied physics and general education at Harvard \cite{LeC-bio}.} labeled self-oscillators as ``motors,'' and passive devices (including forced resonators) as ``transformers''  \cite{LeC-review-Fr,LeC-review-Eng}.  Le Corbeiller also noted that the efficiency of a ``transformer'' (i.e., the ratio of the power received to the power used to drive the output) can be taken to unity ---if nonessential losses are eliminated--- because the power is delivered at the same frequency with which the output moves.  But when the power is inputted at a frequency different from that of the movement of the output, there is an {\it essential} loss of power that cannot be eliminated.  This is the case for ``motors'' in general, for which the input is steady and the output periodic.

In the van der Pol circuit of \Fig{fig:vdP}, the current that passes through the tunnel diode oscillates about $I_0$, while the voltage drop across the diode oscillates about $V_0$.  Thus there is an average power loss
\be
\langle P_{\rm diode} \rangle > V_0 I_0 > 0~,
\la{eq:P-diode}
\ee
in addition to any power dissipated in the resistance $R$.  This essential loss $\langle P_{\rm diode} \rangle $ cannot be eliminated: no device can have a characteristic $I$-$V$ curve with negative slope at the origin ($V_0 I_0 = 0$), since it would require the electrons to move against the electric field.

The same principle is illustrated by the operation of the simplest turbine: the overshot water wheel, shown in \Fig{fig:water-wheel}.  The water flow provides kinetic and gravitational potential energy, which are converted into the wheel's rotation.  But some of the gravitational potential energy {\it must} be wasted, even if the wheel turned without friction, because water begins to spill out of the buckets before they reach the bottom.  This ``tail'' flow is an essential power loss.  That loss vanishes only in the limit of infinite wheel radius, in which case the motion ceases to be periodic.  In an undershot wheel, the paddles must move out of the flow as the wheel turns, and the maximum efficiency is therefore also strictly less than unity for a wheel of finite radius.

\begin{figure} [t]
\begin{center}
	\includegraphics[width=0.4 \textwidth]{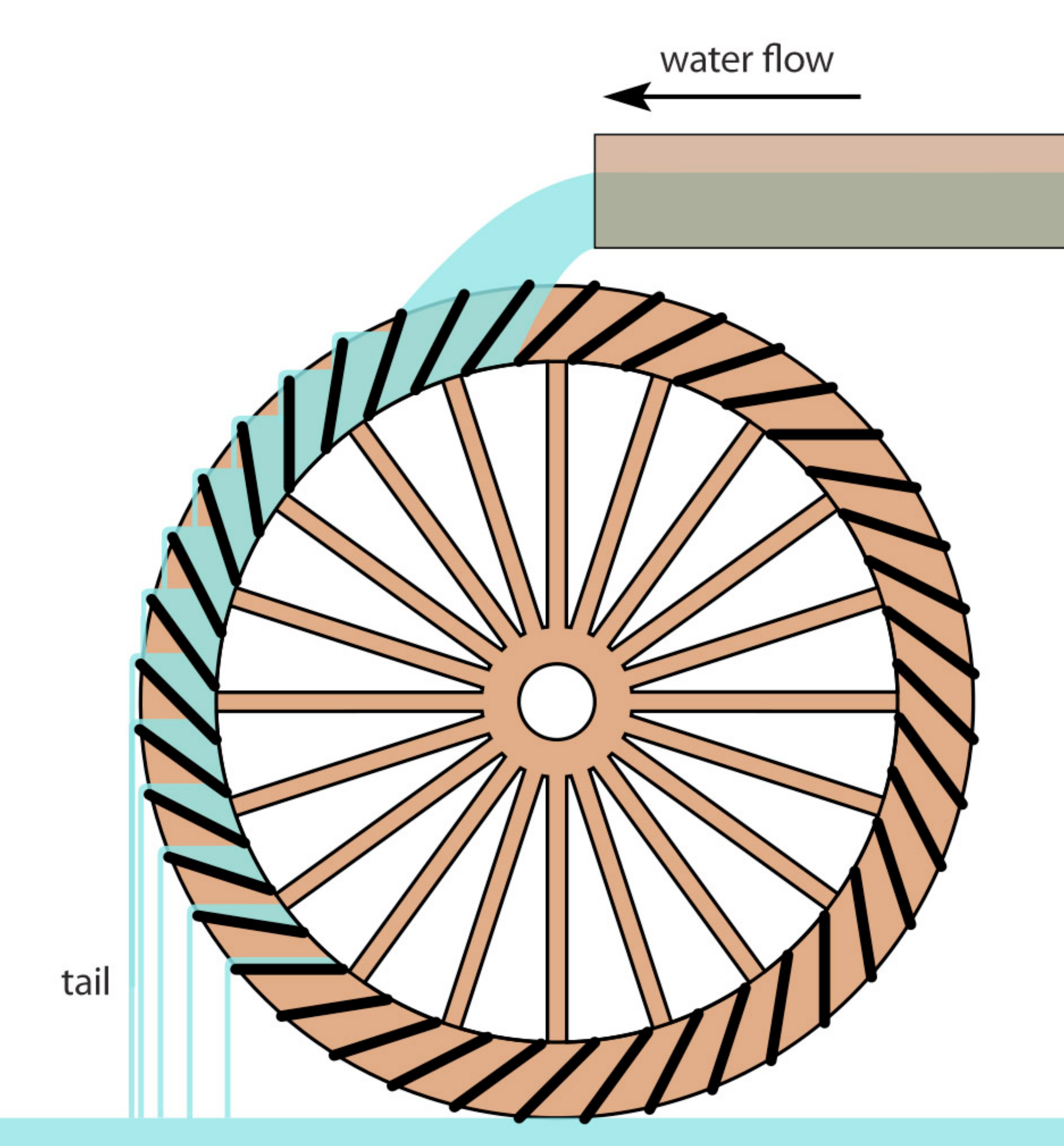}
\end{center}
\caption{\small Overshot water wheel.  The power in the ``tail'' flow is an essential loss, making the efficiency of the wheel strictly less than 1.  The illustration is by Daniel M. Short, available at \url{http://commons.wikimedia.org/wiki/File:Overshot_water_wheel_schematic.svg}\la{fig:water-wheel}}
\end{figure}

Note that the fact that water spills out of the buckets of the wheel before they reach the bottom is what accounts for the modulation of the force of the water in phase with the horizontal velocity of the buckets:  the force is large when the buckets are moving leftwards and downwards, smaller when the buckets are moving rightwards or upwards.  Thus, the loss of power in the tail is necessary to set up the negative damping that causes the wheel to turn (i.e., to self-oscillate).\footnote{In \cite{Denny-buckets}, M.~Denny proposes an overshot water wheel in which the buckets are attached by pivots to the rim of the wheel.  Each bucket is tipped over and emptied at the bottom of the cycle, eliminating the tail flow.  But this design cannot eliminate the power loss, even in principle, because the tipping of a bucket necessarily consumes energy.  Even if each bucket were filled up to the maximum stable level, at which its center of mass coincides with the pivot, the center of mass descends relative to the pivot as the bucket starts to empty.  It therefore would take finite energy to complete the emptying of each bucket, even if all friction and other forms of dissipation were eliminated.}

Modern designs attain about 85\% and 75\% efficiencies for over- and undershot wheels, respectively \cite{wheel-efficiency}.  The most common type of hydraulic turbine used today for industrial applications is the radial inflow (or ``center-vent'') turbine developed by James B.~Francis in 1848, which spins on a horizontal plane and in which the tail flow leaves though the center of the wheel (see, e.g., \cite{Francis-turbine}).  The energy of this tail flow cannot be reduced beyond a certain limit, because the water must drain out at the same rate at which it enters the turbine.  If gravity cannot cause the water at the center of the turbine to drain quickly enough, it will be pushed out by the flow, which implies a waste of kinetic energy.

Another hydraulic instance of this same general principle is seen in a cistern placed under a steady stream of water and connected to a siphon so that the water level in the cistern performs a relaxation oscillation.\footnote{This setup was first used by Andr\'e Blondel in \cite{Blondel-auto} as a pedagogical model of relaxation oscillators in general.  The same example was later taken up by Le Corbeiller \cite{LeC-review-Fr} (who called it {\it le vase de Tantale}, ``Tantalus's cup''), by Herrenden-Harker \cite{AJP-relax} (who pointed out that it is used by chemists in the form of the Soxhlet extractor), and by Pikovsky et al.\ \cite{Pikovsky-relax}.  It is the same mechanism by which a toilet flushes: the water level rises until it reaches the same height as the elbow of the siphon tube, thereupon causing the cistern to flush, until the water level falls below the siphon's intake and the siphon empties.  (If water flowed continuously and quickly into a toilet bowl, the toilet would flush periodically, causing the level of the water to describe a relaxation oscillation.)}  It is easy to show that the maximum efficiency goes to unity only when the frequency of that oscillation vanishes \cite{AJP-relax}.

In an electrical motor, an essential power loss is caused by the ``slip'' (i.e., the relative motion of the rotor with respect to the magnetic field), which induces an electromotive force (the ``back emf'') that opposes the input voltage.  This slip can vanish only when the input voltage has the same frequency as the motion of the rotor.  Similarly, the self-oscillation (``nutation'') of a spinning top, as described in \Sec{sec:gyroscopic}, must be accompanied by a steady precession, which also consumes energy if the tilt of the top's axis is varying.

Le Corbeiller suggested in \cite{LeC-review-Eng} that Carnot's thermodynamic limit
\be
\eta \leq 1 - \frac{T_2}{T_1}
\la{eq:Carnot}
\ee
for the efficiency $\eta$ of a heat engine that absorbs heat from a reservoir at temperature $T_1$ and releases heat into a reservoir at a lower temperature $T_2$, might be an instance of this more general principle.  Note that there is no fundamental obstruction to transforming heat completely into non-periodic motion (e.g., by having an expanding gas push out a piston, or by boiling water to generate a steady flow of steam).  What the second law of thermodynamics forbids is turning the steady heat of a reservoir entirely into the mechanical power that drives a {\it periodic} motion (see \cite{Fermi,Feynman-thermo}).

\subsubsection{Carnot's theorem revisited}
\la{sec:Carnot}

It was Sadi Carnot's uniquely brilliant contribution in \cite{Carnot} to show that no heat engine ---operating between two heat reservoirs each at a constant temperature--- can have greater efficiency than a reversible engine, {\it assuming} that the laws of physics forbid any process that merely converts heat from a source at a single temperature into the mechanical energy of a periodic motion.  This axiomatic approach, later formalized by Clausius and Kelvin, was necessitated at the time by the ignorance of the nature of heat (i.e., by the lack of a {\it microphysical} understanding of thermal processes).  Carnot himself stressed that mechanical engines (such as water wheels) may be completely described by the rules of Newtonian mechanics, without further assumptions.\footnote{Some authors claim that Carnot's work on the efficiency of heat engines was inspired by an analogy between the flow that drives a water wheel ---which his father had investigated in \cite{Lazare}--- and the flow of heat in a steam engine.  There is, however, very little evidence of this in his published essay \cite{Carnot}.  Except for some obscurities associated with the fact that the nature of heat was then totally unknown, Carnot's argument is the same as that given by modern conceptual expositions of the second law of thermodynamics, such as \cite{Fermi} and \cite{Feynman-thermo}.  According to Kelvin, ``nothing in the whole range of Natural Philosophy is more remarkable than [Carnot's] establishment of general laws by such a process of reasoning." \cite{Kelvin-Carnot}}  Carnot then theoretically implemented the reversible heat engine as a cycle of isothermal and adiabatic changes to the volume of an ideal gas, thus allowing him to compute that limiting efficiency in terms of the reservoir temperatures.  Even after the microscopic nature of thermal motion was understood, Carnot's approach has proved highly fruitful, in light of the practical impossibility of keeping track of the state of individual molecules in a heat bath.

Le Corbeiller's conjecture about the efficiency of frequency conversion is easy to show for mechanical systems, in which the input power can be expressed as
\be
P_{\rm in} = {\vv F}_{\rm in} \cdot {\vv v}_{\rm in} ~,
\ee
where ${\vv F}_{\rm in}$ is an instantaneous force and ${\vv v}_{\rm in}$ the velocity of the flow that provides the power.  The power delivered to the output is
\be
P_{\rm out} = \dot {\vv q} \cdot {\vv F}_{\rm out} ~,
\ee
where ${\vv q}$ is the time-dependent coordinate vector that characterizes the output.  By energy conservation, at any instant
\be
P_{\rm out} \leq P_{\rm in}~.
\ee
We may therefore normalize $\vv v_{\rm in}$ so that 
\be
| \vv {F}_{\rm out} | \leq | {\vv F}_{\rm in} | ~,
\ee
with the equality holding only when {\it all} of the flow is applied to the output $\vv q$.  Then we see that unit efficiency ($P_{\rm in} = P_{\rm out}$) is possible only when ${\vv v}_{\rm in}$ moves in phase with $\dot {\vv q}$, which can only be maintained during a complete cycle if the frequency of $\vv q$ matches the frequency of the input.

In particular, if the input and output frequencies do not match, there must be a {\it slip} phase during which the output $\vv q$ moves {\it against} the input, or {\it away} from it, so that $P_{\rm out} \leq 0$.  During this slip phase, $P_{\rm out}$ is non-negative only if the flow is completely diverted away from the output (i.e., if ${\vv F}_{\rm out} = 0$).

If the input and output waveforms have the same form (whether sinusoidal or otherwise), but different periods $\tau_1 < \tau_2$, then the efficiency $\eta$ is limited by
\be
\eta \leq \frac{\tau_1}{\tau_2}~.
\la{eq:AC-slip}
\ee
This can be shown by imagining that the input and output oscillations start in phase at the origin, then letting them run for a sufficient number of cycles until they cross the origin in phase again (or infinitesimally close, if $\tau_1 / \tau_2$ is irrational).  In a steady state in which each cycle consumes the same amount of energy, the difference in the number of input and output cycles can then be equated with the loss of energy from slip.  Equation (\ref{eq:AC-slip}) is familiar from the limiting efficiency of an AC motor in terms of the respective periods of the input voltage and of the rotor's motion (see \cite{LeC-review-Eng}).

When the shapes of the input and output waveform are different (as they must be if the input is non-periodic) the efficiency limit is more difficult to express in generality, but it must still be strictly less than unity if the frequencies are not matched.  For instance, in the case of the stick-slip motion of a violin string, illustrated in \Fig{fig:violin}, during the stick phase $AB$ the violin string moves with the bow ($\dot q = v_{\rm in}$), so that $P_{\rm out} = P_{\rm in}$ (if non-essential power losses are eliminated).  But in order to describe a motion with a finite period, the string must eventually slip against the bow, giving $P_{\rm out} \leq 0$.

Let $v_s$ be the speed of the violin string as it crosses the origin during the slip phase (at point $C$ in the plot of \Fig{fig:violin}).  Since the corresponding kinetic energy cannot exceed the potential energy at the maximum displacement (indicated by $B$ in the plot), for an elastic string
\be
v_s^2 \leq \omega_0^2 A^2~,
\ee
where $\omega_0$ is the string's resonant frequency and $A$ is the amplitude of the displacement.  The time $t_s$ during which the string slips is therefore bounded as:
\be
t_s \geq \frac{2A}{v_s} \geq \frac{2}{\omega_0}~.
\ee
Since the maximum efficiency during the stick phase ($AB$ in the plot) is 1, and the maximum efficiency during the slip phase ($BD$ in the plot) is 0, the total efficiency for the elastic string is limited by
\be
\eta \leq \frac{2 \pi - \omega_0 t_s}{2\pi} = 1 - \frac{1}{\pi} \simeq 0.68~.
\ee
Even for an inelastic string, the limit $t_s \to 0$ requires that the energy of the string (and therefore also its amplitude) diverge, which is impossible with finite power input, except in the limit in which the period of the string's motion also diverges.

Quite in general, we see that the energy wasted during the slip cannot be taken to zero in any system, even under ideal operating conditions, {\it except in a limit in which frequency conversion fails}.  In the familiar case of the thermodynamic Carnot cycle, the slip appears as the phase of isothermal compression of the gas, during which work must be done {\it on} the piston and converted into heat that is released into the colder reservoir at temperature $T_2$.

\begin{figure} [t]
\begin{center}
	\includegraphics[width=0.35 \textwidth]{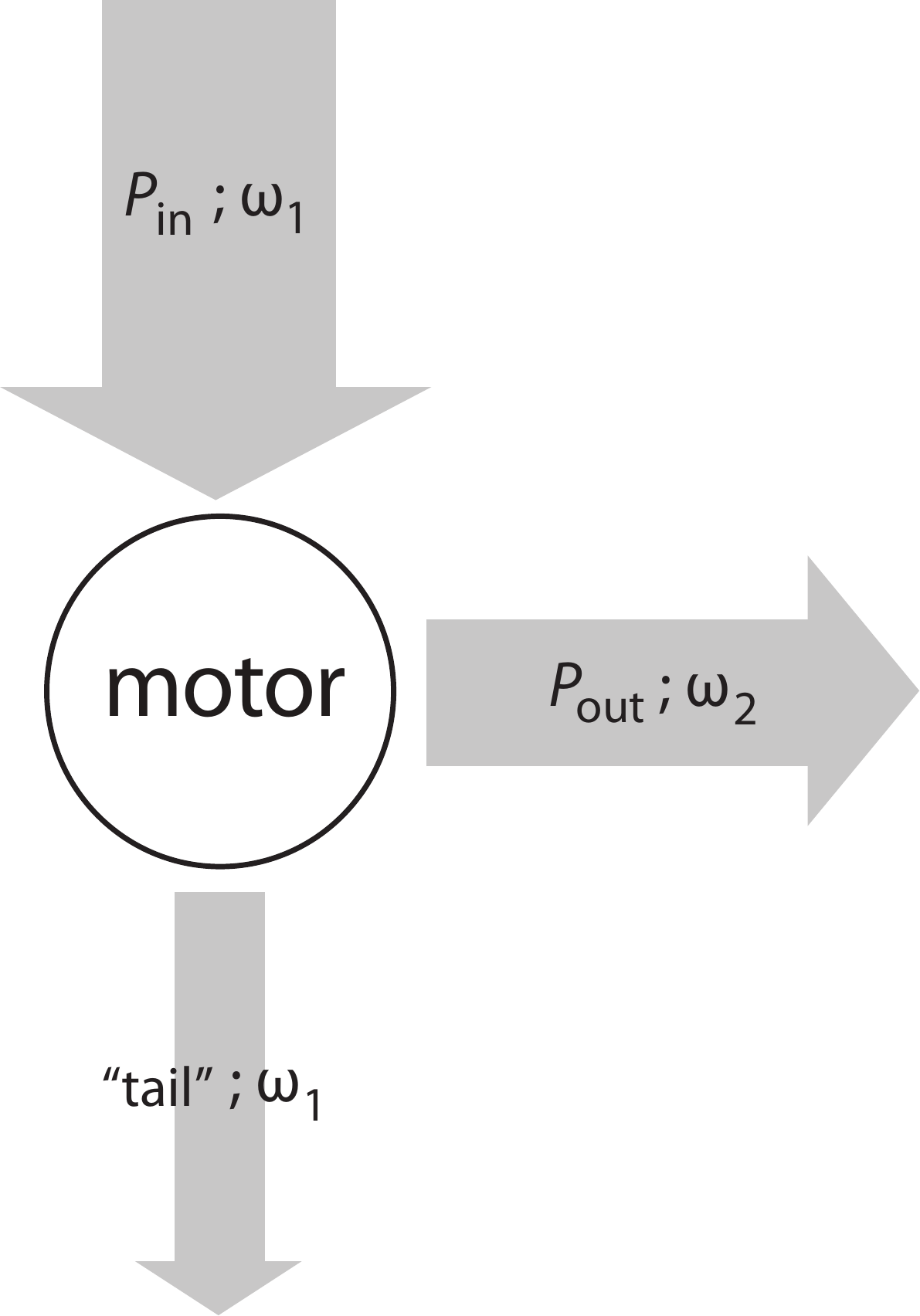}
\end{center}
\caption{\small Schematic representation of the operation of a motor, which takes a power input $P_{\rm in}$, with frequency $\omega_1$, and outputs a power $P_{\rm out}$ at frequency $\omega_2$.  For many motors, including clocks, $\omega_1 = 0$.  The ``tail'' represents the loss of power in the conversion of the two frequencies.\la{fig:Carnot-LeC}}
\end{figure}

This result can then be used to show, by a straightforward generalization of Carnot's argument, that a motor that takes power at frequency $\omega_1$ and outputs it at another frequency $\omega_2$ cannot have greater efficiency than a reversible motor capable of the same conversion; otherwise the more efficient device could drive the other in reverse, making a composite engine capable of slipless frequency conversion.

Thus, we may formalize a {\it Carnot-Le Corbeiller theorem} for the efficiency of motors:
\begin{quote}
The efficiency of any motor converting a power input with frequency $\omega_1$ to an output with frequency $\omega_2$ can approach unity only in the limit $\omega_2 \to \omega_1$.  The maximal efficiency of a motor that converts an input at $\omega_1$ to an output at $\omega_2$ is that of the reversible operation of a given motor design.
\end{quote}
In \Fig{fig:Carnot-LeC}, this is schematically represented by the loss of power in the ``tail'' (which we have named by analogy to the tail flow in \Fig{fig:water-wheel}). 

Note that there is no fundamental obstruction to converting the mechanical power in the tail completely and reversibly from one form to another (e.g., from a flow of water to the motion of buckets), as long as the frequency $\omega_1$ is unchanged.  The power in the tail can also be {\it irreversibly} converted into heat by friction.

Note as well that the $\omega_{1,2}$ in the Carnot-Le Corbeiller theorem are {\it not} analogous to the $T_{1,2}$ in \Eq{eq:Carnot}.  For heat engines, $\omega_1 = 0$ and $\omega_2 > 0$ is arbitrary.  Furthermore, in phenomenological thermodynamics the temperature is {\it defined} in terms of the limit efficiency, as given by the reversible Carnot cycle (see \cite{Fermi,Feynman-thermo}).  All of the details of the system's microphysics that determine the limiting efficiency have been absorbed into the notion of temperature.

Recall that the second law of thermodynamics (in Kelvin's formulation) forbids a machine to return to an initial configuration (described by a coordinate vector $\vv{q^{(0)}}$) after merely having transformed a certain amount of heat into mechanical work.  Similarly, it is clear from the Carnot-Le Corbeiller theorem that {\it no motor} can return to its initial configuration, having merely absorbed power at a frequency different from that of its own motion.  For example, the ``overbalanced'' wheel in \Fig{fig:Worcester-wheel} fails because gravity is steady and therefore pulls down with the same force when the weights are moving down as when they are moving up, so that integrating over one period $\tau$ of the motion of any given weight gives
\be
W_{\rm net} = \int_0^\tau dt \, {\vv F} \cdot \dot {\vv q} = - mg \int_0^\tau dt \, \dot y
= mg \left[ y(0) - y(\tau) \right] = 0 ~,
\ee
where $y(t)$ is the vertical coordinate for the weight and $mg$ is the magnitude of the downward force exerted by gravity.  The water wheel in \Fig{fig:water-wheel} must waste some of the power of the water because it must be diverted into an outgoing tail flow, in order to prevent it from acting on the buckets when they are moving in a direction contrary to that of the flow (i.e., either to the right or upward in the illustration).

\subsubsection{Geometric frequency conversion}
\la{sec:geometric}

Parametric resonance, discussed in \Sec{sec:parametric}, would seem to offer a counterexample to the Carnot-Le Corbeiller theorem, since small angular oscillations can be driven with unit efficiency when the driving frequency is {\it twice} the frequency of angular motion.  Note, however, that it is actually the frequency of the vertical displacement $y$ (which is twice the frequency of the angular motion $\theta$) that should be interpreted as $\omega_2$ when applying the Carnot-Le Corbeiller theorem to the parametrically driven pendulum of \Fig{fig:parametric-pendulum}.

In other words, the lossless conversion of frequencies is purely {\it geometrical}, not dynamical.  What the Carnot-Le Corbeiller theorem establishes is that the dynamical modulation of the input power $P_{\rm in},$ in order to convert it into an output $P_{\rm out}$ with a different frequency, cannot be a lossless process.  In a pendulum that is driven parametrically with unit efficiency there is no such modulation.  As explained in \Sec{sec:parametric}, the conversion between the frequency $\omega$ of the angular displacement $\theta(t)$ and the frequency $2 \omega$ of the vertical displacement $y(t)$ is a purely geometric effect, which can be seen even in an {\it unpowered} pendulum.

We do not know whether, by working in a multidimensional space, geometrical frequency conversion can give a ratio other than 2.\footnote{The instances of geometric frequency conversion that we have been able to identify all involve {\it reflection}: for instance, in a pendulum the conversion is given by the symmetry upon reflection along the vertical axis $x = 0$.  In such cases, it seems reasonable to expect that geometric conversion could only introduce a factor of 2.}  Such conversion cannot, in any case, be applied to devices in which $\omega_1 = 0$ (such as heat engines and most other motors) or to any device in which  $\omega_1 / \omega_2$ can vary continuously.

\subsubsection{Minimal friction in mechanisms}
\la{sec:mechanisms}

The reason why the efficiency of a turbine must be strictly less than unity is intuitively clear: regardless of the design, there is a ``tail flow'' that leaves the turbine with finite energy, as shown in \Fig{fig:water-wheel}.  The energy in this tail was not transferred to the turbine and is therefore wasted.  Let us now imagine a system in which the energy in the tail may be recycled.  If some of the energy of the tail flow in \Fig{fig:water-wheel} could be captured and re-applied to the wheel, then the wheel would speed up relative to the water, so that a steady incoming flow would no longer cause the wheel to turn at a constant rate.

A realistic example of a mechanism in which the tail's energy can be recycled is a pair of gears, in which the teeth of the driver wheel (to which power is applied directly) disengage from the teeth of the follower wheel (which receives its power from the driver), carrying with them a finite kinetic energy, but then come back around after one revolution and reengage.\footnote{I thank Charlie Bennett and Graeme Smith for first calling my attention to this system.}  Clearly, the overall process can be lossless only if the kinetic energy in the driver's teeth just after disengagement (the ``tail'') decreases steadily relative to the kinetic energy of the follower, which implies that the driver must be continually decelerating with respect to the follower.  Not only would this imply that the angular velocities of the wheels could not be constant, it could not be sustained because the gear teeth would become misaligned.

In \cite{LeC-review-Eng}, Le Corbeiller pointed out that an angular velocity $\omega_1$ can be converted to different angular velocity $\omega_2$ by using gears of radii $r_{1,2}$ so that
\be
\omega_1 r_1 = \omega_2 r_2~.
\la{eq:gears}
\ee
He thought that the efficiency of this conversion could, in principle, be taken to 1 by eliminating unessential frictional losses, and that this process was therefore not limited by what we have called the Carnot-Le Corbeiller theorem.  Like many others, both before and after him, Le Corbeiller was unaware that a finite frictional loss is necessary if a driver wheel is to turn a follower wheel steadily.  Actually, Euler had already made this point in his studies of gear teeth, where he showed that the sliding of the teeth is inevitable if the ratio of the angular velocities of the two wheels is to remain constant \cite{Euler-gears1,Euler-gears2}.\footnote{Guido Festuccia gave me valuable assistance in understanding Euler's work on gears, which was published in Latin and has remained untranslated.}  Even though Euler is widely credited with having invented involute gears (today the most common design for gearing), as well as with the first derivation of the important Euler-Savary formula in planar kinematics (see \cite{Euler-preface,Koetsier}), his arguments about friction in gears have received little attention since their publication in the 1760s.

Rather than attempting to reproduce Euler's intricate geometric argument, here we shall give a simple energetic treatment of the same problem.  Let us consider a given tooth on the driver gear that comes in contact with a tooth on the other follower, as shown in \Fig{fig:gears}.  Let $\vv v_{\rm in}$ and $\vv v_{\rm out}$ be, respectively, the instantaneous velocities of the driver tooth and the follower tooth, which by \Eq{eq:gears} are of equal magnitude.\footnote{Strictly, rather than considering entire teeth, we should apply this analysis to mass elements at the point of contact between the teeth.}   If the driver tooth is powered by a tangential force $\vv F_{\rm in}$, then
\be
P_{\rm in} = \vv F_{\rm in} \cdot \vv v_{\rm in} = F_{\rm in} \omega_1 r_1~.
\ee
(There will, of course, be other forces acting on the driver tooth, including a centripetal force that keeps the tooth attached to the wheel, but only an $\vv F_{\rm in}$ parallel to the instantaneous velocity $\vv v_{\rm in}$ can deliver power.)  For $\omega_1$ to be constant, $\vv F_{\rm in}$ must be exactly cancelled by the reaction $- \vv F_{\rm out}$ from its pushing on the follower tooth.

By assumption, all of the positive power on the follower tooth must come from the action of the driver at the point of contact: any work done by the body of the follower wheel on the follower tooth is an unessential {\it loss}. In the steady state, the power delivered to the follower is therefore
\be
P_{\rm out} \leq \vv F_{\rm out} \cdot \vv v_{\rm out} = F_{\rm in} \omega_1 r_1 \cos \theta \leq P_{\rm in}~,
\la{eq:Pin-Pout}
\ee
where $\theta$ is the angle between $\vv v_{\rm in}$ and $\vv v_{\rm out}$.

Thus, some power must be wasted, unless the motions of the teeth are aligned ($\theta = 0$), which occurs only at the configuration of \Fig{fig:gears}(b).  When the teeth are engaging, as in \Fig{fig:gears}(a), or disengaging, as in \Fig{fig:gears}(c), they are moving relative to each other along the direction of a line that joins the centers of the two wheels, and the teeth are therefore {\it slipping}.  Equation (\ref{eq:Pin-Pout}) implies that the limiting efficiency for a given gear design is
\be
\eta \leq \frac{1}{\tau} \int_0^\tau dt \, \cos \theta (t) < 1 ~,
\la{eq:efficiency-theta}
\ee
where $\theta$ is the angle between $\vv v_{\rm in}$ and $\vv v_{\rm out}$, and the time $t$ is measured from the instant $t=0$ when a given pair of teeth engage to the instant $t = \tau$ when they disengage.

\begin{figure} [t]
\begin{center}
	\subfigure[]{\includegraphics[width=0.2 \textwidth]{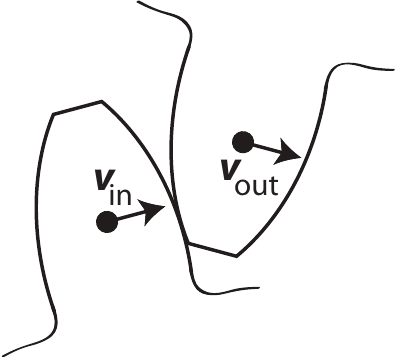}} \hskip 0.4 cm
	\subfigure[]{\includegraphics[width=0.2 \textwidth]{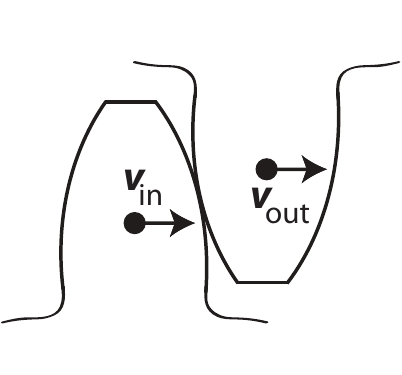}} \hskip 0.4 cm
	\subfigure[]{\includegraphics[width=0.2 \textwidth]{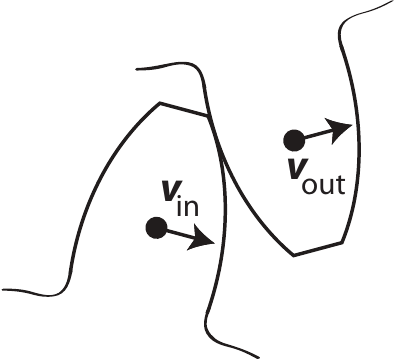}}
\end{center}
\caption{\small At (a), a tooth on the driver gear (below) has begun to engage a tooth on the follower gear (above).  At (b), the two teeth are fully engaged and there is no slippage between them.  At (c), the teeth have begun to disengage.  The label $\vv v_{\rm in}$ indicates the instantaneous velocity of the driver tooth, and $\vv v_{\rm out}$ that of the follower tooth.  Image adapted from Fig.~8 in \cite{wear-gears}.\la{fig:gears}}
\end{figure}

In practice, the difference $P_{\rm in} - P_{\rm out}$ must go into heating the gears by the friction associated with the slipping of the teeth against each other.  If there were no friction at all, so that $P_{\rm out} = P_{\rm in}$, then for $\theta \neq 0$ we would have either \hbox{$F_{\rm out} > F_{\rm in}$} or $\omega_2 r_2 > \omega_1 r_1$, so that slipping would cause the follower gear to {\it speed up} relative to the driver, causing the teeth to become misaligned and finally crush each other.

Note that there must be a loss from slippage even for $r_1 = r_2$, in which case the gears turn with the same angular velocity $\omega_1$, but in opposite directions.  The power loss can only be eliminated in the limit $r_{1,2} \to \infty$, in which case the teeth can be permanently engaged and there is no frequency conversion.  The limiting value of $\eta$ in \Eq{eq:efficiency-theta} can be increased by reducing the size of the teeth, but the limit in which the teeth become infinitely small is obvious unphysical, and very small teeth are impractical because they cannot sustain much force.\footnote{A better solution than reducing the size of the teeth is to cut them obliquely relative to the axis of rotation, giving so-called helical gears.  See, e.g., Willis's discussion in \cite{Willis-helical} (which, however, incorrectly surmised that friction could be wholly eliminated in this way).}

The claim that gears can operate frictionlessly if designed so that the point of contact of the teeth remains collinear with the centers of the two wheels ---thus eliminating the slipping of the teeth against each other--- dates back to Robert Hooke, who referred to it as ``the perfection of Wheel-work; an invention which I made and produced before the Royal Society in 1666'' \cite{Hooke-gears}.  Airy argued that this was possible if the teeth were cut as logarithmic spirals \cite{Airy-gears}, even though Euler had already shown in \cite{Euler-gears1} that such a design is unusable because the follower would accelerate relative to the driver.  Willis, for his part, argued that friction could be eliminated by cutting the teeth obliquely (what are now called ``helical gears'') so that the point of contact always moved in a direction parallel to the axes of the wheels \cite{Willis-helical}.
  
That sustained powered gearing is strictly impossible without finite friction was demonstrated in Willard Gibbs's 1863 doctoral dissertation \cite{Gibbs-frictionless}, ten years before he turned his attention to thermodynamics.\footnote{Gibbs's thesis refers specifically to ``spur gearing'' in the title, but it is clear from the contents that his results extend as well to what now are called helical gears.  This brief and highly abstract dissertation remained unpublished until 1947, when it appeared with a commentary by Yale engineering professor E.~O.~Waters \cite{Waters} that shows no interest in Gibbs's demonstration of the impossibility of frictionless gearing.  It is true that this is more a question of theoretical physics than of practical engineering, because the action of the gear teeth is not usually the principal source of dissipation in real-world mechanisms, which therefore operate far from their limit efficiency.}  Gibbs did stress that, even though any given gear design has a limit efficiency less than 1, ``there is no proper minimum to the friction at the point of contact'' \cite{Gibbs-minimum}, since the limit efficiency can always be increased, in principle, by changing the design (e.g., by reducing the size of the teeth).  This is analogous to Carnot's result for heat engines, since the limit of $\eta$ in \Eq{eq:Carnot} can always be increased by reducing the ratio $T_2 / T_1$.  We have no indication that the significance of Gibbs's result about gear friction has ever been appreciated.\footnote{Because of friction, the condition of constant ratio of angular velocities $\omega_1 / \omega_2$ is not equivalent to the condition that the ratio of the torques exerted by one wheel on the other (i.e., the mechanical advantage) be fixed.  Gibbs clearly makes this point in \cite{Gibbs-torque}, but it is obscured in the commentary by the modern editors of Euler's collected works \cite{Euler-preface}.}

In general, mechanisms can only convert the finite difference between input and output power (the ``tail'' of \Fig{fig:Carnot-LeC}) into frictional heat.  Thus we see that their steady operation is necessarily {\it irreversible}.  If, instead of gears, we consider two smooth wheels rolling against each other without slipping, the reason for this is even clearer:  For one wheel to turn the other, there must be a sufficient attraction between the material in the rims.  Therefore, it costs finite energy to unstick the rims after their point of contact.  It is not possible to get rid of that loss from unsticking without also eliminating the coupling that allows one wheel to turn the other.

A similar analysis can be made for gears connected to chains, or for shafts wrapped by belts.  In the case of chains, there must be some loss as the links engage and disengage from the teeth in the gearing.  Similarly, it must consume non-zero energy to unstick a belt from the surface of the shaft.  Note that if one wheel turns another via a chain or a belt, there is a lossy conversion of power to and from zero frequency (the linear motion of the chain or belt), even if both wheels have the same radius and turn at the same rate and in the same direction.

To quantify the limiting efficiency of the action of smooth wheels or belts, let $\lambda$ be the linear density of the mass along the rims and $ds$ be the infinitesimal length of the region of contact between the rims, as shown in \Fig{fig:rims}.  ``Rim'' is here defined as the finite-width annulus of solid material along the exterior of the wheel or belt that is in effective contact with the other device, and which is therefore involved in the interaction that causes static friction at the point of contact.

\begin{figure} [t]
\begin{center}
	\includegraphics[width=0.6 \textwidth]{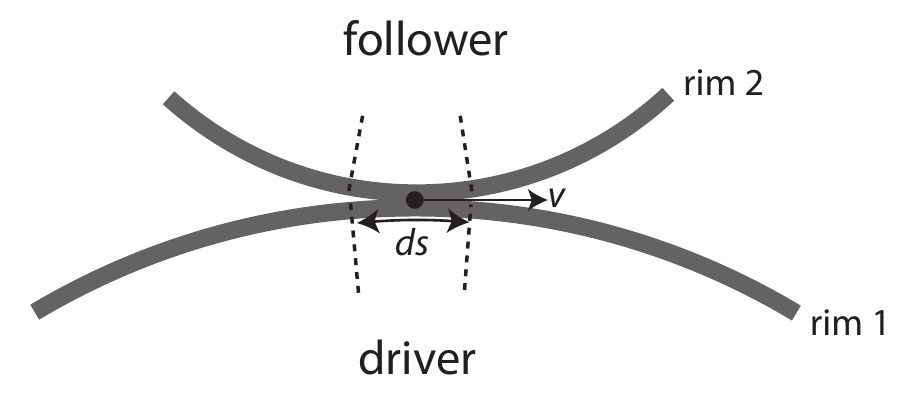}
\end{center}
\caption{\small The lower wheel (the driver) turns the upper one (the follower) by rolling without slipping at the point of contact, where the mass elements on the rim have a common velocity $\vv v$.  Each element has mass $\lambda \, ds$, where $\lambda$ is the linear density of the rims and $ds$ is the tangential length of the region of contact.  (The thickness of the rims is determined by the depth of the layer of material involved in producing static friction at the point of contact.)\la{fig:rims}}.
\end{figure}

For the driver to be able to move the follower at speed $v$ without slipping, a mass element $\lambda ds$ on the driver's rim must be able to come in from infinity with speed $v$, ``collide'' at the point of contact with the corresponding mass element ---initially at rest--- on the follower, and then have the two elements move together with speed $v$.\footnote{Of course, the rim of the follower is not at rest when the wheels are turning steadily.  But this procedure allows us to calculate the condition such that all the energy of the follower comes entirely from the driver.}  Thus, by energy conservation:
\be
\frac{1}{2} (\lambda \, ds) v^2 = (\lambda \, ds) v^2 - dE ~,
\la{eq:rims-conservation}
\ee
where $-dE$ is the potential energy corresponding to the attraction of the rim elements when in close proximity.  It must therefore cost energy $dE$ to unstick the rim elements past their point of contact.  If at least one of the wheels has finite radius, this unsticking occurs over an infinitesimal time
\be
dt = \frac{ds}{v}~.
\ee
By \Eq{eq:rims-conservation}, the minimum power consumed by the rolling friction is therefore
\be
P_{\rm in} - P_{\rm out} = \frac{dE}{dt} = \frac{\lambda v^3}{2}~.
\la{eq:wheel-loss}
\ee
This cannot be taken to zero for $v > 0$, since $\lambda$ must be finite, as otherwise there can be no action between the devices (just as gears cannot operate with massless teeth).  The point is that, even in a conservative system, the energy involved in the unsticking of the rims cannot be transferred from the driver to the follower, because they are {\it moving apart}.

When the wheels are turning steadily, the force that powers the rim of the driving wheel is cancelled by the static friction $f$ from the contact with the follower wheel.  Thus, the input at the rim of the driving wheel is:
\be
P_{\rm in} = fv
\ee
so that the efficiency is constrained by
\be
\eta \equiv \frac{P_{\rm out}}{P_{\rm in}} \leq 1 - \frac{\lambda v^2}{2 f_{\rm max}}~,
\la{eq:eta-wheels}
\ee
where $f_{\rm max}$ is the maximal static friction for the materials on the rims.  Clearly, $f_{\rm max}$ depends on $\lambda$, since
\be
\lim_{\lambda \to 0} f_{\rm max} = 0 ~; ~~~
\lim_{f_{\rm max} \to \infty} \lambda = \infty ~.
\ee
What the precise behavior of $\lambda / f_{\rm max}$ might be lies beyond our personal knowledge of tribology, but \Eq{eq:eta-wheels} suffices to establish that the efficiency of power transfer between wheels turning without slipping (or between a shaft and a belt) cannot be taken to zero, as required by the Carnot-Le Corbeiller theorem.

\subsubsection{Nonlinearity and irreversibility}
\la{sec:entropy}

Non-unit efficiency does not, of course, imply irreversibility: e.g., according to \Eq{eq:Carnot}, a reversible Carnot engine operating between reservoirs with finite temperatures $T_{1,2}$ has efficiency $\eta < 1$.  But the fact that, in practice, self-oscillators robustly attain a unique limit cycle, regardless of initial conditions and of previous transient perturbations, {\it does} make them irreversible.  This irreversibility results from the generation of entropy by the dissipation associated with the nonlinearity that determines the limit cycle, which wipes out information about the oscillator's initial conditions.

Note that it is the nonlinearity of equations of motions with self-oscillating limit cycles that breaks the time-reversal symmetry between the decaying solution to a positively-damped linear oscillator and the growing solution to the same linear equation with the sign of the damping reversed.  Strictly linear self-oscillation is unphysical because it would reverse the thermodynamic arrow of time.  A nonlinear self-oscillator with a limit cycle is compatible with the second law of thermodynamics because it generates entropy.\footnote{Ilya Prigogine used the term ``dissipative structure'' to describe systems ---like self-oscillators and living organisms--- that produce regularity by absorbing energy from the environment and generating entropy, thus remaining far from thermal equilibrium for a very long time.  This work was rewarded with the Nobel Prize in chemistry for 1977 \cite{Prigogine}, but its actual impact on quantitative science, and the grandiose claims that Prigogine and his collaborators later made for it, have been contested: see \cite{Keizer-steady,Anderson,Bricmont}.}

From the point of view of the thermodynamic arrow of time, it is interesting to note that all clocks, being (weakly) nonlinear self-oscillators (see Secs~\ref{sec:clocks} and \ref{sec:weakly-nonlinear}), must themselves generate entropy.  Schr\"odinger made a similar observation in \cite{Schrodinger-clocks}, where he explained that ``if we remember that without a spring the clock is gradually slowed down by friction, we find that [its continued operation] can only be understood as a statistical phenomenon.''

\section{Other applications}
\la{sec:applications}

\subsection{Servomechanisms}
\la{sec:servos}

It is very common in engineering for a system to be driven by a self-oscillating motor, which in turn is controlled by negative feedback so that the motion follows some intended trajectory.  This is called a {\it servomechanism}, and the simplest example is a steam engine whose rate is fixed by the action of a governor, as we mentioned in \Sec{sec:stability}.

The onset of an unwanted self-oscillation about the servomechanism's intended trajectory is sometimes called ``hunting'' in mechanical systems, and ``parasitic oscillation'' in electronics.  The presence of such oscillations can be diagnosed by applying the Routh-Hurwitz criterion to the linearized equation of motion for small perturbations about the intended trajectory.  But for servomechanisms in particular it is usually more practical to use the stability criterion in a different form, proposed by Harry Nyquist in \cite{Nyquist} and based on the form of the gain as a function of frequency.  The Nyquist stability criterion, which is central to modern control theory, is reviewed from a physical point of view in \cite{Pippard-stability}.  For a more thorough mathematical treatment, see \cite{AM-Nyquist}.

An important type of ``hunting'' is the swaying of a railway vehicle when it travels above a critical speed \cite{Carter,Knothe}.  Parasitic oscillations are a common problem in electronic amplifiers, as discussed in \cite{H&H-parasitic}.  ``Intention tremor'' is a neurological disorder seen in patients with injuries to the cerebellum, such that if the patient, for example, tries to pick up a pencil, his hand will overshoot the position of the pencil and then go into an uncontrollable oscillation around it (see, e.g., \cite{tremor}).  Mathematician Norbert Wiener characterized voluntary human motion as a servomechanism and pointed to intention tremor as a parasitic oscillation of that system \cite{Wiener-tremor}.

\subsection{Thermoacoustic self-oscillators}
\la{sec:thermoacoustic}

A thermoacoustic self-oscillator is a device in which the flow of heat between the working substance and its surroundings sustains an oscillation of that working substance's volume and pressure.  The resulting vibration can produce a sound of a well-defined frequency that propagates through a surrounding medium, though of course the generation of sound is incidental and for the most part we will only be interested in the oscillation of the working substance itself.  A thermoacoustic self-oscillator differs from an ordinary heat engine in that a positive feedback is established {\it without any mechanical valve} (or other device) controlling the motion of the working substance (see \Sec{sec:heat}).

\subsubsection{Putt-putt boat}
\la{sec:putt}

An amusing and instructive instance of thermoacoustic self-oscillation is the putt-putt (or pop-pop) boat, a toy whose circulation dates back at least to the early 20th century, when it was being fabricated by Nuremberg toymaker Ernst Plank \cite{Harley}.  The design had been patented in 1891 by Desir\'e Thomas Piot, who described it as a ``steam generator'' well suited to ``the case of toy boats'' \cite{Piot}.  The putt-putt was once very popular but is no longer easily bought, partly because it must be made of metal, while most toys today are plastic.\footnote{Iain Finnie relates how, after becoming a professor of mechanical engineering at UC Berkeley in 1961, he began conducting popular lectures on the operation of various toys, at which the putt-putt (a favorite toy from his own childhood) always attracted the most attention \cite{Finnie-letter}.  The putt-putt featured prominently in the animated Japanese feature film {\it Ponyo}, released in 2008 \cite{Ponyo}.  Toy boats were sold as tie-ins when that movie came out.}

The putt-putt works by heating (usually with a flame) an internal tank partly filled with water and connected to submerged exhausts, as shown in \Fig{fig:boat}.  It is usually easy to adjust the heat so that the water level will self-oscillate.  As water is alternately blown out and sucked in through the exhausts, the boat moves forward with a noisy vibration that gives the toy its name.\footnote{There has been a remarkably enduring confusion in both the research and the popular literature about why the boat does not move backwards when water is drawn into the exhausts, a confusion dating to the first published scientific discussion of the putt-putt boat by J.~G.~Baker in \cite{Baker,PopMech}.  The correct explanation is very straightforward: water flowing out of the boat carries momentum away with it.  By Newton's third law, the boat must pick up an opposite momentum, propelling it forward.  When water is aspirated, there is initially a reaction on the boat that would pull it backwards, but the incoming water soon impinges on the inner walls of the tubes and imparts its forward momentum to the boat.  The initial backwards reaction on the boat is therefore cancelled by the forward pushing of the water on the boat's insides.  This is discussed in detail in \cite{Machian}.  In the French engineering literature, the fact that a tank will recoil if water pours out of it but not if water pours steadily {\it into} the tank is sometimes called the ``paradox of Bergeron,'' after mechanical engineer Paul Bergeron \cite{Bergeron1,Bergeron2}.}

According to the law of ideal gases, for a small displacement $y$ of the liquid level away from its equilibrium, the pressure of the gas inside the tank, at a fixed temperature, is
\be
P = \frac{C N_0}{V_0 - A y} \simeq P_0 + k_1 y~,
\la{eq:Boyle}
\ee
where $N_0$ is the quantity of gas inside the tank (measured in moles), while $P_0 \equiv C N_0 / V_0$ and $k_1 \equiv C N_0 A / V_0^2$ are constants.  The force $A (P_0 - P)$ that acts to restore $y$ to its equilibrium is therefore proportional to $y$.  Thus, small oscillations behave like a damped harmonic oscillator
\be
\ddot y + \gamma \dot y + \omega^2 y = 0
\la{eq:putt-damped}
\ee
(which is what Boyle memorably called the ``spring of the air'' \cite{Boyle}).

\begin{figure} [t]
\begin{center}
	\includegraphics[width=0.55 \textwidth]{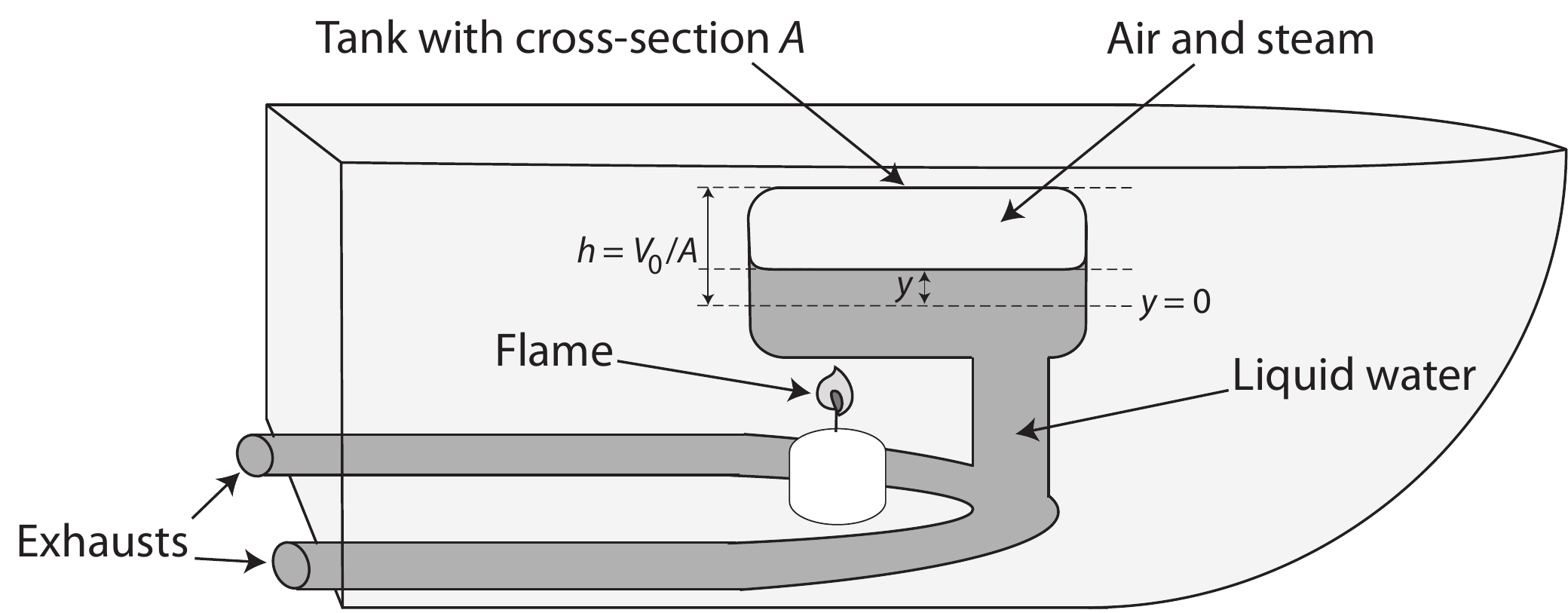}
\end{center}
\caption{\small Schematic of the putt-putt boat: $y$ measures the displacement of the liquid level away from equilibrium, $A$ is the cross-section of the tank, and $h$ the distance between the top of the tank and the level $y=0$, so that the volume of gas in equilibrium within the tank is $V_0 = h A$. \la{fig:boat}}
\end{figure}

Through the heating of the tank, the pressure inside acquires a dependence on $S$, the quantity of steam:
\be
P = \frac{C (N_0 + S)}{V_0 - A y} \simeq P_0 (1 + S) + k_1 y ~.
\la{eq:Boyle-S}
\ee
As explained in \cite{Finnie1, Finnie2}, $S$ increases by a constant evaporation rate and decreases by a condensation rate proportional to the surface area inside the tank that is not covered by liquid.  In equilibrium (which we define as $y=0$) the condensation rate balances the evaporation rate,\footnote{If there is no such equilibrium, the water will be pushed out completely and the putt-putt boat will not run.} so that
\be
\dot S = k_2 \cdot y~.
\la{eq:dotS}
\ee
Thus there is a back-reaction of the oscillation of the water level $y$ on the pressure difference $P_0 - P$ that drives it.  This back-reaction is such that a positive feedback, and therefore a self-oscillation, can be established, as we will see.\footnote{The schematic in \Fig{fig:boat} and the treatment in terms of the level $y$ are somewhat idealized.  Experimentally, it seems that the liquid leaves the tank completely during the high-pressure phase and that the steam only begins to condense when it reaches the tubes that connect to the exhausts \cite{Bindon}.  This is because the walls of the tank are too near the flame and therefore too hot for effective condensation.}

By \Eqs{eq:Boyle-S}{eq:dotS}, the equation of motion for small oscillation of $y$ is
\be
\ddot y + \gamma \dot y + \omega^2 y = - C_2 P_0 \int_0^t dt' \, y ~,
\la{eq:putt-motion}
\ee
where $C_2$ is proportional to the value of $k_2$ in \Eq{eq:dotS}.  For sinusoidal motion, the term on the right-hand side of \Eq{eq:putt-motion} will be in phase with $\dot y$ and therefore can feed energy from the candle into the self-oscillation of $y$, whose amplitude increases until nonlinear effects stabilize it \cite{Finnie2}.\footnote{An important nonlinearity is the dissipative pressure loss of the liquid during inflow, described in detail in sec.\ 5 of \cite{Machian}.  This adds to \Eq{eq:putt-motion} a damping term proportional to $(\dot y + \left| \dot y \right|)^2$.  Since this acts only during the inflow phase ($\dot y > 0$), it can result in an asymmetric limiting cycle for $y$, as reported in \cite{Finnie1,Finnie2}.}  Evidently, the criterion for self-oscillation is
\be
C_2 P_0 > \omega^2 \gamma~,
\la{eq:putt-instab}
\ee
in which case the term on the right-hand side of \Eq{eq:putt-motion} overwhelms the second term of the left-hand side, effectively giving $y$ a negative damping.\footnote{In one of the more popular putt-putt boat designs, the top of the tank is a flexible diaphragm that reverses its concavity during the cycle of the gas pressure.  This is not essential to the operation of the device: its primary function appears to be to reduce the value of $\omega$ so as to make it easier to fulfill the self-oscillation condition of \Eq{eq:putt-instab} \cite{Finnie2}.  The diaphragm also makes the oscillation noisier, an attractive feature in a child's toy.}

The putt-putt, like any other heat engine, must absorb heat from a region with temperature $T_1$ (the evaporating water warmed by the candle) and reject a lesser amount of heat into a region with lower temperature $T_2$ (the tank walls on which the steam condenses), with the difference available as work to move the liquid.  Since steam is generated and re-condensed in the same chamber and therefore at nearly the same temperature,  \Eq{eq:Carnot} implies that the maximum thermodynamic efficiency of the putt-putt is very low.  On top of that, Finnie and Curl found that in the toys they examined only about a tenth of the work on the water was converted into propulsion of the boat, the rest being dissipated by the damping of the motion of the liquid in the tubes \cite{Finnie1,Finnie2}.  Nonetheless, the putt-putt's mechanism is interesting as an instance of a valveless pulse jet engine.  Furthermore, Finnie and Curl point out in \cite{Finnie1} that if the tank were substituted by a large bellows, driven up and down by the oscillation of the internal pressure, the result would be a valveless version of the first steam engine patented by Watt in 1769 (see \cite{Watt-engine}).

\subsubsection{Rijke tube}
\la{sec:Rijke}

An even simpler (and perhaps more striking) demonstration of a thermally-driven self-oscillation is the Rijke tube \cite{Rijke}, shown schematically in \Fig{fig:Rijke}.  If a wire mesh is attached near the bottom of a large tube that is open at both ends and if the mesh is heated with a flame until it glows red, then after the flame is withdrawn the tube will produce a very loud tone, like that of an organ pipe, which dies out when the mesh cools.  The tone can be sustained at will by heating the mesh with an electrical current, rather than a flame. 

\begin{figure} [t]
\begin{center}
	\includegraphics[width=0.2 \textwidth]{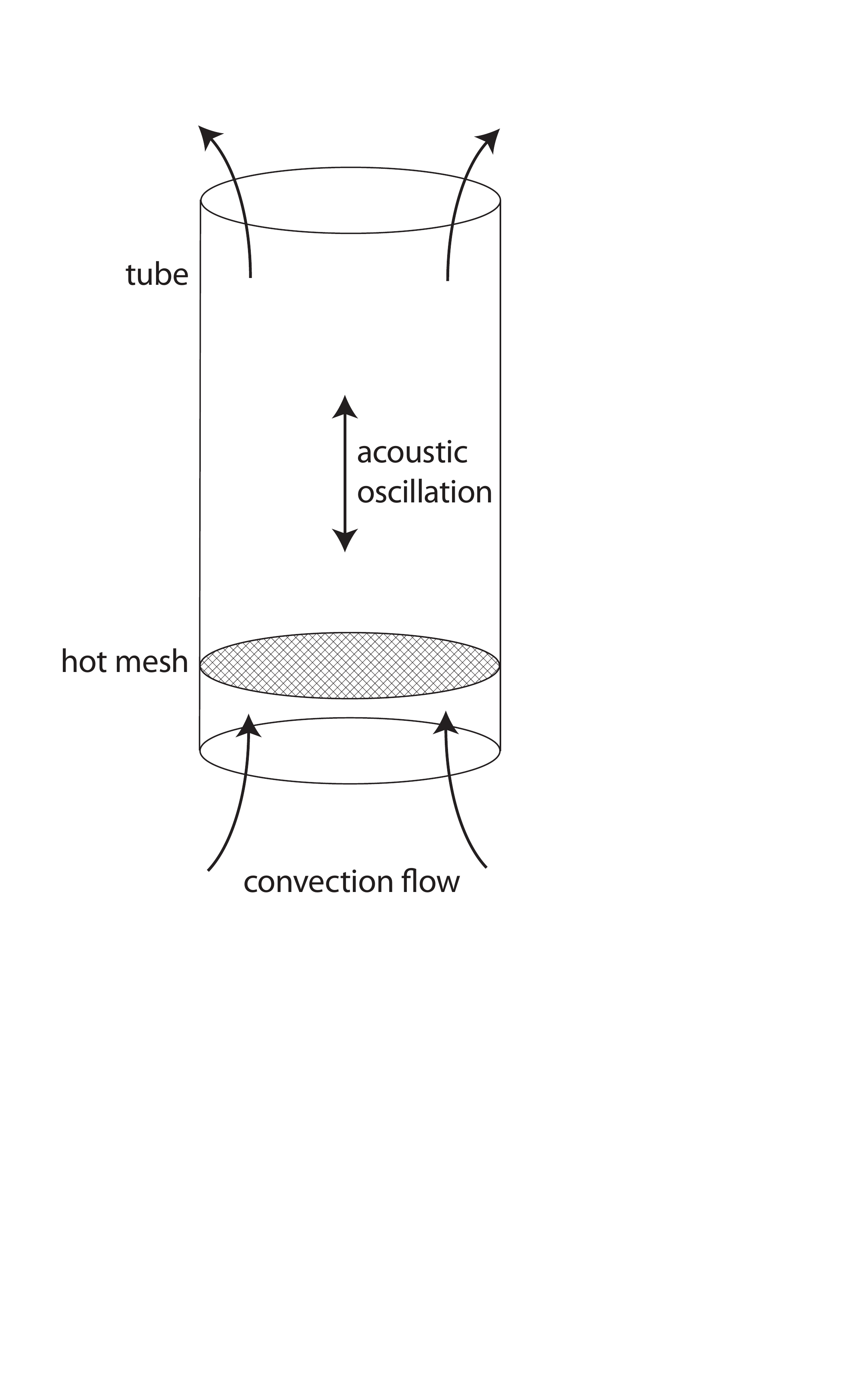}
\end{center}
\caption{\small A Rijke tube is open on both ends and has a heated metal mesh near the bottom.  The movement of the air through the pipe is a combination of a steady upward convection flow and an oscillation given by the periodic variation of the pressure inside the tube.  This acoustic oscillation will produce a loud tone as long as the mesh stays hot.\la{fig:Rijke}}
\end{figure}

The operation of the Rijke tube is explained in \cite{Rayleigh-Rijke,Pippard-Rijke}.  Due to convection, there will be a constant upward flow of air through the tube.  Meanwhile, an oscillation of the pressure within the tube causes a periodic airflow that is superimposed on the steady upward convection.  That oscillation causes air to pass through the hot mesh in alternate directions.  When the velocity of the oscillation is at its maximum, it adds to the steady convection and maximizes the amount of fresh air that is being heated as it passes {\it up} though the mesh.  This air then exerts a further upward push.  After half a period of the oscillation, air is pushed {\it down} through the mesh, but this air is already hot and therefore exerts little upward pushing.  Thus the pushing of the newly-heated air is modulated in phase with the velocity of the acoustic oscillation within the tube, establishing positive feedback.

It should now be clear that the tube can self-oscillate only if the hot mesh is in the bottom half of the tube, and also that the reason why the tube will not ring until the flame is removed is that the flame heats the air {\it before} it reaches the mesh.  The Rijke tube is the simplest thermoacoustic self-oscillator (see \cite{Rijke-thesis} for some recent investigations on the subject).  It can provide a simple and impressive lecture demonstration of self-oscillation in general, since it should be obvious to students that the loud ringing is not being driven by any periodic forcing term external to the system.

Another interesting thermoacoustic self-oscillator is the ``singing flame:'' if a small flame, produced by hydrogen burning as it leaves a narrow tube, is inserted into a larger tube ---which may be either open or closed at the top--- the air inside the larger tube can in some circumstances ring audibly in a sustained way.  This phenomenon was first described in \cite{Higgins-flame} and is discussed by Rayleigh in \cite{Rayleigh-flame}, though his treatment was later found to be somewhat incomplete (see \cite{Jones-flame,Putnam-flame,Howe}).

\subsubsection{Rayleigh criterion}
\la{sec:rayleigh-criterion}  

Thermoacoustic oscillators illustrate a general principle known in mechanical engineering as the Rayleigh criterion:\footnote{This is not to be confused with the better-known Rayleigh criterion for optical angular resolution (see \cite{Rayleigh-optics}), or with the Rayleigh criterion for the stability of Couette flow (see \cite{Couette}).} the acoustic oscillation of a gas is encouraged if heat is injected when the gas is warmed by adiabatic compression, and/or heat is removed when the gas is cooled by adiabatic expansion \cite{Rayleigh-criterion,Rayleigh-criterion-ToS} (see also \cite{Howe,Pippard-Rijke}).  An acoustic oscillation is most encouraged if the transfer of heat {\it out} of the gas varies in phase with the volume, so that the maximum heat is injected when the gas is most compressed and therefore hottest, while the maximum heat is removed when the gas is most rarefied and therefore coldest.  This agrees with Carnot's theorem (\Eq{eq:Carnot}), according to which a heat engine runs more efficiently when it absorbs heat at a higher temperature and/or rejects heat at a lower temperature.

Rayleigh's criterion may also be understood by analogy to parametric resonance: If heat is injected when the air is most compressed, the ``spring of the air'' is thereby stiffened and the subsequent expansion is aided.  Similarly, removing heat when the air is most expanded facilitates the subsequent compression.  The elasticity of the ``spring'' is therefore varied in such a way that the overall oscillation is encouraged.  But, unlike in the parametrically resonant systems described in \Sec{sec:parametric}, in self-oscillators the variation of the elasticity is controlled by the acoustic vibration itself.

Ordinarily, sound waves are thermally {\it damped} because the air loses heat most quickly when its temperature is highest due to compression, while it absorbs the most heat when its temperature is lowest due to expansion.  Newton famously computed a speed of sound \cite{Newton-sound} which turned out to be too low (by about 20\%) in light of subsequent experimental tests.  In effect, he treated the acoustic oscillation as isothermal: his expression is correct in the limit in which the frequency of sound is so low that the vibrating air remains in thermal equilibrium with its surroundings.  Laplace obtained a value more compatible with measurements \cite{Laplace-sound} by approximating the oscillation as adiabatic, which is valid in the limit in which the oscillation is very fast compared to the rate of damping by thermal radiation.\footnote{On the relevance of Rayleigh's criterion to understanding the difference between Newton's and Laplace's computation, see \cite{Rayleigh-criterion,Rayleigh-criterion-ToS}.  On the historic controversy regarding the speed of sound, see \cite{Newton-Laplace}.}

Note that the putt-putt boat (\Sec{sec:putt}) optimizes the Rayleigh criterion, since \Eq{eq:dotS} implies that the heat rejected by the condensation of steam varies in phase with the volume of the working gas inside the tank.  Diesel and gasoline engines ---though not usually conceptualized as thermoacoustic oscillators--- also optimize the Rayleigh criterion, by igniting the fuel (and therefore maximizing the injection of heat) when the working gas is most compressed.

On the other hand, in a Rijke tube the flow of heat into of the tube is modulated in phase with the upward velocity of the air that flows through the heated mesh in \Fig{fig:Rijke}.  The delivery of heat by the mesh to the gas therefore lags the oscillation of the volume by a quarter of a period.  This does not optimize the Rayleigh criterion, but a self-oscillation is sustained because it takes some time for the heat injected by the mesh to reach the tube's center: the injected heat therefore reaches the middle of the tube when the gas there is more compressed (and therefore hotter) than average.

\subsubsection{Cepheid variables}
\la{sec:cepheids}

Cepheid variable stars are a very important instance of naturally occurring thermodynamic self-oscillation.  Their name derives from John Goodricke's observation, in 1784, that the brightness of the star $\delta$ Cephei varied regularly with a period of about five days.  Goodricke and his colleague Edward Pigott then discovered other stars whose brightness also pulsates, usually with a period of a few days.  In the early 20th century, Henrietta Swan Leavitt discovered that the period of a Cepheid's pulsation is tightly related to the star's intrinsic brightness \cite{Leavitt}.  This made Cepheid variables the first reliable ``standard candle'' by which astronomers could measure distances to other galaxies, leading eventually to Edwin Hubble's discovery that the Universe is expanding \cite{Hubble}.  See \cite{Webb} for a review of the history and the science of this subject.

When a yellow supergiant star is running out of hydrogen fuel, it may begin to collapse.  The compression causes an increase in temperature and pressure that slows down the collapse and eventually causes the star to expand and cool.  The resulting cycle of the volume of the star would be damped away relatively quickly if it were not for a mechanism that injects heat into the core of the star during the compression phase and removes it during the expansion phase, in accordance with the Rayleigh criterion (see \Sec{sec:rayleigh-criterion}).

As the core is compressed, the rate of nuclear fusion increases (much like the fuel ignites during compression in a Diesel engine).  The excess heat produced is not radiated away because the core of a Cepheid is surrounded by a layer of ionized helium (He$^+$) and the radiation from the heated core causes that layer of helium to become doubly ionized (He$^{++}$), making it more opaque and preventing the heat from escaping.  During the expansion, the helium layer cools and goes back to being singly ionized, which decreases the opacity, allowing heat to escape.

In 1916, Arthur Eddington became interested in stellar thermodynamics when he realized that the pulsation of the Cepheids presented a theoretical puzzle, namely,
\begin{quote}
to find if possible some cause maintaining the mechanical energy of pulsation against the loss by dissipative forces---some method by which the mechanical energy could be automatically extracted from the abundant supplies of heat at different temperatures in the star without violating the second law of thermodynamics.  \cite{Eddington-appendix}
\end{quote}
Eddington's first paper on the subject \cite{Eddington-Rayleigh} rediscovered the Rayleigh criterion and proposed applying it to the Cepheid problem.  In his 1926 book, {\it The internal constitution of the stars}, he explained that ``we must make the star more heat-tight when compressed than when expanded'' \cite{Eddington-valve}, and he drew an analogy between the corresponding mechanism and the action of the valve in a heat engine.  The realization that this heat-tightening is due to the double ionization of a thin, outer helium layer was due, many years later, to S.~A.~Zhevakin \cite{Zhevakin}.  This mechanism is now called the ``Eddington valve,'' which might be something of a misnomer, because the helium layer controls the flow of heat, rather than the mechanical motion of the stellar material (see \Sec{sec:heat}).

\begin{figure} [t]
\begin{center}
	\includegraphics[width=0.6 \textwidth]{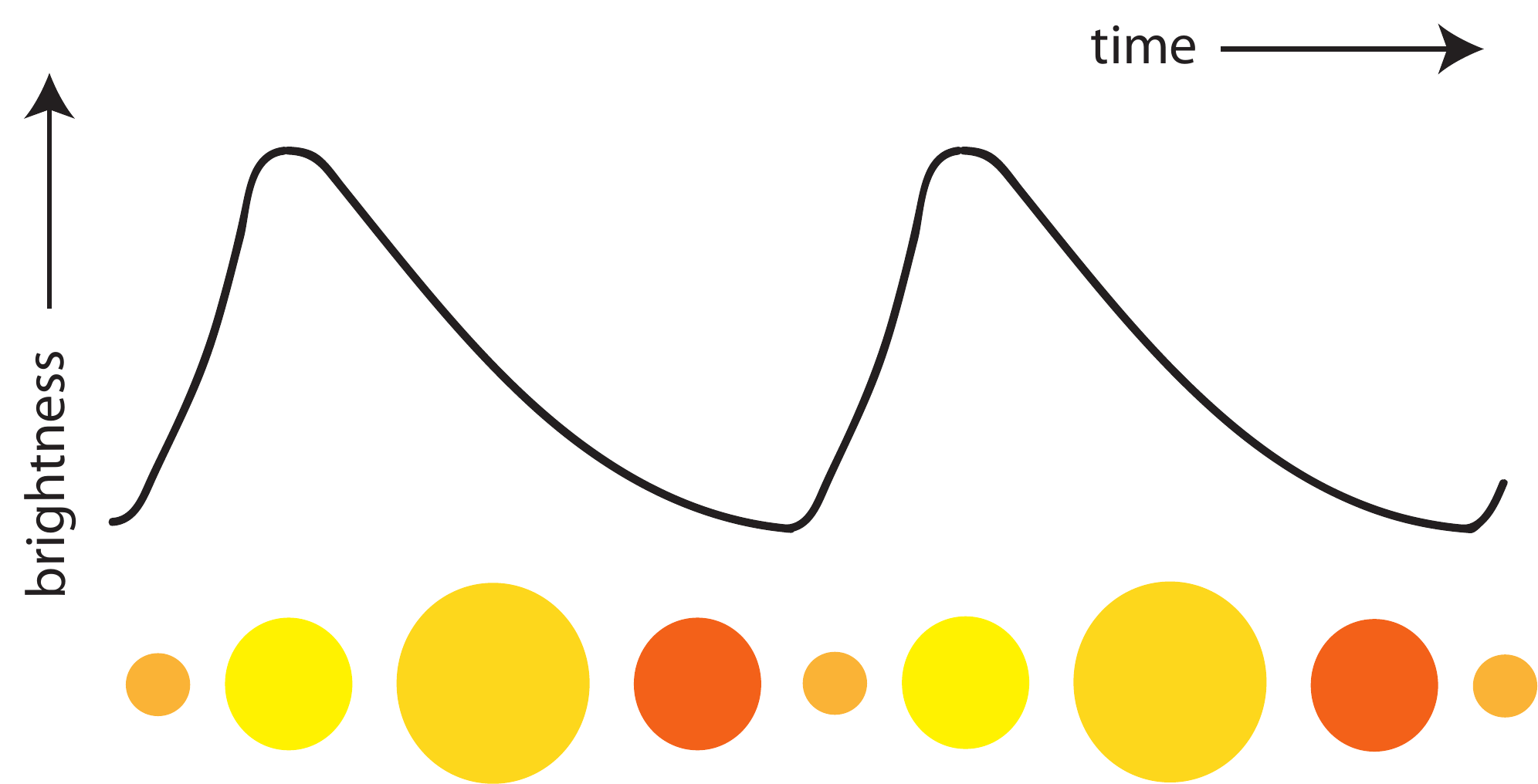}
\end{center}
\caption{\small The waveform above represents the variation in the brightness of a typical Cepheid variable star, which increases rapidly and then fades gradually, with a period of a few days.  The circles below schematically represent the corresponding variations in the size and color of the star, which have been exaggerated for clarity.  The oscillation in brightness leads the oscillation in the volume of the star by a quarter of a period.  This illustration is adapted from \cite{Eddington-S&T}.\la{fig:cepheid}}
\end{figure}

Figure \ref{fig:cepheid} shows a schematic plot for the brightness and the radius of a typical Cepheid variable, as functions of time.  The brightness (which is directly related to the flow of heat {\it out} of the star's core) is in phase with the radial velocity of the star's surface (which gives the rate of the change of the star's volume).  The oscillation of the brightness therefore leads the oscillation of the volume by a phase of $\pi/2$, which is sub-optimal from the point of view the Rayleigh criterion, but adequate to maintain the pulsation.  In this respect Cepheid variables are akin to the Rijke tube, as explained at the end of \Sec{sec:rayleigh-criterion}.

\subsection{Lasers}
\la{sec:lasers}

Masers and lasers may be described as electromagnetic cavities in which the dielectric loss (i.e., the damping) is negative because the molecules or atoms of the dielectric medium have undergone a {\it population inversion}: more of them are in a higher-energy quantum state than in a lower-energy state to which transition via photon emission is possible.  The electromagnetic oscillation within the cavity therefore causes the dielectric medium to feed energy into the oscillation by stimulated emission: the medium is ``active.''  For an excellent introduction to the theory of stimulated emission in two-state quantum systems, see \cite{Feynman-maser}.

A laser is therefore an electromagnetic self-oscillator, even though the mechanism responsible for the negative damping is quantum-mechanical and ---unlike in a classical self-oscillator--- requires a precise tuning of the energy separation $\Delta E$ of the quantum states involved in the stimulated emission, so that the photons emitted have an angular frequency $\omega = \Delta E / \hbar$ (where $\hbar$ is the reduced Planck constant) that is very close to the cavity's resonance (see \cite{Pippard-maser-I}).

On the other hand, the population inversion that sustains the negative damping is achieved {\it without} tuning the frequency of the external power supplied to the active medium.  The ratio of the laser's power output (in the form of photons with energy $E= \hbar \omega$) to the total power input is therefore subject to limitations analogous to those discussed in \Sec{sec:Carnot} for classical self-oscillators.  See \cite{Pippard-inversion} for a basic discussion of the way in which population inversion is accomplished; essential power losses are seen in the production of photons or phonons with energies different from $\hbar \omega$.\footnote{In quantum physics, the origin of the Carnot-Le Corbeiller efficiency limit is very clear, perhaps even clearer than in the classical case.  The probability of photon absorption by the medium goes to 1 only when the frequency of the photon approaches $\Delta E / \hbar$ for a transition.  In practice, the population inversion mechanism involves a variety of transitions, only one of which corresponds to the frequency of the laser light outputted.  The other transitions generate photons and phonons which bounce around the cavity and eventually exit it or else thermalize, heating the dielectric medium.}

Charles Townes, the inventor of the first maser, explained the principle of its operation to laypeople by a simple mechanical analogy:  Imagine a swimming pool, at one end of which a platform has been raised on a wobbly pillar. The platform is loaded with rocks (by analogy to population inversion).  When a rock falls into the water, it produces a wave that travels to the other end of the pool, is reflected, and then hits the wobbly pillar, causing another rock to fall.  If the wobbling of the pillar absorbs negligible energy from the wave, this serves as an analogy of stimulated emission.  The falling rock then produces a new wave, which interferes constructively with the previous one.  The process can continue as long as there are rocks on the platform, giving what Townes humorously called a {\it waser}: water amplification by stimulated emission of rocks; see \cite{Playboy}.

Townes's analogy nicely captures the nature of the laser as a self-oscillator in which the negative damping comes as energy quanta.  The only important element that the waser lacks is an analog of the tuning of the energy of the rocks to the frequency of standing waves in the pool, which is substituted in the waser by timing the fall of the rocks so as to produce a new wave precisely in phase with the previous reflected wave.  (This feature of the analogy is inevitable, since in the classical limit $\hbar \to 0$.)

Borenstein and Lamb showed in detail that a self-oscillating electromagnetic cavity may be built using only the laws of classical electrodynamics \cite{classical-laser}.  They avoid the need to tune the phases of the oscillators in the active medium by making them nonlinear, so that they can be entrained (see Secs.~\ref{sec:entrainment} and \ref{sec:forced-SO}).  On the (wholly academic) possibility of describing classical self-oscillators as quantum lasers, see \cite{Pippard-quantum-SO}.

The applications of the concept of self-oscillation to the semi-classical theory of lasers are treated in detail in \cite{Lasers-vdP,Lasers-semiclassical}, including the relaxation oscillation seen in ruby lasers.  The relevance of entrainment to the description of the operation of lasers is discussed in \cite{Lang-locking,Sargent-locking,Lasers-locking}.  On chaotic behavior in laser systems, see \cite{lasers-chaos}.

An important and instructive kind of macroscopic quantum oscillation is the alternating Josephson current between two superconductors separated by a thin insulator and held at a constant voltage difference by a battery \cite{Josephson}.  Since this is conceptually quite distinct from the operation of a laser, and since its description requires a longer excursion into quantum mechanics, we treat it separately in Appendix \ref{sec:Josephson}.

\subsection{Business cycle}
\la{sec:business}

Pippard mentions geysers and the egg-laying cycle of domestic fowl as naturally occurring relaxation oscillations \cite{Pippard-geyser}.  Sargent, Scully, and Lamb \cite{Lasers-vdP} mention also the business cycle in macroeconomics, i.e., the familiar fluctuation of production and economic activity that has often led to periods of rapid growth followed by recessions.  In fact, interest in modeling the business cycle as a relaxation oscillation was sparked in the 1930s, largely through the influence of Le Corbeiller, who explained to economists that
\begin{quote}
if statistical observation leads us to believe that a given magnitude varies periodically, and if we look for the {\it cause} of those oscillations, we may suppose that that magnitude executes either (a) forced oscillations, or (b) self-oscillations,\footnote{Le Corbeiller writes {\it oscillations autoentretenues}, a term introduced by Blondel in \cite{Blondel-auto}, which would be more faithfully rendered in English as ``self-maintained oscillations.''  On the many names of self-oscillation, see the first footnote in \Sec{sec:intro}.} which may be either (b$\alpha$) sinusoidal or (b$\beta$) of relaxation type. \cite{LeC-econ}
\end{quote}

\subsubsection{Macroeconomic models}
\la{sec:keynesian}

Micha\l~Kalecki \cite{Kalecki}, followed by Alvin Hansen and Paul Samuelson \cite{Samuelson}, proposed simple models of the business cycle as an instability in the relation between macroeconomic income (the total value of all transactions) and capital (the value that has been invested in a durable stock intended to produce goods and services for future consumption).  Kalecki's model invoked a delay in the relation between investment and income.  Hansen and Samuelson's work relied on a linear ``multiplier-accelerator'' model for the relation between income, consumption, and investment.

Le Corbeiller had a great direct influence on the work of Harvard economist Richard M.~Goodwin, who stressed that nonlinearity was necessary to explain how macroeconomic oscillation could be a recurrent phenomenon \cite{Goodwin-cycle}.\footnote{Le Corbeiller, for his part, credited Goodwin with the original insight that two-stroke self-oscillators (see \Sec{sec:asymmetric}) are possible \cite{two-stroke}.}  According to Goodwin,
\begin{quote}
The Great Depression gave rise to two theories, the Kalecki and the Hansen-Samuelson models.  Both were linear but such models are incapable of explaining the continued existence of oscillations.  [Ragnar] Frisch misled a generation of investigators by resolving the problem with exogenous shocks, whereas already in the 1920s van der Pol had shown (as Frisch should have known) that a particular form of nonlinear theory was the appropriate solution.  His solution leads to a limit cycle. \cite{Goodwin-nonlinearity}
\end{quote}

Nonlinearity was incorporated into models of the macroeconomic business cycle by Lord Kaldor \cite{Kaldor}, Sir John Hicks \cite{Hicks}, and others (see \cite{BCycles}).  This is reviewed briefly in \cite{NewSchool}; for a full treatment, see \cite{Gandolfo}.  Chang and Smyth \cite{Chang} and Varian \cite{Varian} have formulated Kaldor's model so that the business cycle corresponds to a relaxation limit cycle on the Li\'enard plane, like that represented in \Fig{fig:Lienard-relax}, with the coordinates $x$ and $y$ interpreted as the total income and the capital stock, respectively.

All of these theories relied on Keynesian macroeconomic constructs ---particularly the IS/LM and ``multiplier-accelerator'' models--- which fell out of favor in the late 20th century because they lack adequate microeconomic foundations (i.e., an explanation in terms of the actual choices made by individuals) and therefore came to be seen as insufficiently explanatory and as vulnerable to the ``Lucas critique'' \cite{Lucas} that if individuals incorporated the cycle into their expectations then it should disappear.  In \cite{Chang}, Chang and Smyth also emphasize that self-oscillation in Kaldor's model occurs only for certain choices of the macroeconomic parameters.

\subsubsection{Leverage cycle and financial instability}
\la{sec:leverage}

That the business cycle may be due, at least in part, to a relaxation oscillation seems plausible in light of John Geanakoplos's recent work on the theory of the financial ``leverage cycle'' \cite{leverage}.  In Geanakoplos's theory, the leverage (i.e., the ratio of the value of an asset bought on credit to the down-payment by the buyer) is not necessarily stabilized by the equilibrium of supply and demand, and may therefore be subject to positive feedback effects that lead to relaxation-type oscillations in asset prices, which in turn can drive the economy as a whole into recurring booms and busts (for commentary on this theory see, e.g., \cite{WSJ,leverage-comment1,leverage-comment2,leverage-discussion}).

This leverage cycle might provide the microeconomic foundation that has been missing from models of the business cycle as a self-oscillation.  See also, e.g., the work by Minksy on the ``financial instability hypothesis'' \cite{Minsky}, of Shleifer on positive feedback investment strategies \cite{Shleifer}, and by Gjerstad and Smith on the role of credit expansion and leverage in the late 2000's financial crisis \cite{VSmith-crisis}.  What principally seems to be lacking at the moment from the theory of the leverage cycle is a solid understanding of just how financial markets evade the Arrow-Debreu welfare theorems \cite{Arrow,Debreu}, thus failing to give consistently efficient outcomes.

\subsubsection{Efficiency of markets}
\la{sec:efficiency}

It should be stressed that the attempt to use control theory and the theory of dynamical systems in economics has a very long history, but has so far yielded few fruitful results.\footnote{Sociologist Niklas Luhmann went as far as to elaborate a ``systems theory'' that sought to replace the sentient individual with the self-maintaining social system as the unit of explanation for cultural phenomena \cite{Luhmann}.  For a cogent ---though abstrusely written--- philosophical critique of this approach, see \cite{Habermas}.}  \r{A}str\"om and Murray suggest that this is in part because in economics ``there are no conservation laws'' \cite{AM-conservation}.  This hardly seems to us a satisfactory explanation, since none of the other systems treated in this article obey conservation laws either.  It is also commonly argued that economic systems are too complex and have too many variables, but this seems equally true of biological systems, in which the theory of dynamical systems has been used more profitably (see Secs.~\ref{sec:heart}, \ref{sec:entrainment}, and \ref{sec:FHN}).  Sometimes it is also pointed out that controlled experiments are not possible in macroeconomics, but the mathematical theory of dynamical systems largely grew out of research in another science without controlled laboratory experiments, namely astronomy.

It seems to us, rather, that the principal complication is that the economic order results from the actions of sentient individuals, whose response depends on the information available to them at any given time.  This information is constantly changing and includes even some awareness of economic theory itself and its predictions.  The Arrow-Debreu theorems establish that under conditions of perfect information and complete markets, the economic system will always be at an equilibrium state in which every individual enjoys the greatest satisfaction that can be obtained without making another individual worse off (what economists call an ``efficient'' outcome).  But most of the interesting, real-world economic phenomena probably take place away from that equilibrium and, in practice, the approach to equilibrium occurs under conditions in which information is far from perfect and markets are far from complete.  From this, economists have drawn strikingly divergent conclusions about policy (cf.\ \cite{Stiglitz,VSmith-rationality}).

This situation is not, after all, so different from that of thermodynamics.  In economics just as much as in physics, the condition of equilibrium allows us to characterize a complex system with only minimal knowledge of its internal dynamics.  But the persistence of non-equilibrium seen in many systems of interest, of which self-oscillation is an obvious example, then presents a major theoretical challenge.  (For a detailed discussion of this issue in theoretical physics, see \cite{Keizer}.)

\subsection{Rayleigh-B\'enard convection}
\la{sec:dd}

Figure \ref{fig:Benard} schematically shows the regular pattern of convection in ``Rayleigh-B\'enard cells,'' seen, under the right conditions, when a horizontal fluid layer is warmed from below and cooled from above \cite{Benard,Rayleigh-Benard}.  This phenomenon has been very widely studied as a paradigmatic example of spontaneous pattern formation in systems far from thermal equilibrium (see \cite{Benard-Scholarpedia} and references therein).

\begin{figure} [t]
\begin{center}
	\includegraphics[width=0.35 \textwidth]{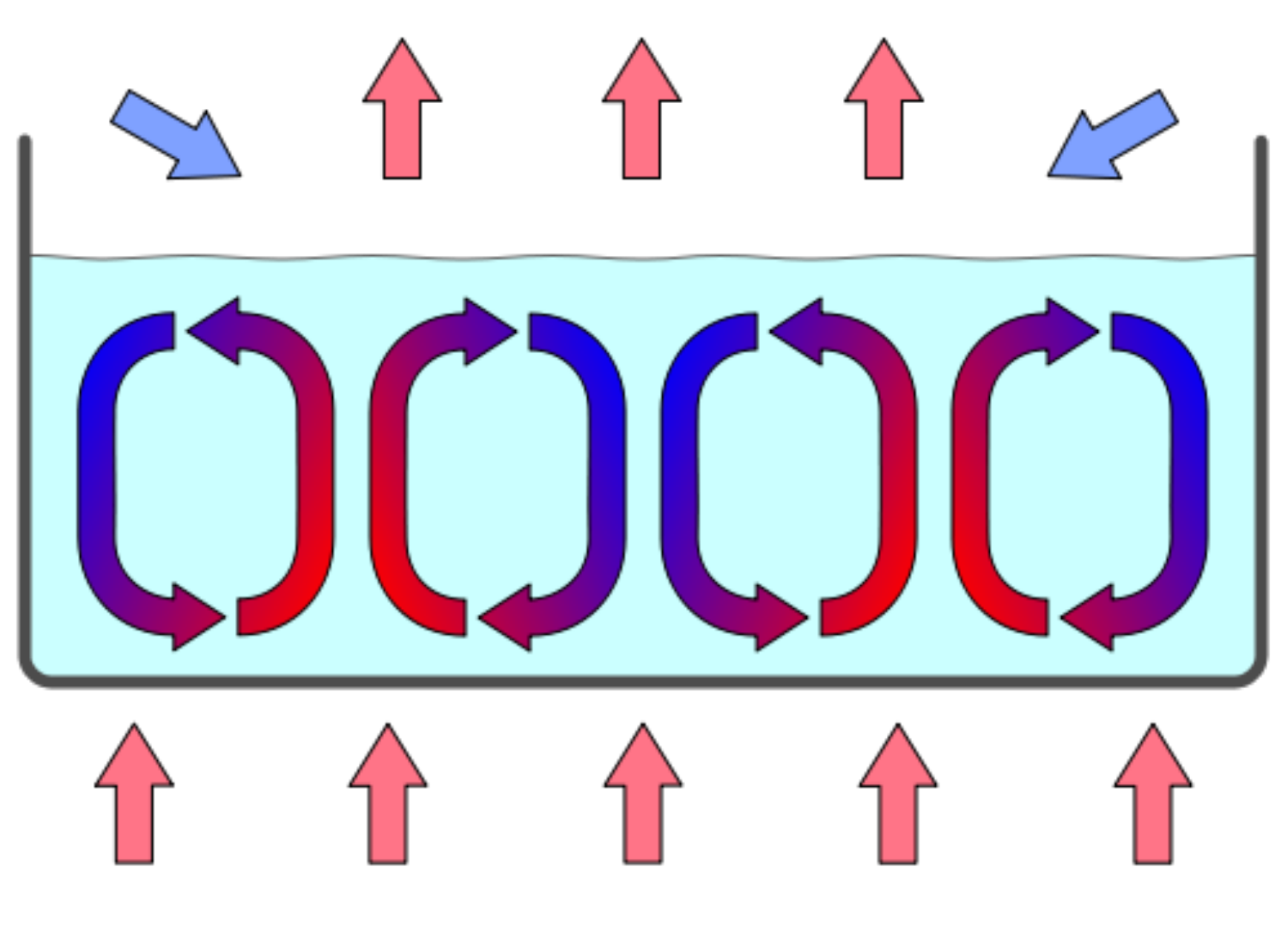}
\end{center}
\caption{\small Schematic representation of Rayleigh-B\'enard convection.  Heat (indicated by the light red, upward pointing arrows) is applied to the bottom of a horizontal fluid layer.  Under the right conditions, the fluid will spontaneously form an array of cells, each containing a regular convection flow.  This circulation can be maintained because more heat is absorbed by the hotter fluid below than is radiated by the colder fluid above. In the curved arrows that represent this flow, higher fluid temperatures are represented in red, and colder ones in blue.  This illustration is taken from \url{http://commons.wikimedia.org/wiki/File:ConvectionCells.svg} \la{fig:Benard}}
\end{figure}

Like in any other self-oscillation, Rayleigh-B\'enard cells begin as a linear instability.  (Chandrasekhar studied this instability in considerable detail in \cite{Chandra-Benard}.)  The subsequent regular convection is a nonlinear, dissipative phenomenon, which takes energy from its environment at a lower entropy and returns it to the environment with higher entropy (see \Sec{sec:entropy}).  Like the action of any other heat engine, Rayleigh-B\'enard convection is maintained because more heat is absorbed by the hotter fluid below than is rejected by the colder fluid above, with the difference available to sustain the macroscopic circulation.

The physics of Rayleigh-B\'enard convection is very lucidly reviewed by Tritton in \cite{Tritton-Benard}.  For a recent treatment of this phenomenon as an example of pattern formation in thermal systems far from equilibrium, see \cite{Cross-Benard}.

\section{Summary and discussion}
\la{sec:summary}

Like the mythical perpetual motion machine, self-oscillation succeeds in driving itself, but does so in a way that is compatible with the known laws of physics. An amusing thought in this regard is that the cranks who insist on building overbalanced wheels and similar devices (see \cite{ToDie}) are simply pursuing what is possible ---indeed, very commonplace--- by means which are already known to be unworkable.\footnote{The na\"ive pursuit of perpetual motion does exert an undeniable fascination, whose precise source is obscure.  In an essay on medieval philosopher Ramon Llull's combinatorial ``thinking machine,'' Borges points out that ``those continual motion devices whose drawings add mystery to the pages of the more effusive encyclopedias do not work either; neither do the metaphysical and theological theories that use to declare who we are and what a thing the world is.'' \cite{Borges}}

Instances of self-oscillation, both useful and destructive, abound in mechanical engineering, music, biology, electronics, and medicine.  Indeed, much of technology ultimately depends on self-oscillation, since only a self-oscillator can turn a steady source of power into a regular periodic motion.

Self-oscillators, as distinct from forced and parametric resonators, can be readily identified by the fact that they sustain large, regular oscillations without an external rate having to be tuned to their frequency: {\it the motion itself} controls the phase of the driving force.  We reviewed, for instance, how a violin works as a self-oscillator, since increasing the velocity with which the bow is drawn simply causes the same note to play more loudly, while the \ae olian harp, on the other hand, is a forced resonator, which rings loudly only when the wind speed happens to give a Strouhal frequency of vortex shedding close to the fundamental tone of the string (or to one of its harmonics).

In his early mathematical modeling of the vocal chords, Airy obtained self-oscillation from a delayed component of the harmonic restoring force.  More generally, self-oscillation can be understood as the result of a component of the driving force that is modulated in phase with the velocity of the displacement.  This gives the device a negative damping, causing the amplitude to grow exponentially with time, until nonlinear effects become significant. 

Clocks and other common self-oscillators work by amplifying the device's vibration and feeding it back in order to drive the oscillator in phase with its velocity.  The amplitude is limited by nonlinearities, as we saw explicitly for the van der Pol equation.  In the regime of large negative damping, the van der Pol oscillator exhibits {\it relaxation oscillation} ---in which the amplitude is fixed and the period is determined, not by the resonant frequency, but rather by nonlinear switching at thresholds--- and it serves as a model of the beating heart and of neuronal firing, among other important phenomena.

We showed that a linear system with more than one degree of freedom can self-oscillate, even if no single mode is negatively damped, as long as the couplings are not symmetric, which is possible only if the degrees of freedom describe perturbations about a non-stationary trajectory.  We then characterized the limit cycles for the simplest nonlinear self-oscillator, the van der Pol equation, and illustrated the phenomena of entrainment, frequency demultiplication, and chaos when a periodic forcing term was added to that equation.

All motors, which take a steady power input and produce a regularly alternating output, are self-oscillators and must waste some of the power, even if they could operate frictionlessly.  We showed how this results from a general Carnot-Le Corbeiller theorem for the limit efficiency of dynamical frequency conversion.  We also showed how the lossless halving of the frequency seen in a parametric resonator is not a violation of this theorem, but rather an instance of a {\it geometric} conversion: in the parametrically-driven pendulum with unit efficiency, the motor pulls up and down in phase with the vertical velocity of the mass, whose frequency is twice that of the pendulum's angular displacement.

We mentioned how the approach of self-oscillators to a nonlinear limit cycle implies that they irreversibly erase information about their initial conditions, thus generating entropy.  This explains why the negative damping of the linearized self-oscillators near their equilibrium does not imply a reversal of the thermodynamic arrow of time.  Indeed, all clocks, being weakly-nonlinear self-oscillators, must themselves generate entropy.  

We briefly reviewed the concept of a servomechanism, emphasizing both their intended operation, based on applying negative feedback to the action of a motor, as well as the possible unwanted self-oscillation about the intended trajectory.  We worked out the Rayleigh criterion for thermoacoustic self-oscillators and applied it to several systems, including Cepheid variable stars.  We discussed how lasers can be described as self-oscillating electromagnetic cavities, and reviewed some ideas (both old and new) on describing the macroeconomic business cycle as a relaxation oscillation.

Self-oscillation is both theoretically interesting and practically useful.  Furthermore, it naturally connects with the mathematics and the history of control theory, since self-oscillation corresponds to the presence of positive feedback (and therefore of a dynamical instability).  We see no excuse for the fact that the subject is hardly taught to physics students and that it remains, for most physicists, in the shadow of the notions of forced and parametric resonance.

Three major open theoretical problems raised in this article obviously call for further investigation as well: a fuller characterization of the operation of self-oscillators (including clocks) from the point of view of nonequilibrium thermodynamics, a general theory of the limits on the efficiency of frequency conversion that incorporates geometrical conversion and extends to both classical and quantum systems, and a better understanding of the mechanism of macroeconomic self-oscillations, including the business cycle.  We also hope that greater conceptual clarity about the energetics of self-oscillation might motivate other useful investigations in both pure and applied physics.

\begin{acknowledgements}

Despite many years first as a physics student and then as a researcher in theoretical physics, I owe my awareness of self-oscillation to a recent series of accidents too complicated to relate here, but which is reflected in the eccentricities of this exposition.  I especially thank Giancarlo Reali for communicating his enthusiasm for the putt-putt boat, for sharing with me the correspondence on the putt-putt that he had received from the late Iain Finnie \cite{Finnie-letter}, and for pointing me to the discussion of the van der Pol oscillator in \cite{Lasers-vdP} and of the ``paradox of Bergeron'' in \cite{Bergeron1}.

I thank Charlie Bennett, Tom Hayes (who also helped me to procure copies of \cite{LeC-review-Fr,Euler-gears1,Euler-gears2}, and \cite{Euler-preface} from the Harvard libraries), Tristan McLoughlin, Take Okui, and Graeme Smith for extended discussions of self-oscillation and related matters, and also Bob Jaffe and Carl Mungan for their critiques of early drafts of this manuscript.  Bob also kindly sponsored my electronic access to the MIT library resources, which greatly facilitated the bibliographical research for this article.

I thank Sharon and Viviana Zlochiver for their guidance on the cardiology of \Sec{sec:heart}, and Andr\'es Marroqu\'{i}n and \'Alvaro Ramos for commenting on the discussion of the business cycle in \Sec{sec:business}.  I also thank Guido Festuccia for his assistance with Euler's Latin in \cite{Euler-gears1,Euler-gears2} and for discussions about the geometry and dynamics of gears.

I thank Sarah Stacey for permission to reproduce the drawing of \Fig{fig:Worcester-wheel}, Paul Horowitz for providing the image for \Fig{fig:LC}, and Olivier Doar\'e and Emmanuel de Langre for the photographs shown in \Fig{fig:gardenhose}.  I also thank Prof.\ de Langre for assistance in comparing the editions of \cite{Bergeron1}.  I thank Greg Miller, of the University of Washington's Department of Civil and Environmental Engineering, for permission to use the photograph in \Fig{fig:Tacoma}, as well as Louis Hand and Janet Scheel ({\it n\'ee} Finch) for permission to use \Fig{fig:top}.

I thank Jenny Hoffman, Bob Jaffe, Howard Georgi, and Take Okui for their encouragement.  I also thank Jonathan Betts, George Levin, Markus Luty, and John McGreevy for discussions, and Igor Bargatin, Mariano Echeverr\'{\i}a, Paul O'Gorman, Nitin Rughoonauth, Wati Taylor, and Ricardo Trujillo for comments, questions, and other sundry assistance.

I thank the NBI Summer Institute, in Copenhagen, for hospitality while some of this work was being completed. This work was supported in part by the US Department of Energy under contract DE-FG02-97IR41022.

\end{acknowledgements}

\appendix
\addappheadtotoc

\section{Josephson effect}
\la{sec:Josephson}

\begin{figure} [t]
\begin{center}
	\includegraphics[width=0.45 \textwidth]{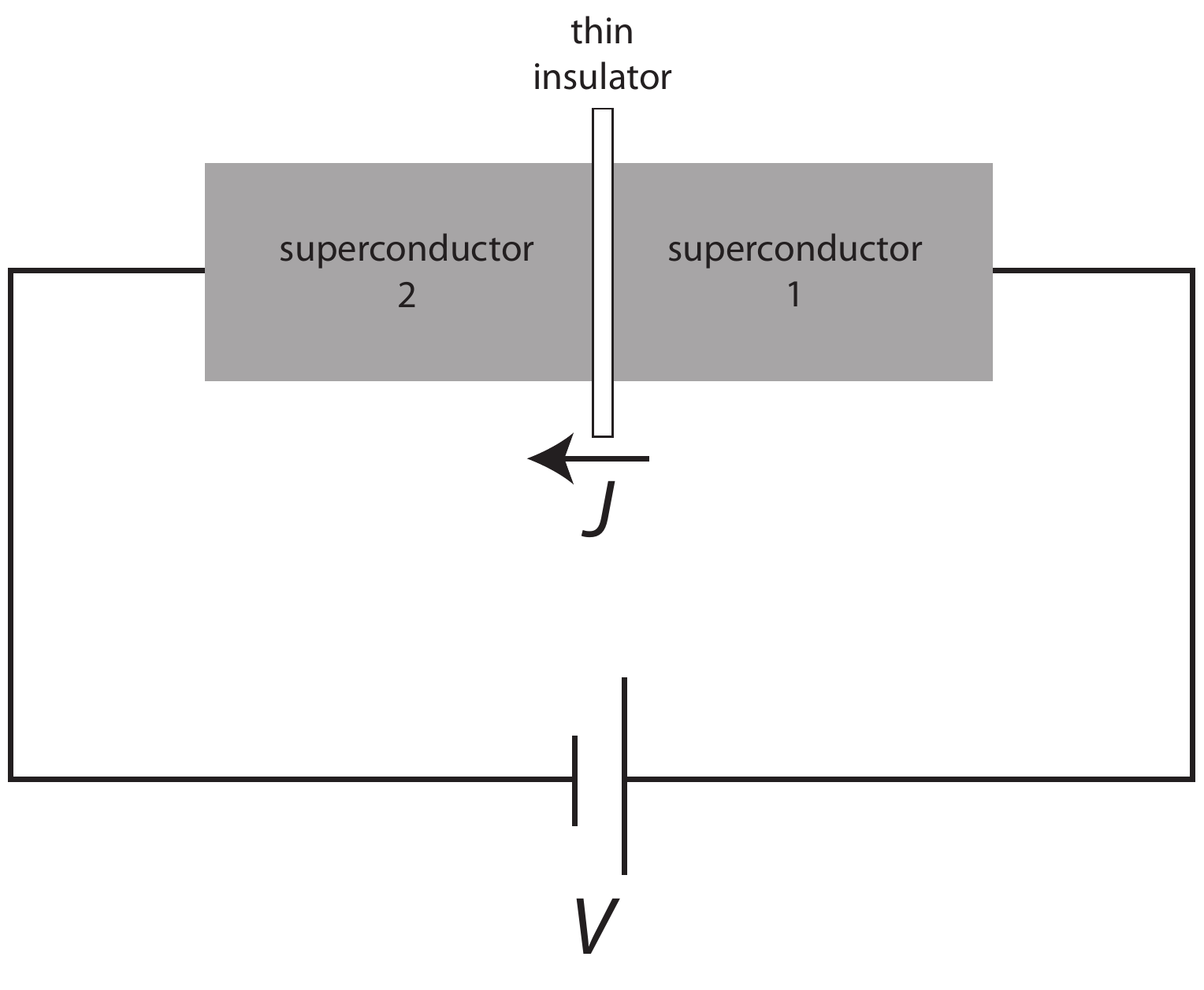}
\end{center}
\caption{\small In a Josephson junction \cite{Josephson}, two superconductors are separated by a thin insulator and kept at a constant voltage difference $V$ by a battery.  For $V=0$ a steady current $J$ can tunnel across the insulator if the relative phase of the quantum states of the superconductors is neither 0 nor $\pi$.  For a fixed $V > 0$ the current $J$ alternates with angular frequency $2eV/\hbar$.\la{fig:josephson}}
\end{figure}

If two superconductors are separated by a thin insulator (a ``Josephson junction,'' shown schematically in \Fig{fig:josephson}), a current
\be
J = J_0 \sin \delta
\la{eq:Josephson-J}
\ee
tunnels across the insulator, where $J_0$ is the maximum current attainable (whose value characterizes the junction) and $\delta$ is the relative phase between the two wavefunctions that describe the collective quantum state of each superconductor. \cite{Josephson}

For a constant voltage difference $V$ between the two superconductors, the relative phase varies at a fixed rate
\be
\dot \delta = \frac{2 e V}{\hbar}~,
\la{eq:Josephson-freq}
\ee
where $e$ is the electron's charge.  Thus, when $V=0$ there will be a non-zero, direct current (DC) as long as the relative phase is neither $0$ nor $\pi$.  For $V > 0$ the alternating current (AC) will oscillate with angular frequency $2 e V / \hbar$.

Feynman beautifully reviews the Josephson effect in \cite{Feynman-Josephson}, where he shows that it is a general feature of the weak coupling of two coherent quantum states.  Although the Josephson effect for $V > 0$ satisfies Andronov's definition of self-oscillation as a macroscopic periodic motion (the AC across the Josephson junction) generated at the expense of a non-periodic source of energy (the battery that maintains the voltage difference between the superconductors), it is qualitatively different from all other systems that we have discussed in this article (including lasers), since its mechanism is strictly quantum-mechanical.  This is obvious in the fact that the expression for the AC frequency as a function of $V$ contains a factor of Planck's constant, $\hbar$.

Nonetheless, Pippard (who was Josephson's graduate advisor at the time of Josephson's discovery) offers a useful classical analogy in \cite{Pippard-Josephson}: Consider two coupled linear oscillators (for instance, two pendula joined by a spring).  If they have the same frequency $\omega$, then no energy will flow between them when they move either in phase ($\delta = 0$) or antiphase ($\delta = \pi$), since those are normal modes of the coupled system (see \cite{Georgi-modes}).  But if the phase takes another value, then energy will slosh back and forth between the oscillators.

In the case of the Josephson effect, each superconductor is held at a constant energy by the battery, so for constant $\delta$ the flow of power between them is steady.  If the oscillators have different frequencies $\omega_{1,2}$, then the relative phase $\delta$ increases at a constant rate $\left| \omega_1 - \omega_2 \right|$, which by \Eq{eq:Josephson-J} is also the angular frequency of the flow of power across the junction.

What quantum mechanics adds to this classical picture is the conversion between frequencies and energies. Indeed, one may think of the DC Josephson effect for $V=0$ as an exchange of quanta of energy
\be
E_0 = \hbar \omega~,
\la{eq:Josephson-E0}
\ee
where $\omega$ is the frequency of the oscillators.  For $V > 0$ the energies of the two superconductors are not matched and the exchange of $E_0$ requires another process in which the energy of the whole system is adjusted by the emission of a quantum 
\be
E' = \hbar \left( \omega_1 - \omega_2 \right) = \hbar \dot \delta~.
\la{eq:Josephson-omega12}
\ee
Physically, this emission can only correspond to the change in energy of a superconducting Cooper pair (with charge $2e$) as it moves across the junction voltage, so that
\be
E' = 2 e V~,
\la{eq:Josephson-2eV}
\ee
leading to the result of \Eq{eq:Josephson-freq}.

Thus we see that the AC Josephson effect is more akin to quantum emission (in which a constant energy splitting $\Delta E$ is turned into radiation of frequency $\Delta E / \hbar$) than it is like the other self-oscillations described in this article; on this point, see also \cite{Lasers-Josephson}.  The Josephson AC is a macroscopic effect because a superconductor is in a macroscopic, coherent quantum state.  Note as well that the AC across the junction turns on immediately when the external $V$ is applied, without the exponential growth of small oscillations followed by an approach to a nonlinear limit cycle that we saw in classical self-oscillators.  (Indeed, since the Schr\"ondinger equation is linear, no purely quantum process can be described by a limit cycle.)

On the other hand, the emission described by \Eq{eq:Josephson-omega12} appears to be somewhat analogous to the ``tail'' loss that we argued had to accompany the classical interconversion of frequencies; see \Sec{sec:Carnot}.  The fact that for $\omega_1 = 2 \omega_2$ we have $\dot \delta = \omega_2$ suggests that lossless parametric down-conversion in quantum optics (see \cite{down-conversion}) might not be forbidden if $\omega_1 / \omega_2 = 2$.  Whether this is so, and the possible connection to the classical arguments about ``geometric conversion'' made in Secs.~\ref{sec:parametric} and \ref{sec:Carnot}, deserve further investigation.

\section{Historical aper\c{cu}}
\la{sec:history}

The practical uses of self-oscillators are very ancient and, as we emphasized in \Sec{sec:motors}, are an intrinsic aspect of human technology.  But the theoretical question of how a steady source of power can produce and maintain a periodic motion seems not to have been posed in the context of Newtonian mechanics until the late 1820s, with the work of Robert Willis \cite{Willis-larynx} and G.~B.~Airy \cite{Airy-perpetual} on the operation of the human voice.

Even though Airy went on to become a long-serving Astronomer Royal, and Willis the Jacksonian Professor of Natural Philosophy at Cambridge, this early work was totally ignored.  As far as we have been able to discover, only the anatomical component of Willis's research on the operation of the larynx attracted any attention, while Airy's paper was cited only by Henry Dircks, a self-taught engineer, in the context of his exhaustive but eccentric investigation into the history of perpetual motion research \cite{Dircks-Airy}, published in 1861.\footnote{Dircks cited Airy as a last-minute addition to the manuscript and without evidence of understanding his argument; he dropped all reference to Airy from the ``second series,'' published in 1870 \cite{Dircks2}.  It was presumably through Dircks's book that R.~T.~Gould learned of Airy's paper (see \Sec{sec:overbalanced}).}  Amusingly, in \cite{T&T-perpetual} Kelvin and Tait originally dismissed the possibility of linear self-oscillation as an unphysical perpetual motion (see \Sec{sec:linear}).

The theory of self-oscillation begins anew with Maxwell's research into the stability of machines controlled by governors \cite{Maxwell}.  Maxwell posed the problem as a mathematical puzzle, which was later solved in generality by Routh \cite{Routh} and Hurtwitz \cite{Hurwitz}, leading to a criterion for the stability of linear systems (see \Sec{sec:stability}).  This line of investigation leads eventually to modern control theory, primarily via the work of engineers and mathematicians.  Its focus has always been on the conditions for a servomechanism to keep to its intended trajectory.  The connection to the problem of how the motor that drives such a mechanism can run in the first place was not pursued.  If it had been, then perhaps the study of self-oscillation might have influenced the development of thermodynamics.  As it was, the link was not made explicitly until 1936 by Le Corbeiller \cite{LeC-review-Eng} and his tentative ideas were then left unexplored (see \Sec{sec:motors}).\footnote{It is intriguing that the young Willard Gibbs should have investigated the efficiency of power transfer in gears \cite{Gibbs-frictionless} (see \Sec{sec:mechanisms}) and hydraulic turbines \cite{Gibbs-turbine}, as well as the stability of steam engine governors \cite{Gibbs-governor} (see \Sec{sec:linear}), before he started to work on thermodynamics in the 1870s.  Gibbs's retiring personality and the highly abstract and parsimonious style of his writing make it difficult to determine to what extent his early work in engineering motivated or influenced his mature research on thermodynamics.  See \cite{Gibbs-Wheeler} for a discussion of this question.}

The first edition of Lord Rayleigh's {\it Theory of Sound}, whose two volumes were published in 1877 and 1878 respectively, contains no discussion of self-oscillation, except for the brief comment on the role of delayed action in the operation of Helmholtz's ``fork-interrupter'' \cite{Rayleigh-fork}, and of velocity-dependent friction on the powering of the vibration of a violin string by the bow \cite{Rayleigh-violin}.  In his review of the second volume of Rayleigh's treatise, Helmholtz called attention to this deficiency:
\begin{quote}
There is still an important chapter wanting, viz., that on the theory of reed-pipes, including the human voice [...] Altogether, the whole of this important class of motions, where oscillatory movements are kept up through a cause which acts constantly, deserves detailed theoretical consideration. \cite{Helmholtz-review}
\end{quote}
It might well have been Helmholtz's urging that led Rayleigh to his investigations into self-oscillation \cite{Rayleigh-vdP,Rayleigh-criterion} and parametric resonance \cite{Rayleigh-parametric}, which were eventually incorporated into the second edition of the {\it Theory of Sound} \cite{Rayleigh-ToS-maintained}, published in 1894--1896.\footnote{Even though Helmholtz \cite{Helmholtz-KH} and Kelvin \cite{Kelvin-KH} had worked out the instability of the surface of separation between two fluid layers with a sufficient relative velocity (see \Sec{sec:flow-induced}), thus explaining the self-oscillation of the waves formed by the wind on the surface of the sea and other bodies of water, the connection was not made by either of them to ``maintained oscillators'' such as clocks and wind musical instruments.}

It is odd that Rayleigh's work on self-oscillation should have attracted so little attention for so long, even among acousticians, who have continued to cite Rayleigh's book even to this day, but not usually on that subject.\footnote{It is also telling that Eddington should have had to rediscover Rayleigh's criterion when he became interested in the thermodynamics of the pulsation of Cepheid variables (see \Sec{sec:cepheids}).}  Van der Pol's first paper on electrical self-oscillation \cite{vdP}, published in 1920, cites Rayleigh in a surprisingly narrow context, considering that van der Pol's nonlinear equation of motion is equivalent to the one proposed by Rayleigh in \cite{Rayleigh-clocks,Rayleigh-vdP}.

In the late 19th century and early 20th centuries, the development of electrical amplifiers for telegraphy and radio provided a new impetus for the theory of self-oscillation.  A major pioneer in this field was the French engineer Andr\'e Blondel (1863--1938), who coined the term {\it oscillations autoentretenues} (``self-maintained oscillations'') and who explicitly invoked the Routh-Hurwitz criterion in order to identify systems capable of self-oscillation \cite{Blondel-auto}.  On Blondel's work on this subject, its context, and its influence on later research, see \cite{Blondel-history}.

Active electric circuits provided scientists like Balthasar van der Pol with a convenient way of simulating nonlinear equations of motion whose analytic solution could only be obtained laboriously and approximately.  The work of van der Pol and his collaborators on self-oscillators attracted much attention, in the 1930s and beyond, from mathematicians, electrical engineers, physiologists, and even economists, but not from physicists.  The research by Andronov (who coined the term ``self-oscillation'') and his associates in the Soviet Union was largely mathematical ---in the tradition of Poincar\'e and Lyapunov--- and had its greatest impact in the theory of ordinary differential equations and in control engineering (see, e.g., \cite{Andronov-history,history-chaos,Arnold}).

Research on electric circuits with amplifiers also gave rise to the concept of {\it feedback}.  Already in 1909, wireless communication pioneer Ferdinand Braun casually used that term to describe the possible disruption of an electric oscillation made to drive a resonator like the one in \Fig{fig:LC} \cite{Braun}.  Harold Black's development of the negative feedback amplifier in the 1930s \cite{op-amp} greatly increased the importance of the subject in electrical engineering, and Nyquist's formulation of the stability criterion in terms of the feedback loop gain as a function of frequency \cite{Nyquist} became a central idea in control engineering: see \cite{AM-Nyquist} and references therein.  As a broad theoretical concept, feedback was much emphasized in the 1950s by mathematician Norbert Wiener, in the context of what he termed ``cybernetics'' \cite{Wiener-feedback1,Wiener-feedback2}.\footnote{Wiener's cybernetics also motivated work on a ``general systems theory,'' mainly by theoretical biologists and sociologists.  This grew into an active but diffuse academic pursuit in the second half of the 20th century: see \cite{Luhmann,Francois}, and the references therein.  The intellectual rigor and fruitfulness of this approach has been challenged from various different perspectives (cf.\ \cite{Anderson,Bricmont,Habermas,Berlinski}).}  Qian Xuesen's application of cybernetics to the problem of missile control then helped lay the conceptual groundwork for robotics.  According to Qian:

\begin{quote}
A distinguishing feature of this new science [i.e., cybernetics] is the total absence of considerations of energy, heat, and efficiency, which are so important in other natural sciences.  In fact, the primary concern of cybernetics is on the qualitative aspects of the interrelations among the various components of a system and the synthetic behavior of the complete mechanism. \cite{Qian}
\end{quote}

The issue of the flow of energy in self-oscillators seems to have received comparatively little attention.  Initially, this might have been due to a widespread impression that the study of nonlinear equations of motion was a technically challenging problem, unlikely to shed light of the sort of fundamental questions that might interest a theoretical physicist.  What is perhaps more puzzling is that the explosion in interest in nonlinear dynamical systems that was engendered by the advent of fast and cheap digital computers and by the ascendance of chaos theory after the 1970s should have largely failed to call the attention of physicists back to these questions, except in the certain narrow contexts, such as the mathematical characterization of limit cycles and their entrainment (see \Sec{sec:limits}).

\section{Note on sources}
\la{sec:sources}

In this review of self-oscillation, we have made an effort to give all relevant references, both to the original research reports and to textbooks.  For a physics student wishing to learn the subject systematically, further guidance is in order.

Lord Rayleigh's justly celebrated {\it Theory of Sound} \cite{ToS} covers almost all topics of importance in mechanical vibrations and acoustics.  Unfortunately, the age, length, and organization of Rayleigh's treatise may limit its usefulness for modern students.  Note that, as explained in Appendix \ref{sec:history}, the theory of self-oscillation (which he calls ``maintained vibration'') is covered in the second edition, but not in the first.  

At this time, there does not appear to be any textbook treatment of self-oscillation suitable for undergraduate physics students, and even leading contemporary textbooks in acoustics (e.g., \cite{Pierce,Sound}) ignore the subject altogether.  Perhaps the best introductory sources for physics students are the review of ``self-sustaining oscillators'' in \cite{Pikovsky-review}, as well as the pedagogical articles by Billah and Scalan \cite{Billah}, and by Green and Unruh \cite{Unruh} on the Tacoma Narrows bridge collapse, which make a clear conceptual distinction between forced resonance and self-oscillation.  As noted in \Sec{sec:Rijke}, the Rijke tube provides an impressive demonstration of self-oscillation that can be easily conducted during a classroom lecture.  (Pippard gives a clear and complete explanation of the operation of the Rijke tube in \cite{Pippard-Rijke}.)

Self-oscillation is covered by Sargent, Scully, and Lamb in \cite{Lasers-vdP} (where it is called ``sustained'' oscillation), by Pippard in \cite{Pippard-Maintained,Pippard-locking} (where it is called ``maintained'' oscillation), and by Fletcher and Rossing in \cite{F&R-nonlinear} (where it is called ``self-excited'' oscillation).  These treatments all use the method of slowly varying amplitude and phase (or ``method of averaging''), developed by van der Pol \cite{vdP-review} and by Krylov and Bogolyubov \cite{K&B,B&M} for the study of nonlinear vibrations; see also \cite{Andronov-nonlinear}.

{\it Selected Papers on Mathematical Trends in Control Theory} \cite{ControlThy} reproduces several important primary documents directly related to self-oscillation, including the work of Maxwell \cite{Maxwell}, Hurwitz \cite{Hurwitz}, Nyquist \cite{Nyquist}, and van der Pol \cite{vdP-locking}, each prefaced by a short explanation of its significance.  It also includes Bateman's thorough historico-mathematical review of linear stability analysis \cite{Bateman}, followed by a brief overview of the work of Poincar\'e and Lyapunov on the stability of the trajectories of nonlinear dynamical systems \cite{Lyapunov} (see \Sec{sec:Lyapunov}).  For a modern and thorough textbook treatment of control theory, see \cite{Astrom-Murray}.

The review articles written in the 1930s by van de Pol \cite{vdP-review} and Le Corbeiller \cite{LeC-review-Fr,LeC-review-Eng} are still worth reading.  Also valuable is the brief overview presented by Le Corbeiller before the first meeting of the Econometric Society \cite{LeC-econ}, convened at Lausanne in 1931, from which we quoted in \Sec{sec:business}.  (Unfortunately, \cite{LeC-review-Fr} and \cite{LeC-econ} are only available in French.)  Groszkowski compiled a very comprehensive bibliography of the research on self-oscillation published up to 1962, with emphasis on electrical engineering \cite{Groszkowski-bibl}, but, unfortunately, those works focus on electrical oscillators with vacuum tube amplifiers, which are now obsolete.

Soviet scientists A.~A.~Andronov, A.~A.~Vitt, and S.~\`E.~Kha\u{\i}kin produced one of the fullest available accounts of self-oscillation in the second edition of their monumental {\it Theory of Oscillators}.  This is available in English in an edition revised and abridged by Wilfred Fishwick in 1966 \cite{Andronov}.  (The first edition appeared in Russian in 1937 without any acknowledgment of Vitt's contribution, as Vitt had been imprisoned and executed during Stalin's Great Purge \cite{Arnold-Vitt,Andronov-history}.)  V.~I.~Arnol'd's {\it Ordinary Differential Equations} \cite{Arnold} admirably combines mathematical rigor with physical intuition and can serve as an excellent point of contact with the mathematical and engineering literature based on Poincar\'e's ``qualitative theory of differential equations,'' Lyapunov stability, and differential topology (see \Sec{sec:Lyapunov}).  {\it Nonlinear Oscillations}, by Nayfeh and Mook, is a standard modern textbook on the subject for engineers \cite{Nayfeh}, in which self-oscillators are identified as ``self-sustaining systems.''

Norbert Wiener's {\it Cybernetics} \cite{Wiener-feedback1} offers a fascinating perspective on some of the concepts covered in this article, though it is hardly a systematic treatment and appears to have been most influential in disciplines other than physics.  Sir Brian Pippard presented another interesting and idiosyncratic treatment of these topics in {\it Response and stability} \cite{Pippard-Response}.  The description of lasers as self-oscillators by Willis Lamb and his collaborators, cited in \Sec{sec:lasers}, are also pedagogically quite valuable.

Many authors have written books for lay readers on aspects of the nonlinear theory of dynamical systems, especially chaos theory.  Steven Strogatz's {\it Sync} \cite{Sync} emphasizes some of the topics covered in this article, particularly about entrainment and chaos in relaxation oscillators (see Secs.~\ref{sec:entrainment}, \ref{sec:chaos}, and \ref{sec:forced-SO}).

%%%%%%%%%%
%%% REFERENCES
%%%%%%%%%%

\bibliographystyle{aipprocl}   % if natbib is available

\end{document}